\documentclass[journal]{IEEEtran}

\ifCLASSINFOpdf

\else

\fi

\usepackage{cite}
\usepackage{graphicx}
\usepackage{amsmath}
\usepackage{caption}
\usepackage{subcaption}
\usepackage{flushend}
\usepackage{algpseudocode}
\usepackage[linesnumbered,ruled,vlined,commentsnumbered]{algorithm2e}
\usepackage{lipsum}
\usepackage[font=footnotesize]{caption}
\usepackage{graphicx}
\usepackage{subcaption}
\usepackage{amssymb}
\usepackage{gensymb}
\usepackage{wasysym}
\usepackage{textcomp}
\usepackage{color}
\usepackage{soul}
\usepackage[table]{xcolor}
\usepackage{multirow}
\usepackage{pifont}
\usepackage{longtable}
\usepackage{quantikz}
\usetikzlibrary{shapes.geometric, arrows}
\usepackage[bottom]{footmisc}
 \usepackage{tabularx}

\usepackage{multirow}
\usepackage{array}
\newcolumntype{P}[1]{>{\centering\arraybackslash}p{#1}}
\usepackage{longtable}

\usepackage{blindtext} % to generate dummy text

\usepackage{mathtools}
\DeclarePairedDelimiterX{\norm}[1]{\lVert}{\rVert}{#1}

\makeatletter
\renewcommand{\@algocf@capt@plain}{above}% formerly {bottom}

\newcommand{\stkout}[1]{\ifmmode\text{\sout{\ensuremath{#1}}}\else\sout{#1}\fi}
\usepackage[normalem]{ulem}

\begin{document}

\title{Quantum Computing in Next-Gen Smart Grid Operations: A Comprehensive Review}

\author{ Md Habib Ullah,~\IEEEmembership{Member,~IEEE} \\
\thanks{Md Habib Ullah is with the Dept. of Electrical Engineering at Penn State Harrisburg, Middletown, PA 17057, USA. Email: mzu5071@psu.edu. }
}

\maketitle
\thispagestyle{plain}
\pagestyle{plain}

\begin{abstract}
The rapid proliferation of grid-edge distributed energy resources has significantly increased the operational complexity of modern power systems. Consequently, conventional computational techniques face growing scalability and computational-efficiency challenges in addressing large-scale optimization and control, uncertainty management, nonlinear dynamics, and combinatorial decision-making in smart grid operations. Quantum computing has therefore emerged as a promising computational paradigm that can complement classical methods in addressing selected computationally intensive problems. In this context, this paper presents a comprehensive structured review of quantum computing applications in smart grid operations. Following a transparent keyword-based literature search, the paper classifies and assesses existing studies across monitoring and estimation, system planning, operation and control, security, reliability and resilience, stability assessment, data-driven intelligence, and digital twin technologies. The paper also describes fundamental quantum-computing concepts and key algorithms, highlighting their relevance for power system applications. Furthermore, it reviews the current state of quantum hardware, software frameworks, simulators, cloud services, and emerging hardware-agnostic ecosystems that support cross-platform application development and deployment. The reviewed studies are examined by implementation environment, benchmarking practices, application scale, and evidence of computational advantage. Finally, the principal challenges associated with practical implementation are discussed, and future research directions are outlined. This paper provides a consolidated and evidence-calibrated perspective on the current state and future potential of quantum computing for next-generation smart grid operations.
\end{abstract}

\IEEEpeerreviewmaketitle
\vspace{-6pt}
\section{Introduction}
The modern electric power grid is undergoing a profound transformation toward a decentralized, digitized, and intelligent cyber–physical infrastructure, commonly referred to as the smart grid \cite{liu2025smart}. This transition is primarily driven by global decarbonization goals, the electrification of transportation and industry, and the rapid deployment of renewable energy resources. As a result, distributed energy resources (DERs), including photovoltaic systems, wind turbines, battery storage, electric vehicles, and demand-side technologies, are being integrated into the grid at an unprecedented scale \cite{cavus2025advancing, kurella2025comprehensive}. While this evolution enhances sustainability, resilience, and operational flexibility, it simultaneously introduces significant complexity. Unlike traditional centralized generation, DERs are inherently distributed, intermittent, and often coordinated through decentralized control, leading to increased uncertainty, variability, and interdependencies across the network. These characteristics give rise to highly stochastic, nonlinear, and spatially distributed system behaviors, which complicate critical tasks such as monitoring, control, optimization, and multi-scale decision-making \cite{khan2025comprehensive}. As DER penetration continues to grow, the associated computational burden is expected to expand substantially. Core functions, including optimal power flow, unit commitment, contingency analysis, state estimation, and others, now involve large-scale, high-dimensional, and nonlinear problems that often scale combinatorially with system size. In addition, emerging smart grid applications, including peer-to-peer energy trading, demand response coordination, digital twins, and cyber–physical security, require fast, reliable, and scalable computational frameworks that operate in real time under uncertainty. However, conventional computational methods are increasingly strained in their ability to scale to meet these growing demands, underscoring the need for fundamentally new computational approaches \cite{assad2022smart, patari2021distributed, pavon2023review}.

Driven by these growing computational challenges, quantum computing has emerged as a promising and fundamentally distinct paradigm for smart grid operations. By leveraging principles of quantum mechanics, most notably superposition, entanglement, and interference, quantum systems can represent and process information in ways that are unattainable with classical approaches \cite{preskill2018quantum}. In particular, qubits enable the encoding of exponentially large state spaces, allowing quantum algorithms to explore multiple solution pathways simultaneously \cite{nielsen2010quantum}. This capability offers substantial computational advantages for key problem domains in power and energy systems. For instance, quantum algorithms such as the Harrow–Hassidim–Lloyd (HHL) method \cite{harrow2009quantum} provide accelerated solutions for large systems of linear equations arising in applications such as power flow analysis, state estimation, and contingency analysis. In parallel, algorithms tailored for combinatorial optimization, particularly the Quantum Approximate Optimization Algorithm (QAOA) \cite{farhi2014quantum}, are well-suited for addressing discrete decision-making problems in smart grids. These include unit commitment, network reconfiguration, resource placement, demand response, and generation scheduling, where the solution space grows combinatorially with system size. Moreover, the emergence of hybrid quantum–classical frameworks has enabled the practical use of quantum techniques on current noisy intermediate-scale quantum (NISQ) devices, facilitating early-stage applications in smart grid analytics and operations.

The increasing interest in quantum computing is rooted in its rich historical development, which spans several decades of theoretical and technological progress. The concept of using quantum systems for computation was first proposed by Feynman in the early 1980s, who suggested that they could simulate physical processes more efficiently than classical computers \cite{feynman2018simulating}. Subsequent theoretical advances, including Deutsch’s model of a universal quantum computer, laid the groundwork for algorithmic breakthroughs in the 1990s \cite{deutsch1985quantum}. Notably, Shor’s algorithm demonstrated exponential speedup for integer factorization, and Grover’s algorithm provided quadratic speedup for unstructured search problems, establishing the computational potential of quantum systems \cite{shor1994algorithms, grover1996fast}. More recently, the transition to the NISQ era has marked a significant milestone, characterized by the availability of programmable quantum processors with tens to thousands of qubits. Although these systems are limited by noise and decoherence, they have catalyzed the development of variational and hybrid algorithms designed to operate under realistic hardware constraints. Ongoing research efforts in fault-tolerant quantum computing are expected to further unlock large-scale, practical applications.

\begin{figure}[t!] % "[t!]" placement specifier just for this example
\centering
\begin{subfigure}{0.456\textwidth}
\includegraphics[width=1\linewidth]{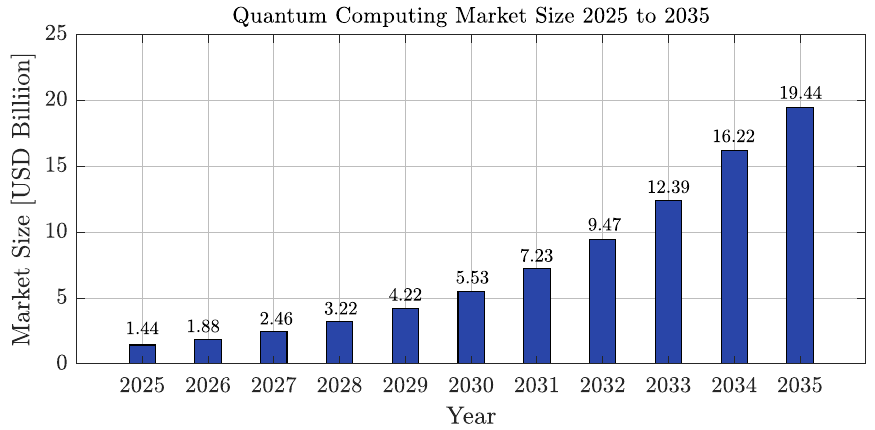}
\caption{} \label{fig:bl23_speeda}
\end{subfigure}\hspace{\fill}
\begin{subfigure}{0.456\textwidth}
\includegraphics[width=1\linewidth]{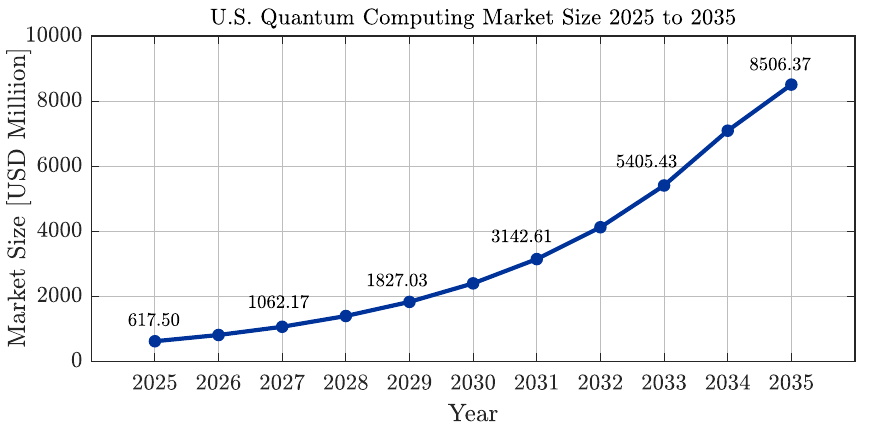}
\caption{} \label{fig:bl23_speedb}
\end{subfigure}
\caption{Quantum computing market size projections \cite{Precedence_Research}. (a) Global market. (b) U.S. market.}
\label{Market_projection}
\vspace{-16pt}
\end{figure}

In parallel with these technological advancements, the global quantum computing ecosystem has experienced rapid growth and strategic investment. Governments, academic institutions, and industry leaders are heavily investing in quantum technologies, recognizing their potential impact across multiple sectors, including energy, finance, healthcare, and national security \cite{nstc2018quantum, eu2018flagship, mckinsey2026quantum}. Market projections, as shown in Fig. \ref{Market_projection}, indicate a substantial expansion of the quantum computing industry over the next decade, with both global and U.S. markets expected to grow significantly due to increasing commercialization and adoption. Major quantum technology companies, such as IBM, Google, Microsoft, Rigetti, D-Wave, IonQ, and others, are actively developing a range of quantum hardware platforms, including superconducting circuits, trapped-ion systems, photonic devices, and neutral-atom architectures \cite{gyongyosi2019survey}. At the same time, the emergence of cloud-based quantum computing platforms, such as IBM Quantum, AWS Braket, and Azure Quantum, has democratized access to quantum resources to a certain extent, enabling researchers and practitioners to experiment with quantum algorithms without requiring direct access to specialized hardware \cite{oukaira2025quantum}. Complementary software frameworks, including Qiskit, Cirq, and PennyLane, further support the development and deployment of quantum applications across different domains \cite{shafique2024quantum}.

    %\begin{figure*}[htbp]
    %\centering
    %\includegraphics[scale=0.75]%{Figures/TQE_Paper_Organization.drawio_New_Version.pdf} 
    %\caption{Paper organization.}
    %\label{fig. 2}
    %\vspace{-10pt}
    %\end{figure*}

Despite these promising advancements, it is important to recognize that applying quantum computing to smart grids remains an emerging field with several unresolved challenges. Limitations related to hardware noise, limited qubit counts, problem-encoding complexity, and algorithmic maturity continue to constrain the scalability and reliability of quantum solutions. In addition, the practical deployment of these methods in real-world power systems will require seamless integration with classical computing frameworks, robust error mitigation techniques, and standardized benchmarking methodologies to accurately evaluate performance gains. Nevertheless, the rapid pace of progress in both hardware and algorithm development suggests that quantum computing could play an increasingly important role in future power system applications. Realizing this potential will require sustained interdisciplinary collaboration among power engineers, computer scientists, and physicists. Against this backdrop, there is a pressing need for a comprehensive and structured review that consolidates existing knowledge, identifies research gaps, and outlines future directions in this interdisciplinary domain. Although several review studies on quantum computing for smart grid applications have been reported in the early literature \cite{golestan2023quantum, li2025application, ullah2022quantum, ganeshamurthy2024next, morstyn2024opportunities, zhou2022quantum, eskandarpour2020quantum, giani2021quantum, zhang2024review, chen2025review, jang2024review, tightiz2026next, strata2025quantum, corli2025quantum}, they tend to emphasize specific aspects, such as quantum algorithms, quantum machine learning, or particular power and energy system applications. Furthermore, many of these studies capture only a limited portion of the rapidly growing literature and, therefore, do not fully reflect recent advancements across diverse smart grid domains. These limitations may restrict the development of a unified perspective that seamlessly integrates algorithms, hardware platforms, software ecosystems, and the full spectrum of modern smart grid applications, thereby reinforcing the need for a more cohesive overview of the field.

In this context, this paper presents a comprehensive structured review of quantum computing applications in smart grid operations. The primary contributions of this work are as follows. \textit{First}, it provides a detailed overview of quantum computing fundamentals and algorithms relevant to power system applications. \textit{Second}, it provides a structured survey of the application of quantum techniques across a wide range of smart grid problems, including system monitoring and planning, operation and control, security, reliability and resilience, stability assessment, data-driven intelligence, and digital twin technologies. \textit{Third}, it analyzes the current state of the quantum ecosystem, including global market trends, hardware development, simulators, software frameworks, and hardware-agnostic platforms. \textit{Fourth}, it identifies key challenges and limitations that must be addressed to enable practical deployment. \textit{Finally}, it outlines future research directions and a roadmap for integrating quantum technologies into next-generation (next-gen) smart grid operations. By bridging the gap between quantum computing theory and power system applications, this paper aims to serve as a foundational reference for researchers and practitioners working at the forefront of this rapidly evolving field. 

    \begin{figure}[t!]
    \centering
    \includegraphics[scale=0.65]{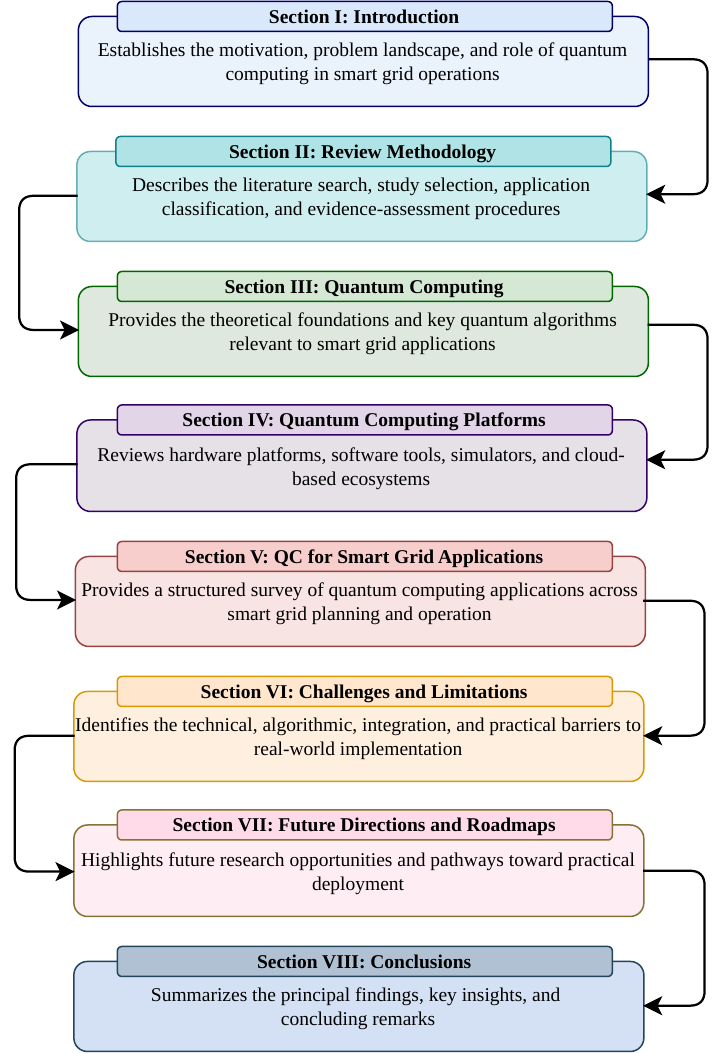} 
    \caption{Paper organization. QC: Quantum Computing.}
    \label{Paper_Organization}
    \vspace{-8pt}
    \end{figure}

The remainder of the paper is organized as follows: Section \ref{Review Methodology} describes the review methodology, including the literature search, study selection, data extraction, application classification, and evidence-assessment procedures. Section \ref{Quantum Computing Fundamental} presents the fundamentals of quantum computing and the associated algorithms for smart grid operations. Section \ref{Quantum Computing Platforms} reviews quantum computing platforms, including hardware technologies, software tools, simulators, and cloud-based ecosystems. Section \ref{QC Smart Grid Applications} provides a comprehensive survey of quantum computing applications in smart grid planning and operation. Section \ref{Challenges and Limitations} discusses the challenges and limitations associated with current quantum technologies, and Section \ref{Future Directions} outlines future research directions and development roadmaps. Finally, Section \ref{Conclusion} concludes the paper with key insights and findings. The overall organization of the paper is illustrated in Fig. \ref{Paper_Organization}. The frequently used acronyms in this paper and their abbreviations are presented in Table \ref{tab:nomenclature}.

\begin{table}[t!]
\caption{List of acronyms and abbreviations.}
\label{tab:nomenclature}
\centering
\renewcommand{\arraystretch}{1.1}
\begin{tabular}{ll}
\hline
\textbf{Acronym} & \textbf{Full Name} \\
\hline
ADMM & Alternating Direction Method of Multipliers \\
ANN & Artificial Neural Network \\
AQC & Adiabatic Quantum Computing \\
BiLSTM & Bidirectional Long Short-Term Memory \\
BMS & Battery Management System \\
BQM & Binary Quadratic Model \\
CNN & Convolutional Neural Network \\
DA & Digital Annealer \\
DAE & Differential-Algebraic Equation \\
DER & Distributed Energy Resource \\
DR & Demand Response \\
DT & Digital Twin \\
EMS & Energy Management System \\
EMT & Electromagnetic Transient \\
EMTP & Electromagnetic Transients Program \\
EV & Electric Vehicle \\
EVCSP & Electric Vehicle Charging Station Placement \\
FDIA & False Data Injection Attack \\
GRU & Gated Recurrent Unit \\
GSA & Grover's Search Algorithm \\
HHL & Harrow--Hassidim--Lloyd \\
LSTM & Long Short-Term Memory \\
MILP & Mixed-Integer Linear Programming \\
MINLP & Mixed-Integer Nonlinear Programming \\
MIQP & Mixed-Integer Quadratic Programming \\
MPC & Model Predictive Control \\
NISQ & Noisy Intermediate-Scale Quantum \\
NR & Newton--Raphson \\
OPF & Optimal Power Flow \\
P2P & Peer-to-Peer \\
PF & Power Flow \\
PMU & Phasor Measurement Unit \\
PQC & Parameterized Quantum Circuit \\
QA & Quantum Annealing \\
QAE & Quantum Amplitude Estimation \\
QAOA & Quantum Approximate Optimization Algorithm \\
QCNN & Quantum Convolutional Neural Network \\
QDC & Quantum Distributed Control \\
QEC & Quantum Error Correction\\
QFT & Quantum Fourier Transform \\
QKD & Quantum Key Distribution \\
QML & Quantum Machine Learning \\
QNN & Quantum Neural Network \\
QPCA & Quantum Principal Component Analysis\\
QPE & Quantum Phase Estimation \\
QRAO & Quantum Random Access Optimization \\
QRL & Quantum Reinforcement Learning \\
Q-SAVLR & Quantum Surrogate Absolute-Value Lagrangian Relaxation\\
QSVT & Quantum Singular Value Transformation \\
QUBO & Quadratic Unconstrained Binary Optimization \\
%QVNN & Quantum Variational Neural Network \\
RQBN & Restricted Quantum Bayesian Network\\
RL & Reinforcement Learning \\
RNN & Recurrent Neural Network \\
RTDS & Real-Time Digital Simulator \\
SA & Simulated Annealing \\
SDK & Software Development Kit \\
SoH & State of Health \\
SVM & Support Vector Machine \\
VQC & Variational Quantum Circuit \\
VQE & Variational Quantum Eigensolver \\
VQNN & Variational Quantum Neural Network \\
VQLS & Variational Quantum Linear Solver \\
\hline
\end{tabular}
\vspace{-16pt}
\end{table}

\section{Review Methodology} \label{Review Methodology}
This paper employs a comprehensive structured-review methodology to identify, classify, and assess studies on quantum computing applications in smart grid planning and operation. Because this is a new and rapidly evolving research area, the search was keyword-based, with no predefined lower publication-year limit. The search was updated through July 2026, and the earliest eligible application-oriented study identified was published in 2015. However, most included studies were published from 2020 onward, reflecting the field’s rapid growth in recent years.

\subsection{Literature Search and Study Selection}
The literature search used Google Scholar, IEEE Xplore, Scopus, Web of Science, ScienceDirect, SpringerLink, and arXiv. Quantum-related terms, including “quantum computing,” “quantum annealing,” “quantum optimization,” “quantum machine learning,” “hybrid quantum-classical computing,” “QAOA,” “VQE,” “VQLS,” and “HHL,” were combined with “smart grid,” “power system,” “electric grid,” and the individual application names examined in this paper. The database searches were supplemented by examining the references of relevant reviews and primary studies.
In this paper, journal articles, conference papers, preprints, theses/dissertations, and book chapters published in English were considered. Studies were included when they presented a substantive quantum, hybrid quantum-classical, quantum-annealing, or explicitly quantum-inspired method for a power or energy system application. Studies were excluded when they were outside the scope of power and energy systems, contained no substantive quantum method, duplicated another publication, lacked sufficient technical information, or had no accessible full text. Because detailed screening counts were not retained, this paper is characterized as a comprehensive structured review rather than a formal systematic review.

\subsection{Data Extraction and Application Classification}
A structured data-extraction procedure was applied to maintain consistency across the selected studies. For each publication, the extracted information included the publication year, principal smart grid application, problem formulation, implementation type, computing platform, quantum algorithm, classical comparison method, test-system size, and principal reported finding. When available, information concerning solution accuracy, computational time, convergence behavior, resource requirements, circuit depth, qubit count, and robustness to noise was also examined.
The implementation environment was classified as real quantum hardware, quantum simulation, classical emulation, or theoretical analysis. Studies executed on physical gate-based quantum processors or quantum annealers were identified as hardware implementations, whereas experiments performed using quantum software on classical computers were classified as simulation-based studies. Digital annealers and other quantum-inspired platforms are distinguished from physical quantum processors because they employ specialized classical hardware or algorithms, despite solving quadratic unconstrained binary optimization (QUBO) or Ising formulations similar to those used by quantum annealers.
The studies were subsequently organized according to their primary smart grid applications. When a publication addressed multiple applications, it was assigned to the category corresponding to its principal objective, while relevant secondary applications were discussed in the accompanying text. This classification supports consistent comparisons among studies addressing similar problems and provides a structured overview of their implementation status, algorithmic maturity, application scale, and reported performance across different smart grid functions.

\subsection{Evidence and Quality Assessment}
The strength of evidence reported by each study was assessed based on its implementation environment, theoretical or empirical nature, classical benchmark, test-system scale, and reproducibility. Particular attention was given to whether the study demonstrated an end-to-end quantum advantage or only reported theoretical potential, improved accuracy, faster convergence, reduced model complexity, or performance under a limited experimental configuration. The completeness of the problem formulation, algorithm description, implementation settings, and experimental procedure was also considered.
Claims of quantum speedup were treated as theoretical unless they were supported by matched end-to-end comparisons that accounted for problem encoding, state preparation, circuit execution, measurement, solution decoding, and classical processing. Quantum machine learning improvements were interpreted as application-specific empirical findings rather than general quantum advantage. Similarly, quantum-annealing results were assessed based on the classical benchmark and inclusion of embedding and preprocessing costs.

\vspace{-2pt}

\section{Quantum Computing} \label{Quantum Computing Fundamental}
\subsection{Fundamentals}
\subsubsection{Qubit, Superposition, and Entanglement}
Quantum computing differs fundamentally from classical computing in both its theoretical foundation and operational behavior. In classical systems, a given input deterministically produces a fixed output. In contrast, a quantum computing system may yield different outcomes for the same input due to its inherent probabilistic nature. While classical computing relies on bits as the fundamental unit of information, quantum computing utilizes quantum bits (qubits), which exhibit significantly richer behavior. A classical bit takes one of two distinct values, 0 or 1. A qubit, however, can exist not only in these basis states, denoted as $|0\rangle$ and $|1\rangle$, but also in a linear combination of both states simultaneously. This property, known as superposition, enables a qubit to encode more information than a classical bit. To formally describe this behavior, the general state of a qubit can be expressed as
\begin{equation}
|\psi\rangle = \alpha |0\rangle + \beta |1\rangle,
\end{equation}
where $\alpha$ and $\beta$ are complex probability amplitudes satisfying the normalization condition
\begin{equation}
|\alpha|^2 + |\beta|^2 = 1.
\end{equation}

A convenient geometric representation of a qubit is provided by the Bloch sphere, where any pure qubit state can be expressed in terms of two real parameters, $\theta$ and $\phi$ \cite{nielsen2010quantum}, as 
\begin{equation}
|\psi\rangle = \cos\left(\frac{\theta}{2}\right) |0\rangle + e^{i\phi} \sin\left(\frac{\theta}{2}\right) |1\rangle.
\end{equation}
In this representation, the state $|0\rangle$ lies at the north pole, whereas $|1\rangle$ lies at the south pole of the sphere.
The probabilistic nature of quantum mechanics becomes evident during measurement. When a qubit is measured in the computational basis, the state collapses to either $|0\rangle$ or $|1\rangle$. The probabilities of these outcomes are determined by the squared magnitudes of the corresponding amplitudes, i.e.,
\begin{equation}
P(0) = |\alpha|^2, \quad P(1) = |\beta|^2.
\end{equation}
%
%In quantum computing, quantum operations are performed via quantum gates, which represent unitary transformations acting on qubit states. %For an $n$-qubit system, these operations are represented by $2^n \times 2^n$ unitary matrices $U$ satisfying
%\begin{equation}
%U U^\dagger = U^\dagger U = I,
%\end{equation}
%where $U^\dagger$ denotes the conjugate transpose of $U$.
%
For multi-qubit systems, the overall state space grows exponentially through the tensor product structure. Given two qubits $|\psi\rangle$ and $|\phi\rangle$, their combined state is represented as
\begin{equation}
|\psi\rangle \otimes |\phi\rangle.
\end{equation}
A general two-qubit state can thus be expressed as
\begin{equation}
|\psi\rangle = \alpha |00\rangle + \beta |01\rangle + \gamma |10\rangle + \delta |11\rangle,
\end{equation}
with the normalization constraint
\begin{equation}
|\alpha|^2 + |\beta|^2 + |\gamma|^2 + |\delta|^2 = 1.
\end{equation}

A key feature of multi-qubit systems is quantum entanglement, which represents non-classical correlations between qubits. Entangled states cannot be decomposed into independent single-qubit states, and the measurement of one qubit instantaneously influences the combined quantum state regardless of spatial separation. This phenomenon allows quantum systems to encode information across $2^n$ basis states simultaneously, enabling parallelism unattainable in classical systems.
The combination of superposition and entanglement enables quantum computers to explore large solution spaces more efficiently than classical computers. These properties form the foundation of quantum algorithms and are responsible for the potential exponential speedups observed in various computational problems.

\subsubsection{Quantum Gates and Circuits}
Basic quantum gates serve as the fundamental building blocks for manipulating quantum information and implementing quantum algorithms. Unlike classical logic gates that operate on 0s and 1s, quantum gates act on qubits; however, their functionality is governed by the principles of quantum mechanics and is therefore described by unitary transformations. A matrix $U$ is unitary if it satisfies the condition $U^\dagger U = U U^\dagger = I$, where $U^\dagger$ denotes the conjugate transpose of $U$, and $I$ represents the identity matrix of appropriate dimension. This property ensures that quantum operations are reversible and preserve the normalization of quantum states \cite{nielsen2010quantum}. Single-qubit gates, such as the Pauli-$X$, Pauli-$Y$, and Pauli-$Z$ operators, perform bit-flip and phase-flip operations, while the Hadamard gate enables the creation of superposition states by transforming a computational basis state into a balanced linear combination of $|0\rangle$ and $|1\rangle$. In addition, parameterized rotation gates, including $R_X(\theta)$, $R_Y(\theta)$, and $R_Z(\theta)$, provide continuous transformations of qubit states on the Bloch sphere, thereby allowing precise control over quantum state evolution \cite{mermin2007quantum}.

Beyond individual gate operations, quantum circuits provide a systematic framework for implementing quantum computations through sequences of quantum gates. A quantum circuit is composed of qubit wires that track the evolution of quantum states over time, with gates applied in a prescribed order along these wires. Multi-qubit gates, such as the Controlled-NOT (CNOT), Controlled-Z (CZ), and SWAP gates, enable interactions between qubits and facilitate the creation of entangled states, a key resource underlying the power of quantum computation. The action of a quantum circuit is mathematically described by the composition of the unitary operators associated with its constituent gates, resulting in an overall unitary transformation on the multi-qubit system. In addition to their mathematical formulation, quantum circuits provide an intuitive graphical representation of quantum algorithms, clearly illustrating the flow of quantum information and the sequence of operations. Quantum gates and circuit architectures form a universal model of quantum computation capable of realizing arbitrary unitary transformations. Further details on the theory and implementation of quantum gates and circuits are available in \cite{nielsen2010quantum, mermin2007quantum, crooks2020gates}.

\subsection{Quantum Algorithms}

\subsubsection{Quantum Fourier Transform}
The Quantum Fourier Transform (QFT) is a unitary quantum transformation that applies the discrete Fourier transform to the amplitudes of a quantum state. For an $n$-qubit system, the QFT operates on a computational basis state $|x\rangle$, where $x \in \{0,1,\dots,2^n -1\}$, according to
\begin{equation}
\text{QFT}|x\rangle = \frac{1}{\sqrt{N}} \sum_{k=0}^{2^n -1} e^{2\pi i xk / N} |k\rangle, \quad N = 2^n.
\end{equation}
This operation maps the amplitudes of a quantum state into a Fourier-domain representation. When the input amplitudes contain a suitable periodic structure, the transformed state can produce measurement outcomes associated with that periodicity. A classical fast Fourier transform applied to a length-$N$ vector requires $O(N\log N)$ arithmetic operations, whereas an exact QFT on $n=\log_{2}N$ qubits can be implemented using $O(n^{2})$ quantum gates. However, this comparison does not include the costs of quantum-state
preparation, measurement, and classical output extraction.

The quantum circuit for QFT consists of a structured sequence of single-qubit and controlled operations applied to $n$ qubits. For each qubit, a Hadamard gate is applied together with a sequence of controlled phase-rotation gates involving the remaining qubits. The phase-rotation gate is defined as
\begin{equation}
R_k =
\begin{pmatrix}
1 & 0 \\
0 & e^{2\pi i / 2^k}
\end{pmatrix},
\end{equation}
which introduce conditional phase shifts depending on the states of other qubits. Specifically, each qubit interacts with subsequent qubits through controlled rotations with progressively smaller phase angles, ensuring that the phase information is accurately encoded. The transformation can be expressed in a factorized form as
\begin{equation}
\text{QFT} |x_1 x_2 \dots x_n\rangle = \frac{1}{2^{n/2}} \bigotimes_{j=1}^{n} 
\left( |0\rangle + e^{2\pi i \, 0.x_j x_{j+1}\dots x_n} |1\rangle \right),
\end{equation}
where $0.x_j x_{j+1}\dots x_n$ denotes the binary fractional expansion. The standard circuit produces the output qubits in reversed order. SWAP gates may be used to restore the conventional ordering, although the same correction can be handled by relabeling the output qubits.

In power systems, QFT can be utilized for harmonic analysis, frequency estimation, and power quality monitoring, where accurate identification of frequency components is critical for grid stability and performance. It is particularly useful in systems with high penetration of power electronic converters and renewable energy sources, where non-sinusoidal waveforms and harmonics are prevalent. Additionally, QFT can support fault detection and disturbance analysis by enabling rapid identification of transient events and anomalies in voltage and current signals

\subsubsection{Quantum Phase Estimation}
Quantum Phase Estimation (QPE) is a fundamental quantum computing algorithm used to determine the eigenvalue (phase) associated with a unitary operator. Given a unitary operator $U$ and its eigenstate $|\psi\rangle$, the relation is expressed as
\begin{equation}
U |\psi\rangle = e^{2\pi i \phi} |\psi\rangle,
\end{equation}
where $\phi \in [0,1)$ represents the unknown phase to be estimated. The standard QPE algorithm utilizes two quantum registers: an $n$-qubit estimation register initialized in the computational basis state $|0\rangle^{\otimes n}$, and a target register prepared in the eigenstate $|\psi\rangle$. The procedure begins with the application of Hadamard gates to all qubits in the estimation register, producing a uniform superposition:
\begin{equation}
\frac{1}{\sqrt{2^n}} \sum_{k=0}^{2^n-1} |k\rangle |\psi\rangle.
\end{equation}
The next stage encodes the phase information through a sequence of controlled unitary operations. Specifically, for $j=0,1,\ldots,n-1$, the $j$-th estimation qubit controls the application of $U^{2^j}$ to the target register. The correspondence between $j$ and the physical qubit position depends on the bit-ordering convention used. These controlled operations produce phase kickback, resulting in
\begin{equation}
\frac{1}{\sqrt{2^n}} \sum_{k=0}^{2^n-1} e^{2\pi i k \phi} |k\rangle |\psi\rangle,
\end{equation}
where $\phi$ is the eigenphase associated with the eigenstate 
$|\psi\rangle$. In the final stage, the inverse Quantum Fourier Transform (QFT$^\dagger$) is applied to the estimation 
register, mapping the phase-encoded state to
\begin{equation}
\frac{1}{\sqrt{2^n}}
\sum_{k=0}^{2^n-1}
e^{2\pi i k\phi}|k\rangle
\xrightarrow{\operatorname{QFT}^{\dagger}}
\sum_{y=0}^{2^n-1}c_y|y\rangle,
\end{equation}
where
\begin{equation}
c_y=
\frac{1}{2^n}
\sum_{k=0}^{2^n-1}
e^{2\pi i k\left(\phi-y/2^n\right)}.
\end{equation}
If $\phi=m/2^n$ for some integer $m$, the inverse QFT produces the computational basis state $|m\rangle$ exactly. Otherwise, it produces a superposition whose measurement probabilities are concentrated around values of $y$ for which $y/2^n$ closely approximates $\phi$. In QPE, a measurement of the estimation register produces an integer $y$, from which the phase is estimated as: $\tilde{\phi}=\frac{y}{2^n}$. QPE accuracy improves with the number of qubits in the estimation register. Modern implementations introduce optimized variants, such as iterative phase estimation and adaptive measurement strategies, which can reduce the number of ancillary qubits by repeatedly using a single estimation qubit. However, these variants do not necessarily reduce the total number or maximum power of the controlled-$U$ operations required for a specified precision.

QPE serves as a fundamental subroutine in prominent quantum algorithms and has potential applications in power and energy systems, particularly in spectral analysis and system dynamics. Specifically, QPE enables the assessment of power system stability by analyzing the eigenvalues of system matrices, which determine stability margins and dynamic behavior. Additionally, QPE applies to state estimation and parameter identification, where accurate evaluation of system characteristics is essential.

\subsubsection{Harrow-Hassidim-Lloyd}
The Harrow-Hassidim-Lloyd (HHL) \cite{harrow2009quantum} is a quantum algorithm for solving systems of linear equations of the form $A\mathbf{x}=\mathbf{b}$ by preparing a quantum state proportional to the solution vector $\mathbf{x}$. Instead of explicitly computing all components of $\mathbf{x}$, HHL encodes the normalized right-hand side $\mathbf{b}$ into a quantum state $|b\rangle$ and operates within the eigenbasis of the matrix $A$. Assuming $A$ is Hermitian, nonsingular, sufficiently sparse, and well-conditioned, it can be decomposed as $A = \sum_i \lambda_i |u_i\rangle \langle u_i|$, where $\lambda_i$ and $|u_i\rangle$ denote its eigenvalues and eigenvectors. A non-Hermitian linear system can instead be embedded into a larger Hermitian system. The input state is therefore expressed as a superposition
\begin{equation}
|b\rangle = \sum_i \beta_i |u_i\rangle,
\end{equation}
which allows the inverse operation $A^{-1}$ to act straightforwardly in this basis. In a quantum circuit, this process begins with state preparation followed by QPE that extracts the eigenvalues of $A$ by applying controlled unitary operations of the form $U = e^{iAt}$ \cite{harrow2009quantum, nielsen2010quantum}. Since $U|u_i\rangle = e^{i\lambda_it}|u_i\rangle$, the estimated phases can be used to infer the eigenvalues of A after accounting for the evolution time and phase convention. The QPE circuit uses an ancillary register initialized to $|0\rangle$, where $|0\rangle$ denotes the all-zero state of the estimation register, Hadamard gates to create superpositions, and controlled-$U^{2^k}$ operations to encode phase information. After the inverse Quantum Fourier Transform (QFT$^\dagger$), the eigenvalue information is represented in the estimation register. For notational simplicity, the following derivation assumes ideal phase estimation and denotes the encoded eigenvalue by $\lambda_i$; practical implementations produce a finite-precision approximation. The resulting transformation is
\begin{equation}
\sum_i \beta_i |u_i\rangle |0\rangle \xrightarrow{\text{QPE}} \sum_i \beta_i |u_i\rangle |\lambda_i\rangle.
\end{equation}
Following eigenvalue estimation, the algorithm performs a controlled rotation on an additional ancilla qubit to encode the reciprocals of the eigenvalues into quantum amplitudes, effectively implementing the matrix inversion step. This transformation is given by
\begin{equation}
|\lambda_i\rangle |0\rangle \rightarrow |\lambda_i\rangle 
\left( \sqrt{1 - \frac{C^2}{\lambda_i^2}}\,|0\rangle + \frac{C}{\lambda_i}|1\rangle \right),
\end{equation}
where $C$ is a normalization constant, which is chosen such that $0 < C \leq \min_i |\lambda_i|$, ensuring that the controlled-rotation amplitudes are valid. The combined system after this step becomes
\begin{equation}
\sum_i \beta_i |u_i\rangle |\lambda_i\rangle 
\left( \sqrt{1 - \frac{C^2}{\lambda_i^2}}\,|0\rangle + \frac{C}{\lambda_i}|1\rangle \right).
\end{equation}
The final stage of the HHL circuit applies inverse QPE to uncompute the eigenvalue register, disentangling it from the system and leaving the state
\begin{equation}
\sum_i \beta_i |u_i\rangle 
\left( \sqrt{1 - \frac{C^2}{\lambda_i^2}}\,|0\rangle + \frac{C}{\lambda_i}|1\rangle \right).
\end{equation}
Measurement of the ancilla qubit followed by post-selection of the outcome $|1\rangle$ collapses the system register into the desired solution state
\begin{equation}
|x\rangle
=
\frac{A^{-1}|b\rangle}
{\left\|A^{-1}|b\rangle\right\|}
\propto A^{-1}|b\rangle.
\end{equation}

The overall circuit structure of the HHL algorithm, therefore, consists of four main components: state preparation, phase estimation using controlled Hamiltonian simulation, controlled rotations for eigenvalue inversion, and inverse QPE for uncomputation. While HHL offers the potential for exponential speedup in the system dimension (scaling as $O(\log N)$ under ideal assumptions), its practical performance depends critically on factors such as the sparsity and condition number $\kappa$ of $A$, the efficiency of implementing $e^{iAt}$, and the probability of successful post-selection \cite{harrow2009quantum, childs2017quantum}. Consequently, although the algorithm is theoretically powerful, current implementations are constrained by circuit depth, noise, and resource requirements, making HHL primarily significant in quantum algorithm research and future fault-tolerant quantum computing.

The HHL algorithm has various potential applications in power and energy systems, particularly for solving large-scale linear algebra problems arising in grid analysis and optimization. Many power system computations, such as power flow analysis, state estimation, and stability assessment, involve solving linear or linearized systems of equations, which can be computationally demanding for large networks. In addition, it can support real-time monitoring, contingency analysis, and optimization of grid operations, all of which require rapid solutions to large sparse matrices.

\subsubsection{Variational Quantum Linear Solver}
The variational quantum linear solver (VQLS) is a hybrid quantum-classical algorithm designed to solve systems of linear equations of the form $A\mathbf{x}=\mathbf{b}$ by preparing a quantum state proportional to the solution vector $\mathbf{x}$ \cite{bravo2023variational}. Similar to the HHL algorithm, VQLS aims to construct the quantum state $|x\rangle \propto A^{-1}|b\rangle$, where the right-hand side vector $\mathbf{b}$ is encoded into a normalized quantum state $|b\rangle$. However, unlike the HHL algorithm, which relies on quantum phase estimation and explicit eigenvalue inversion, VQLS avoids deep quantum subroutines and instead employs a variational optimization framework that is perhaps better suited to NISQ devices. The desired solution state can be expressed as
\begin{equation}
|x\rangle = \frac{A^{-1}|b\rangle}{\|A^{-1}|b\rangle\|}.
\end{equation}
Rather than explicitly working in the eigenbasis of $A$, VQLS introduces a parametrized trial state of the form
\begin{equation}
|x(\boldsymbol{\theta})\rangle = U(\boldsymbol{\theta})|0\rangle,
\end{equation}
where $U(\boldsymbol{\theta})$ is a quantum circuit with tunable parameters $\boldsymbol{\theta}$. The objective is to optimize these parameters such that the resulting quantum state approximates the target solution state.
To enable efficient implementation on a quantum computer, the matrix $A$ is expressed as a linear combination of unitary operators,
\begin{equation}
A = \sum_{l} c_l A_l,
\end{equation}
where $c_l\in\mathbb{R}$ are scalar coefficients and $A_l$ are unitary matrices. In power systems, the matrix $A$ often corresponds to the Jacobian or admittance matrix, which is sparse and structured, making it well suited to efficient linear combinations of unitary-based representations. Instead of inverting the matrix explicitly, VQLS defines a cost function that measures how well the trial state satisfies the linear system. A natural choice is the residual-based cost function
\begin{equation}
C(\boldsymbol{\theta}) = \left\|A|x(\boldsymbol{\theta})\rangle - |b\rangle\right\|^2,
\end{equation}
which expands as
\begin{equation}
C(\boldsymbol{\theta}) = \langle x(\boldsymbol{\theta})|A^\dagger A|x(\boldsymbol{\theta})\rangle 
- 2\,\mathrm{Re}\{\langle b|A|x(\boldsymbol{\theta})\rangle\} + 1.
\end{equation}

To improve stability and remove dependence on normalization, a commonly used cost function is
\begin{equation}
C_{\text{norm}}(\boldsymbol{\theta}) = 1 - \frac{|\langle b|A|x(\boldsymbol{\theta})\rangle|^2}{\langle x(\boldsymbol{\theta})|A^\dagger A|x(\boldsymbol{\theta})\rangle}.
\end{equation}
Minimizing this cost function drives $A|x(\boldsymbol{\theta})\rangle$ toward a state collinear (proportional) to $|b\rangle$, thereby ensuring convergence toward the desired solution. Using the linear combination of unitary representations, these expectation values are decomposed as
\begin{equation}
\langle x|A^\dagger A|x\rangle = \sum_{l,l'} c_l^* c_{l'} \langle x|A_l^\dagger A_{l'}|x\rangle,
\end{equation}
and
\begin{equation}
\langle b|A|x\rangle = \sum_{l} c_l \langle b|A_l|x\rangle.
\end{equation}
These quantities can be estimated on a quantum processor using techniques such as the Hadamard test, overlap measurements, or other expectation-value estimation procedures. The VQLS algorithm proceeds iteratively through a hybrid optimization loop. For a given parameter vector $\boldsymbol{\theta}$, the quantum processor evaluates the cost function, while a classical optimizer updates the variational parameters to minimize the measured value. At convergence,
\begin{equation}
|x(\boldsymbol{\theta}^*)\rangle \approx |x\rangle,
\end{equation}
with approximation error
\begin{equation}
\epsilon = \left\| |x(\boldsymbol{\theta}^*)\rangle - |x\rangle \right\|.
\end{equation}
Thus, VQLS consists of three main components: parametrized state preparation, quantum evaluation of cost functions, and classical optimization.
In contrast to HHL, which relies on quantum phase estimation, controlled rotations, and inverse quantum Fourier transforms, VQLS reformulates the linear system problem as a variational optimization task, making it much better suited to NISQ-era quantum hardware. In summary, while HHL can provide stronger asymptotic complexity guarantees under specific assumptions, including sparse and well-conditioned matrices along with efficient state preparation, VQLS offers a more hardware-efficient framework tailored for NISQ devices by transforming the linear system problem into a variational optimization problem, trading asymptotic guarantees for practical implementability \cite{harrow2009quantum, bravo2023variational, cerezo2021variational}.

The VQLS is particularly relevant to power and energy system applications, where large-scale linear systems arise naturally in power flow analysis, state estimation, network sensitivity calculations, and optimization problems. Since many practical power-system applications can tolerate approximate solutions provided within strict computational time constraints, the hybrid nature of VQLS offers a potentially attractive balance between solution quality and computational efficiency for future quantum-enhanced grid analytics.

\subsubsection{Quantum Amplitude Estimation}
The Quantum Amplitude Estimation (QAE) \cite{brassard2000quantum} is a fundamental quantum algorithm for estimating the probability amplitude of a specific target state in a quantum system. It plays a significant role in applications such as quantum finance, optimization, and probabilistic simulations. Compared to classical approaches, it provides a quadratic speedup in estimation accuracy \cite{brassard2000quantum, maronese2023quantum}.
The algorithm begins by preparing a quantum state using a unitary operator $\mathcal{A}$ such that
\begin{equation}
\mathcal{A}|0\rangle = \sqrt{1-a}\,|\psi_0\rangle + \sqrt{a}\,|\psi_1\rangle,
\end{equation}
where $a \in [0,1]$ represents the unknown probability associated with the designated good subspace. The states $|\psi_1\rangle$ and $|\psi_0\rangle$ are normalized and mutually orthogonal components of the bad and good subspaces, respectively. To encode this probability through a Grover-type rotation, the operator is defined as
\begin{equation}
\mathcal{Q} = -\mathcal{A} S_0 \mathcal{A}^{-1} S_{\psi},
\end{equation}
where $S_0 = I - 2|0\rangle\langle 0|$ and $S_{\psi} = I - 2\Pi_{\mathrm{good}}$ are reflection operators and $\Pi_{\mathrm{good}}$ projects onto the good subspace. In canonical QAE, an $m$-qubit estimation register controls the powers $\mathcal{Q}^{2^j}$, where $j = 0, 1, ..., m-1$ within a quantum phase-estimation framework.
In the two-dimensional subspace spanned by $|\psi_0 \rangle$ and $|\psi_1 \rangle$ the operator $\mathcal{Q}$ has eigenvalues
\begin{equation}
e^{\pm 2i\theta}, \quad \text{with} \quad a = \sin^2(\theta),
\end{equation}
where $\theta \in [0, \pi/2]$ encodes the amplitude information. The inverse QFT is then applied to the estimation register to recover an approximation of the corresponding eigenphase.
Measuring the estimation register produces an integer
$y\in\{0,1,\ldots,M-1\}$, where $M=2^m$. The probability $a$ is then
estimated as $ \hat{a} = \sin^2\left(\frac{\pi y}{M}\right)$.
Because $\mathcal{Q}$ has eigenphases $\pm 2\theta$, the outcomes $y$
and $M-y$ produce the same estimate of $a$. The estimation error
decreases as $O(1/M)$ in terms of the number of applications of the
Grover operator. Variants such as iterative amplitude estimation remove the QPE and inverse-QFT stages and can reduce the number of ancillary qubits. However, they still require repeated measurements and may require high powers of the Grover operator, so their practical performance depends on circuit depth, hardware noise, and oracle complexity. 

QAE has potential applications in smart grids involving probabilistic and statistical analysis. In particular, QAE may provide a quadratic improvement in the oracle-query complexity required to estimate suitably encoded expectations, event probabilities, and selected distributional quantities. Possible applications include uncertainty quantification, risk assessment, reliability analysis, and stochastic optimization involving renewable generation, electricity demand, equipment outages, and operating contingencies \cite{egger2020credit}. QAE could also support Monte Carlo-based power-system studies when the desired quantity can be efficiently encoded as a quantum amplitude. However, the potential advantage depends on the costs of state preparation, coherent power-system model evaluation, Grover-operator implementation, and classical output extraction. Therefore, these applications should currently be viewed as prospective quantum or hybrid quantum–classical approaches.

%\vspace{-6pt}
\subsubsection{Variational Quantum Eigensolver}
The Variational Quantum Eigensolver (VQE) \cite{peruzzo2014variational} is a hybrid quantum-classical algorithm designed to approximate the ground-state eigenvalue of a Hermitian operator \( H \in \mathbb{C}^{2^n \times 2^n} \), where $n$ is the number of qubits. Based on the variational principle, any normalized parameterized state \( |\psi(\boldsymbol{\theta})\rangle \) satisfies
\begin{equation}
E(\boldsymbol{\theta}) = \langle \psi(\boldsymbol{\theta}) \rvert H \lvert \psi(\boldsymbol{\theta}) \rangle \geq E_0,
\end{equation}
where \(E_0\) is the true ground-state energy. From an optimization perspective, VQE can be interpreted as
a constrained functional minimization problem over a nonconvex
parameter space defined by a parameterized quantum circuit
(ansatz), where the feasible set is implicitly determined by
the unitary group $U(2^n)$. The ansatz state is generated by applying a parameterized unitary transformation \( U(\boldsymbol{\theta}) \) to a reference state \( \lvert 0 \rangle^{\otimes n} \), such that
\begin{equation}
\lvert \psi(\boldsymbol{\theta}) \rangle = U(\boldsymbol{\theta}) \lvert 0 \rangle^{\otimes n},
\end{equation}
where $\boldsymbol{\theta} = [\theta_1,\theta_2,\ldots,\theta_p]^T$. Accordingly, the objective function becomes
\begin{equation}
E(\boldsymbol{\theta}) = \langle 0 \rvert U^\dagger(\boldsymbol{\theta}) H U(\boldsymbol{\theta}) \lvert 0 \rangle.
\end{equation}
In practical implementations, the Hamiltonian is decomposed into a weighted sum of Pauli operators,
\begin{equation}
H = \sum_{k=1}^{M} h_k P_k,
\end{equation}
where \( h_k \in \mathbb{R} \) and \( P_k \in \{I, X, Y, Z\}^{\otimes n} \). The expectation value is then evaluated as
\begin{equation}
E(\boldsymbol{\theta}) = \sum_{k=1}^{M} h_k \langle \psi(\boldsymbol{\theta}) \rvert P_k \lvert \psi(\boldsymbol{\theta}) \rangle,
\end{equation}
which is estimated via repeated projective measurements on quantum hardware. This formulation enables the quantum processor to estimate individual expectation values of Pauli strings through repeated measurements, while the classical processor aggregates these measurements to compute the total energy. The expectation value of each Pauli term is obtained as
\begin{equation}
\langle P_k \rangle
=
\langle
\psi(\boldsymbol{\theta})
|
P_k
|
\psi(\boldsymbol{\theta})
\rangle,
\end{equation}
such that
\begin{equation}
E(\boldsymbol{\theta})
=
\sum_{k=1}^{M}
h_k
\langle P_k \rangle.
\end{equation}
The optimal parameters are obtained by solving
\begin{equation}
\boldsymbol{\theta}^{*}
=
\arg \min_{\boldsymbol{\theta}}
E(\boldsymbol{\theta}),
\end{equation}

From an optimization standpoint, VQE can be interpreted as a stochastic, sample-based empirical risk minimization problem in which the loss function $E(\boldsymbol{\theta})$ is not directly accessible but is instead approximated through finite-shot measurements, introducing statistical noise.
The optimization procedure proceeds iteratively. Initially, a parameter vector $\boldsymbol{\theta}^{0}$ is selected and loaded into the quantum circuit. The quantum processor prepares the state $|
\psi(\boldsymbol{\theta}^{t}) \rangle$ and evaluates the corresponding energy $E(\boldsymbol{\theta}^{t})$. For a gradient-based optimizer, one possible update rule is
\begin{equation}
\boldsymbol{\theta}^{(t+1)} = \boldsymbol{\theta}^{(t)} + \alpha^{(t)} \mathbf{d}^{(t)},
\end{equation}
where \( \mathbf{d}^{(t)} = - \nabla_{\boldsymbol{\theta}} E(\boldsymbol{\theta}^{(t)}) \) denotes the search direction and \( \alpha^{(t)} \) is the step size. VQE can also employ gradient-free optimizers, in which case the parameter update does not explicitly require the energy gradient. The process is repeated until a convergence criterion is satisfied
\begin{equation}
\left|
E(\boldsymbol{\theta}^{(t+1)})
-
E(\boldsymbol{\theta}^{(t)})
\right|
<
\epsilon,
\end{equation}
where $\epsilon$ is a user-defined convergence tolerance. 

VQE has shown strong applicability in power and energy systems, particularly for optimization and modeling challenges involving continuous and discrete variables. VQE can be used for optimal power flow, economic dispatch, and energy management by formulating these problems as Hamiltonian systems and minimizing the corresponding energy function. It is also well suited for power system stability analysis, state estimation, and parameter identification, where complex nonlinear relationships can be represented using a variational quantum circuit (VQC). Furthermore, VQE has been explored in QML as a subroutine for variational classifiers and generative models, where parameterized circuits serve as trainable quantum feature maps.

\subsubsection{Quantum Approximate Optimization Algorithm}
The \textit{Quantum Approximate Optimization Algorithm (QAOA)} is a hybrid quantum--classical variational framework developed to address combinatorial optimization problems through parameterized quantum circuits (PQCs). It is specifically designed to produce approximate solutions to discrete optimization tasks by integrating quantum state preparation with classical parameter optimization \cite{farhi2014quantum}. In this approach, the optimization problem is mapped onto a cost Hamiltonian $(H_C)$, whose ground state encodes the optimal solution. The algorithm then constructs a quantum state that approximates this ground state by applying a sequence of alternating unitary operators, with tunable parameters updated through a classical feedback loop.

Consider a combinatorial optimization problem defined by a cost function, $C(z),$ where \( z \in \{0,1\}^n \) denotes a candidate solution. QAOA begins by initializing the system in a uniform superposition state
\begin{equation}
|\psi_0\rangle = |+\rangle^{\otimes n}
= \frac{1}{\sqrt{2^n}}
\sum_{z\in\{0,1\}^n}|z\rangle,
\end{equation}
using the Hadamard gates applied to all qubits that serve as the starting point for the variational procedure, on which layers of parameterized transformations are applied to gradually bias the quantum state toward high-quality low-cost solutions. The algorithm then alternates between the problem Hamiltonian $H_C$ and a mixing Hamiltonian $H_M$, typically defined as \cite{farhi2014quantum, zhou2020quantum}
\begin{equation}
H_M=\sum_{i=1}^{n}X_i,
\end{equation}
where $X_i$ is the Pauli-X operator acting on qubit $i$. In QAOA, the parameter $p$ denotes the circuit depth (or number of QAOA layers), where each layer consists of one application of the problem unitary and one application of the mixer unitary. For a circuit depth $p$, the variational quantum state can be generated through a sequence of parameterized unitary operations \cite{farhi2014quantum},
\begin{equation}
|\psi(\boldsymbol{\gamma},\boldsymbol{\beta})\rangle
=
\prod_{k=1}^{p}
e^{-i\beta_k H_M}
e^{-i\gamma_k H_C}
|\psi_0\rangle,
\end{equation}
where $\boldsymbol{\gamma}=(\gamma_1,\gamma_2,\ldots,\gamma_p)$ and $\boldsymbol{\beta}=(\beta_1,\beta_2,\ldots,\beta_p)$ are variational parameters. In the quantum circuit, the operator $e^{-i\gamma_k H_C}$ is implemented using controlled-phase and rotation gates that encode the optimization objective, while $e^{-i\beta_k H_M}$ is realized through single-qubit $R_x(2\beta_k)$ rotations. Consequently, the QAOA circuit consists of $p$ alternating layers of cost and mixer operators acting on the initial superposition state.
The performance of the generated quantum state is evaluated through the expectation value of the cost Hamiltonian
\begin{equation}
F(\boldsymbol{\gamma},\boldsymbol{\beta})
=
\langle\psi(\boldsymbol{\gamma},\boldsymbol{\beta})|
H_C
|\psi(\boldsymbol{\gamma},\boldsymbol{\beta})\rangle,
\end{equation}
which serves as the objective function for a classical optimizer. The classical optimization routine iteratively updates the parameters $\boldsymbol{\gamma}$ and $\boldsymbol{\beta}$ to maximize (or minimize) this expectation value depending on the problem formulation. For example, for the Max-Cut problem in graphs, one of the most common applications of QAOA, the cost Hamiltonian is expressed as \cite{zhou2020quantum, kashapogu2024exploring}

\begin{equation}
H_C=
\frac{1}{2}
\sum_{(i,j)\in E}
w_{ij}
\left(I-Z_iZ_j\right),
\end{equation}
where $E$ denotes the set of graph edges, $w_{ij}$ represents the edge weights, and $Z_i$ is the Pauli-Z operator acting on qubit $i$. The corresponding phase-separation unitary is given by $U_C(\gamma_k)=e^{-i\gamma_k H_C}$, while the mixer unitary is expressed as $U_M(\beta_k)=e^{-i\beta_k H_M}$. Thus, the complete QAOA state after $p$ layers can be written as $|\psi_p\rangle = U_M(\beta_p)U_C(\gamma_p)
\cdots U_M(\beta_1)U_C(\gamma_1) |\psi_0\rangle$. The optimization objective can alternatively be formulated as
\begin{equation}
(\boldsymbol{\gamma}^{*},\boldsymbol{\beta}^{*})
=
\arg\max_{\boldsymbol{\gamma},\boldsymbol{\beta}}
\langle\psi_p|H_C|\psi_p\rangle,
\end{equation}
where the optimal parameters $(\boldsymbol{\gamma}^{*},\boldsymbol{\beta}^{*})$ are determined using a classical optimization routine \cite{zhou2020quantum}. Following the convergence of the classical optimization process, repeated measurements of the optimized quantum state
\begin{equation}
|\psi^*\rangle
=
|\psi(\boldsymbol{\gamma}^{*},\boldsymbol{\beta}^{*})\rangle
\end{equation}
produce bitstrings with high probability of corresponding to near-optimal or optimal solutions. The QAOA quantum circuit therefore combines parameterized quantum evolution with classical optimization, enabling approximate solutions to large-scale discrete optimization problems while maintaining a circuit structure suitable for NISQ hardware \cite{zhou2020quantum}. As the circuit depth $p$ increases, the expressive power of the ansatz generally increases. For sufficiently large $p$, QAOA can approximate the optimal solution arbitrarily well for many problem instances, although the required depth may scale unfavorably with problem size \cite{zhou2020quantum}. Recent studies examining circuit depth under realistic noisy conditions indicate that the best performance is generally achieved at intermediate depths \cite{pellow2024effect}. Increasing circuit depth beyond this range often yields diminishing returns, as noise begins to outweigh any additional computational benefits. 

In power and energy systems, the QAOA can be employed for unit commitment, economic dispatch, and optimal power flow by mapping these problems into Ising or QUBO formulations and approximating their optimal solutions. It can also be used for network reconfiguration, distributed energy resource scheduling, and contingency analysis, where large-scale binary decision variables are involved.

\subsubsection{Quantum Annealing}
Quantum Annealing (QA) is a quantum computational paradigm that solves combinatorial optimization problems by exploiting quantum-mechanical phenomena such as tunneling and adiabatic evolution. Unlike gate-based quantum algorithms, which employ discrete quantum logic operations, QA encodes an optimization problem into a time-dependent Hamiltonian \( H_P \) and gradually evolves a quantum system from an easily prepared ground state toward the ground state of a problem Hamiltonian. Typically, \( H_P \) is expressed in Ising form:
\begin{equation}
H_P = \sum_{i} h_i \sigma_i^z + \sum_{i<j} J_{ij} \sigma_i^z \sigma_j^z,
\end{equation}
where \( \sigma_i^z \) denotes the Pauli-\(Z\) operator acting on qubit \(i\), and \(h_i, J_{ij} \in \mathbb{R}\) represent local fields and coupling strengths, respectively. The optimal solution corresponds to a ground state, or one of the degenerate ground states, of \( H_P \).
The QA algorithm operates by initializing the system in the ground state of a driver Hamiltonian \( H_D \), often chosen as a transverse-field Hamiltonian:
\begin{equation}
H_D = - \sum_i \sigma_i^x,
\end{equation}
where \( \sigma_i^x \) is the Pauli-\(X\) operator. The ground state of this driver Hamiltonian is $|\psi(0)\rangle = |+\rangle^{\otimes N}$. The overall time-dependent Hamiltonian that governs the annealing process is expressed as
\begin{equation}
H(t) = A(t) H_D + B(t) H_P, \quad t \in [0,T],
\end{equation}
where $T$ denotes the total annealing time and the annealing schedules \(A(t)\) and \(B(t)\) satisfy \( A(0) \gg B(0) \) and \( A(T) \ll B(T) \), ensuring a gradual transition from \( H_D \) to \( H_P \). The quantum state evolves according to the Schrödinger equation:
\begin{equation}
i \hbar \frac{d}{dt} |\psi(t)\rangle = H(t) |\psi(t)\rangle.
\end{equation}

According to the adiabatic theorem, if the total evolution time \(T\) is sufficiently large compared to the inverse square of the minimum spectral gap \( \Delta_{\min} \), the system remains close to its instantaneous ground state:
\begin{equation}
|\psi(T)\rangle \approx |\psi_0\rangle,
\end{equation}
where \( |\psi_0\rangle \) is the ground state of the problem Hamiltonian. The adiabatic condition is formally given by:

\begin{equation}
T \gg \max_{t \in [0,T]} \frac{ \left| \langle \psi_1(t) | \frac{dH(t)}{dt} | \psi_0(t) \rangle \right| }{\Delta(t)^2}.
\end{equation}
From an algorithmic perspective, QA can be interpreted as an
analog optimization process governed by continuous quantum
dynamics, in which the computational trajectory is determined
by Hamiltonian evolution rather than discrete parameter updates. This distinguishes QA from variational methods by eliminating the need for classical optimization loops. However, practical realizations are subject to decoherence, thermal noise, and non-adiabatic transitions, resulting in probabilistic outputs characterized by:
\begin{equation}
P(m) = |\langle m | \psi(T) \rangle|^2,
\end{equation}
where \(m\) denotes a computational basis state. In QA, multiple measurements are required to identify the optimal solution with high confidence. The efficiency and scalability of QA are fundamentally governed by the Hamiltonian's spectral properties, annealing schedules, and environmental interactions. 

Many practical optimization problems can be reformulated as QUBO models. A generic QUBO problem is expressed as
\begin{equation}
\min_{x} \quad
x^{T}Qx
,
\label{eq:qubo}
\end{equation}
where $x_i \in \{0,1\}$, $i=1,\ldots,N$,
%
%\begin{equation}
%x_i \in \{0,1\},
%\qquad
%i=1,\ldots,N,
%\end{equation}
%
and $Q$ is a symmetric cost matrix. The binary variables can be transformed into Ising spin variables using
\begin{equation}
x_i
=
\frac{1-z_i}{2},
\qquad
z_i \in \{-1,+1\},
\label{eq:qubo_to_ising}
\end{equation}
thereby converting the QUBO formulation into an equivalent Ising Hamiltonian suitable for QA hardware.

For power-system applications, optimization objectives such as unit commitment, economic dispatch, optimal power flow, PMU placement, EV charging station deployment, facility location allocation, contingency analysis, and network reconfiguration can be reformulated as QUBO or Ising models. The resulting problem Hamiltonian encodes both the objective function and system constraints. QA subsequently searches for the ground-state configuration corresponding to the optimal solution, providing a promising framework for addressing large-scale optimization problems with exponentially growing solution spaces.
\subsubsection{Grover's Search Algorithm}
Grover’s search algorithm (GSA) is a quantum amplitude amplification procedure designed to locate a marked item within an unsorted database of size \(N = 2^n\) with quadratic speedup over classical exhaustive search \cite{grover1996fast, nielsen2010quantum}. The problem is formulated using an oracle function \(f(x): \{0,1\}^n \rightarrow \{0,1\}\), where \(f(x)=1\) if \(x\) is a solution (marked state) and \(f(x)=0\) otherwise. For the basic formulation, it is assumed that exactly one input $w$ satisfies $f(w)=1$. The case of multiple marked inputs can be handled using the more general amplitude-amplification formulation. The quantum oracle is implemented as a unitary operator \(U_f\) that encodes the solution via phase inversion:
\begin{equation}
   U_f |x\rangle = (-1)^{f(x)} |x\rangle, 
\end{equation}
which flips the phase of marked states while leaving unmarked states unchanged \cite{nielsen2010quantum}. The algorithm begins by preparing an equal superposition over all computational basis states using Hadamard gates:
\begin{equation}
|\psi_0\rangle = H^{\otimes n} |0\rangle^{\otimes n} = \frac{1}{\sqrt{N}} \sum_{x=0}^{N-1} |x\rangle.
\end{equation}
This uniform state ensures equal probability amplitudes for all candidates before amplitude amplification \cite{nielsen2010quantum}.

In GSA, Grover’s iteration operator \(G\) is constructed as the product of the oracle \(U_f\) and the diffusion operator \(D\), i.e., \(G = D U_f\). The diffusion operator performs inversion about the mean amplitude and is defined as
\begin{equation}
D = 2|\psi_0\rangle\langle \psi_0| - I,
\end{equation}
where \(I\) is the identity operator. Each application of \(G\) amplifies the probability amplitude of the marked state while suppressing all others. After \(k\) iterations, the quantum state evolves as
\begin{equation}
|\psi_k\rangle = G^k |\psi_0\rangle.
\end{equation}
Geometrically, the evolution can be interpreted as a rotation in a two-dimensional Hilbert subspace spanned by the marked state \(|w\rangle\) and the unmarked subspace \(|w^\perp\rangle\), with the rotation angle determined by \(\theta \approx \arcsin(1/\sqrt{N})\) \cite{boyer1998tight}. The optimal number of iterations required to maximize the probability of measuring the marked state is
\begin{equation}
k \approx \left\lfloor \frac{\pi}{4}\sqrt{N} \right\rfloor,
\end{equation}
after which a projective measurement yields the solution with high probability \cite{grover1996fast, boyer1998tight}. This quadratic speedup, from \(O(N)\) classically to \(O(\sqrt{N})\) quantumly, arises from coherent amplitude amplification governed by repeated reflections about the oracle-defined subspace and the initial uniform superposition \cite{grover1996fast, nielsen2010quantum}. As GSA provides quadratic speedup for unstructured search, it has potential applications in power-system problems that can be formulated as combinatorial search tasks, such as fault location identification, feature selection, contingency screening, unit commitment, network reconfiguration, and discretized optimal power flow.

\begin{table*}[t!]
\centering
\footnotesize
\caption{Overview of leading quantum computing platforms worldwide. The table is organized by technology type. }
\begin{tabular}{l l l l l l}
\hline
\textbf{Technology Type}         & \textbf{Computation Type}   & \textbf{Platform} & \textbf{Current Model}            & \textbf{Qubit No.} & \textbf{Origin} \\ \hline
\multirow{9}{*}{Superconducting} & \multirow{2}{*}{Annealing}  & D-Wave            & Advantage2                        & 4,400+                     & Canada          \\  
                                 &                             & Fujitsu           & -                                 & 256                        & Japan           \\ \cline{2-6} 
                                 & \multirow{7}{*}{Gate-based} & IBM               & Heron r3                          & 156                        & USA             \\  
                                 &                             & Google            & Willow                            & 105                        & USA             \\  
                                 &                             & Rigetti           & Cepheus                           & 107                        & USA             \\  
                                 &                             & IQM               & Radiance                          & 150                        & Finland         \\  
                                 &                             & OQC               & Toshiko                           & 32                         & UK              \\  
                                 &                             & Anyon Systems     & Qube\texttrademark & 6-54                       & Canada          \\  
                                 &                             & Origin Quantum    & Wukong 180                        & 180                        & China           \\ \hline
\multirow{3}{*}{Photonic}        & \multirow{2}{*}{Gate-based} & Xanadu            & Aurora                            & 12                         & Canada          \\  
                                 &                             & PsiQuantum        & -                                 & N/A                        & USA             \\ \cline{2-6}
                                 & Optical Transformations                      & ORCA Computing             & PT-2                            & -                        & UK \\ \hline
                                 
\multirow{4}{*}{Trapped-ion}     & \multirow{4}{*}{Gate-based} & IonQ              & Forte                             & 36                         & USA             \\  
                                 &                             & Quantinuum        & H2                                & 56                         & USA/UK          \\  
                                 &                             & Oxford Ionics     & -                                 & 16–64                      & UK              \\  
                                 &                             & AQT               & MARMOT                            & 20                         & Austria         \\ \hline
\multirow{4}{*}{Neutral-atom}    & \multirow{3}{*}{Gate-based} & Pasqal            & Orion Gamma                       & 140+                       & France          \\  
                                 &                             & Atom Computing    & AC1000                            & 1,200+                     & USA             \\  
                                 &                             & Infleqtion        & Sqale                             & 100+                       & USA             \\ \cline{2-6} 
                                 & Analog                      & QuEra             & Aquila                            & 256                        & USA             \\ \hline
\end{tabular}
\label{tab:QC_Platforms}
\end{table*}

\subsubsection{Quantum Machine Learning}
Quantum machine learning (QML) integrates quantum algorithms with classical learning frameworks, leveraging quantum computational primitives such as superposition, entanglement, and amplitude amplification to improve learning efficiency \cite{biamonte2017quantum}. Within this paradigm, quantum algorithms serve as \textit{modular subroutines} that accelerate computationally intensive tasks underlying classical machine learning. For example, the HHL algorithm efficiently solves linear systems, forming the basis for regression and support vector machine formulations \cite{rebentrost2014quantum}. Similarly, GSA offers a quadratic speedup for unstructured search and can support feature selection and optimization tasks \cite{schuld2015introduction}. These subroutines enable efficient linear algebra operations, similarity evaluation, and kernel estimation when data can be effectively encoded into quantum states. In general, a QML model is constructed by embedding classical data \(x\) into a quantum state \( |\psi(x)\rangle = U_{\mathrm{enc}}(x)|0\rangle^{\otimes n} \), where $U_{\mathrm{enc}}(x)$ is the data encoding (or feature map) unitary operator in QML models. The learning model is typically implemented using a PQC, also known as a VQC, denoted by $U(\theta)$. The PQC acts on the encoded state as
\begin{equation}
|\psi(x,\boldsymbol{\theta})\rangle = U(\boldsymbol{\theta})\, U_{\text{enc}}(x)\, |0\rangle^{\otimes n}.
\end{equation}
Here, $U(\boldsymbol{\theta})$ consists of trainable rotation gates and entangling layers that define the hypothesis space of the quantum model. The expectation value of an observable $\hat{O}$ is measured to define a cost function:
\begin{equation}
f(x,\boldsymbol{\theta}) = \langle \psi(x,\boldsymbol{\theta}) | \hat{O} | \psi(x,\boldsymbol{\theta}) \rangle,
\end{equation}
which is minimized (or maximized) using a classical optimizer in a hybrid loop. This expectation value represents a prediction, score, or learned response rather than the complete training cost. For a supervised dataset
\begin{equation}
\mathcal{D}
=
\{(x_j,y_j)\}_{j=1}^{M},
\end{equation}
a problem-dependent loss function may be defined as
\begin{equation}
\mathcal{L}(\boldsymbol{\theta})
=
\frac{1}{M}
\sum_{j=1}^{M}
\ell\!\left(
f(x_j,\boldsymbol{\theta}),y_j
\right),
\end{equation}
where $\ell(\cdot,\cdot)$ is an appropriate sample-level loss, such as 
mean-squared error for regression or cross-entropy for classification. The 
trainable parameters are obtained through the hybrid optimization problem
\begin{equation}
\boldsymbol{\theta}^{*}
=
\arg\min_{\boldsymbol{\theta}}
\mathcal{L}(\boldsymbol{\theta}).
\end{equation}
The quantum processor prepares the parameterized states and estimates the 
required observables, while the classical processor evaluates the loss and 
updates $\boldsymbol{\theta}$.
The inclusion of VQC/PQC enables expressive feature mappings in Hilbert space, potentially achieving an advantage over classical neural architectures for certain structured learning problems.

Within this framework, variational quantum algorithms (VQE and QAOA) can also be used, representing a class of hybrid QML models in which the learning task is formulated as an optimization problem over PQC \cite{cerezo2021variational}. Despite the theoretical advantages of QML, its practical implementation remains constrained by hardware and algorithmic limitations, including noise, decoherence, limited circuit depth, and the challenge of implementing nonlinear activation functions within linear quantum operations. In power and energy systems, QML enables advanced data-driven solutions such as resource forecasting (e.g., load and renewable prediction), fault diagnostics and anomaly detection, and battery management, including state-of-charge and state-of-health estimation. Furthermore, QML supports optimization-driven applications such as economic dispatch, unit commitment, and network reconfiguration, highlighting its potential to address nonlinear, high-dimensional, and combinatorial challenges in next-gen intelligent power grids.

\section{Quantum Computing Platforms}\label{Quantum Computing Platforms}
\subsection{Existing Quantum Computing Real Hardware}

Quantum computing platforms are broadly classified according to the physical realization of qubits, including \textit{superconducting circuits, trapped-ion systems, photonic architectures, neutral-atom platforms}, and alternative paradigms such as \textit{quantum annealing and analog quantum computing}. These technologies differ fundamentally in coherence time, qubit scalability, and operational fidelity, which directly influence their applicability to large-scale quantum computation.

\textbf{Superconducting qubits}, based on Josephson junction circuits operating at millikelvin temperatures, are currently the most technologically mature platform. Typical coherence times for superconducting qubits are in the range of $20~\mu s$ to $200~\mu s$, with recent state-of-the-art devices demonstrating coherence up to approximately $1~ms$ \cite{kjaergaard2020superconducting, somoroff2023millisecond}. These relatively short coherence times are compensated by fast gate operations (on the order of nanoseconds) and compatibility with semiconductor fabrication processes. 
In contrast, \textbf{trapped-ion quantum computing} platforms encode qubits in the internal states of ions confined in electromagnetic traps and manipulated using laser fields. These systems exhibit significantly longer coherence times, typically ranging from $1~s$ to $10^3~s$ (i.e., up to several minutes), with extreme cases reported beyond this range under optimized conditions \cite{haffner2008quantum,wang2017single}. Such long coherence times enable very high gate fidelities, although scaling to large qubit numbers remains challenging.

\begin{table*}[htbp]
\centering
\caption{Comprehensive comparison of quantum computing frameworks and simulators. S: Simulator, L: Library/SDK, H: Hybrid. Support levels: Strong, Moderate, Limited, None. Q+C: Quantum-Classic implementation.}
\label{tab:full_quantum_frameworks}
\footnotesize

\begin{tabularx}{\textwidth}{p{3.2cm} c c c c c c c c p{1.3cm}}
\hline
\textbf{Toolkit} & \textbf{Type} & \textbf{Hybrid} & \textbf{Noise} & \textbf{Tensor} & \textbf{Optimization} & \textbf{Linear} & \textbf{Quantum} & \textbf{QML} & \textbf{Language} \\
 &  & \textbf{Q+C} & \textbf{Simulation} & \textbf{Network} &  & \textbf{System} & \textbf{Simulation} &  &  \\
\hline

Qiskit \cite{javadi2024quantum} & H & Yes & Strong & Moderate & Strong & Moderate & Strong & Strong & Python \\
Cirq \cite{isakov2021simulations} & H & Yes & Moderate & Limited & Limited & Limited & Limited & Limited & Python \\
ProjectQ \cite{steiger2018projectq} & L & Limited & Limited & Limited & Limited & Limited & Moderate & Limited & Python \\
PyQuil \cite{smith2016practical,karalekas2020quantum} & L & Yes & Moderate & Limited & Limited & Limited & Limited & Limited & Python \\
PennyLane \cite{bergholm2018pennylane} & H & Yes & Moderate & Moderate & Strong & Moderate & Strong & Strong & Python \\
TensorFlow Quantum \cite{broughton2020tensorflow} & H & Yes & Moderate & Limited & Strong & Limited & Moderate & Strong & Python \\
Q\# (QDK) \cite{svore2018q} & H & Yes & Moderate & Limited & Moderate & Moderate & Strong & Limited & Q\#, Python \\
QuTiP \cite{lambert2026qutip} & L & No & Strong & Limited & Limited & Limited & Strong & Limited & Python \\
Qulacs \cite{suzuki2021qulacs} & S & No & Moderate & Limited & Moderate & Limited & Moderate & Limited & Python/C++ \\
Intel QSim \cite{guerreschi2020intel} & S & No & Moderate & Limited & Limited & Limited & Moderate & Limited & C++/Python \\
QX Simulator \cite{khammassi2017qx} & S & No & Moderate & Limited & Limited & Limited & Moderate & Limited & C++/Python \\
Yao.jl \cite{luo2020yao} & L & Yes & Moderate & Moderate & Moderate & Limited & Moderate & Moderate & Julia \\
Strawberry Fields \cite{killoran2019strawberry} & H & Yes & Moderate & Limited & Moderate & Limited & Strong & Strong & Python \\
Quimb \cite{gray2018quimb} & L & No & None & Strong & Limited & Moderate & Strong & Limited & Python \\
ITensor \cite{fishman2022itensor} & L & No & None & Strong & Limited & Moderate & Strong & Limited & C++/Julia \\
QTensor \cite{lykov2021performance} & L & No & None & Strong & Limited & Moderate & Moderate & Limited & Python \\
Quantum++ \cite{gheorghiu2018quantum++} & L & No & Limited & Limited & Limited & Limited & Limited & Limited & C++ \\
QCGPU \cite{kelly2018simulating} & S & No & None & Limited & Limited & Limited & Limited & Limited & Python \\
Tequila \cite{kottmann2021tequila} & H & Yes & Moderate & Moderate & Strong & Limited & Strong & Moderate & Python \\
Qibo \cite{efthymiou2022qibo} & H & Yes & Moderate & Moderate & Strong & Strong & Strong & Strong & Python \\
ExaTN \cite{lyakh2022exatn} & L & No & None & Strong & Moderate & Moderate & Strong & Limited & C++ \\
QFlex \cite{villalonga2019flexible} & S & No & Limited & Strong & Limited & Limited & Moderate & Limited & Python/C++ \\
QuEST \cite{jones2019quest} & S & No & Moderate & Limited & Limited & Limited & Strong & Limited & C/C++ \\
Qrack \cite{strano2023exact} & S & No & Moderate & Limited & Limited & Limited & Moderate & Limited & C++ \\
TorchQuantum \cite{wang2022quantumnas} & H & Yes & Moderate & Limited & Strong & Limited & Moderate & Strong & Python \\
NetKet \cite{carleo2019netket} & L & No & Moderate & Moderate & Moderate & Limited & Strong & Moderate & Python \\
Qadence \cite{seitz2025tool} & H & Yes & Moderate & Moderate & Strong & Limited & Strong & Strong & Python \\
Qrisp \cite{seidel2024qrisp} & H & Yes & Moderate & Limited & Moderate & Moderate & Strong & Moderate & Python \\
QODE & L & No & None & Moderate & Limited & Limited & Strong & Limited & Python \\
Staq \cite{amy2020staq} & L & No & Limited & Limited & Limited & Limited & None & Limited & C++ \\
QuantumOptics.jl \cite{kramer2018quantumoptics} & L & No & Strong & Moderate & Limited & Limited & Strong & Limited & Julia \\
QuantumInformation.jl \cite{gawron2018quantuminformation} & L & No & Moderate & Moderate & Limited & Limited & Moderate & Limited & Julia \\
CUDA-Q \cite{cudaq} & H & Yes & Strong & Moderate & Strong & Limited & Strong & Moderate & Python/C++ \\
TKET & L & Yes & Moderate & Limited & Moderate & Moderate & Moderate & Limited & Python/C++ \\
QTensorAI \cite{qtensorai_github} & L & No & None & Strong & Limited & Limited & Moderate & Strong & Python \\
Ocean SDK \cite{dwave_ocean_sdk} & L & Yes & None & None & Strong & None & Limited & Limited & Python \\
\hline
\end{tabularx}
\vspace{-10pt}
\end{table*}

\textbf{Photonic quantum computing} utilizes photons as qubits, typically encoded in polarization, time-bin, or spatial modes. Due to the extremely weak interaction between photons and the environment, photonic qubits exhibit effectively negligible intrinsic decoherence. In practical implementations, coherence is limited by optical losses and system imperfections rather than true decoherence. Reported effective coherence times correspond to photon propagation and delay-line storage durations, typically ranging from $1~\mu s$ to several ms, depending on the optical medium and device architecture \cite{o2009photonic}. In integrated photonic circuits and fiber-based systems, coherence can be preserved over long distances, making photonic platforms particularly attractive for quantum communication and distributed quantum computing.
\textbf{Neutral-atom platforms}, which rely on atoms trapped in optical tweezers and interacting via Rydberg states, typically exhibit coherence times in the range of $10~\mu s$ to $1~ms$ \cite{saffman2010quantum}. These systems support highly scalable and reconfigurable qubit arrays, with demonstrated implementations exceeding a thousand qubits.
\textbf{Quantum annealing systems}, implemented using superconducting flux qubits, generally exhibit shorter coherence times in the range of $1~\mu s$ to $10~\mu s$ due to design trade-offs that favor connectivity and large-scale integration over coherence preservation \cite{weber2017coherent}. Despite this limitation, such systems can scale to thousands of qubits for solving optimization problems.

Building on these technological characteristics, current quantum computing platforms vary widely in qubit counts, computational models, hardware implementations, and geographic distribution. Superconducting gate-based platforms dominate industrial development, with systems from IBM, Google, Rigetti, IQM, and Origin Quantum reaching qubit counts from tens to over 150. Trapped-ion platforms, including IonQ and Quantinuum, typically operate with tens of qubits but offer superior coherence and fidelity. Photonic platforms, such as Xanadu and PsiQuantum, remain in relatively early stages, with limited or evolving qubit counts but strong potential for scalability via integrated optics. In contrast, neutral-atom platforms, including Atom Computing, QuEra, and Pasqal, demonstrate large-scale qubit arrays ranging from hundreds to over one thousand qubits. Furthermore, QA systems, most notably those developed by D-Wave, achieve qubit counts in the thousands, albeit within a specialized computational framework. These platforms are distributed across North America, Europe, and Asia, reflecting the global and competitive nature of quantum hardware development. A consolidated comparison of these platforms is presented in Table \ref{tab:QC_Platforms}.
\vspace{-10pt}

\subsection{Quantum Simulators and Software}
%PennyLane, Qiskit, Paddle Quantum, and so on

\begin{table*}[htbp]
\centering
\footnotesize
\caption{Comparison of quantum platforms classified by hardware‑agnosticity levels across infrastructure (hardware‑agnostic), middleware (hybrid), and software abstraction layers.}
\begin{tabular}{p{2.8cm}p{2.5cm}p{1.9cm}p{2.9cm}p{3.0cm}p{2.2cm}}
\hline
\textbf{Agnostic Level} & \textbf{Platform} & \textbf{Category} & \textbf{Quantum Providers / Backends} & \textbf{Programming Languages / SDKs} & \textbf{Access Model} \\
\hline

\multirow{10}{*}{Hardware-Agnostic} 
& Strangeworks \cite{strangeworks} & Quantum Cloud Platform & IBM, IonQ, IQM, QuEra, Rigetti, D-Wave, and others & Python, Qiskit, Cirq, Q\#, OpenQASM & Subscription, Usage \\
\cline{2-6}
& qBraid \cite{Hill_qBraid-SDK_Platform-agnostic_quantum_2026} & Quantum Cloud Platform & AWS Braket, IBM, IonQ, Rigetti, QuEra, IQM, others & Python (qBraid SDK, Qiskit, Cirq, Braket SDK) & Subscription, Usage \\
\cline{2-6}
& AWS Braket \cite{braket2019} & Quantum Cloud Platform & IonQ, Rigetti, IQM, QuEra, OQC & Python (Braket SDK, PennyLane) & Pay-per-use \\
\cline{2-6}
& Azure Quantum \cite{Azure_Quantum} & Quantum Cloud Platform & IonQ, Quantinuum, Rigetti, Pasqal & Q\#, Python, Qiskit, Cirq & Subscription, Usage \\
\hline

\multirow{9}{*}{Hybrid (Middle Layer)} 
& Orquestra \cite{Orquestra_Zapata} & Workflow Orchestration Platform & D-Wave, IBM, IonQ, Quantinuum, Rigetti, and simulators & Python & Subscription \\
\cline{2-6}
%& QC Ware Forge & Quantum Algorithm / Application Platform & IBM, Rigetti, D-Wave (integrated backends) & Python (proprietary libraries) & Enterprise subscription \\
& Superstaq \cite{Superstaq_Infleqtion} & Quantum Compiler / Optimization Platform & Infleqtion, IBM, AQT, and others & Python (Qiskit, Cirq) & Subscription, Usage \\
\cline{2-6}
& Q-CTRL \cite{Q-CTRL_Fire_Opal} & Quantum Error Mitigation Platform & IBM, IonQ, AWS Braket & Python & Subscription \\
%\cline{2-6}
%& Multiverse Computing & Quantum Software Platform & IBM, IonQ, and others & Python & Enterprise subscription \\
\hline

\multirow{9}{*}{Software-Agnostic$^*$} 
& PennyLane \cite{pennylane_website} & Quantum Software Framework & IBM, AQT, AWS Braket, IonQ, Rigetti, OQC etc., and simulators & Python & Open-source, and cloud integrations \\ \cline{2-6}
& Classiq \cite{classiq_platform} & Quantum Software Platform & IBM, IonQ, Rigetti (via AWS Braket, Azure Quantum) & Python, Qmod & Subscription \\
\cline{2-6}
& CUDA-Q \cite{cudaq} & Quantum Programming Framework & QPU (IQM, IonQ, Infleqtion, and others) backends, simulators & C++, Python & Open-source, and cloud integrations \\
\hline
\end{tabular}
$*$: Indirectly hardware-agnostic via plugins or cloud integrations
\label{hardware_agnostic_classification}
\vspace{-10pt}
\end{table*}

The presented comparison in Table \ref{tab:full_quantum_frameworks} provides a detailed technical landscape of open-source quantum computing frameworks by jointly evaluating their architectural design, simulation fidelity, and algorithmic capabilities. A clear distinction emerges between fully integrated hybrid environments and standalone simulation engines. Hybrid platforms such as Qiskit, Cirq, and PennyLane demonstrate comprehensive functionality by supporting both circuit construction and quantum–classical feedback loops, which are essential for variational algorithms. These platforms consistently exhibit strong capabilities in optimization and QML, reflecting their design focus on PQCs and iterative training workflows. In contrast, standalone simulators such as QuEST, Qulacs, and Intel Quantum Simulator prioritize computational efficiency and are typically implemented in lower-level languages such as C or C++, enabling faster state evolution but offering limited integration with higher-level algorithmic pipelines.

The ability to model realistic hardware imperfections varies significantly across frameworks and is a major differentiator. Platforms such as Qiskit and QuTiP provide robust support for noise-aware simulations, incorporating advanced quantum channels and density-matrix evolution to emulate decoherence and operational errors. Several other tools, including Cirq, PyQuil, and Strawberry Fields, offer moderate noise modeling capabilities suitable for approximate NISQ analysis. In contrast, many high-performance simulators and lightweight libraries provide limited or no noise support, as they focus on ideal state-vector evolution to maximize computational speed. Similarly, only a subset of frameworks incorporates tensor network techniques; tools such as TensorFlow Quantum, QuantumOptics.jl, and TensorNetwork-based implementations leverage structured decompositions to mitigate exponential scaling under restricted-entanglement conditions, whereas most frameworks rely on conventional full-state representations.

From an algorithmic perspective, optimization and problem-specific capabilities reveal strong heterogeneity. Frameworks such as PennyLane, Tequila, Qibo, and Qiskit provide robust support for integrating classical optimization routines with quantum circuit execution, enabling efficient implementation of variational quantum algorithms. However, support for solving linear systems remains relatively limited across most platforms, indicating that algorithms such as HHL and its variants are not yet broadly optimized in practice. In contrast, quantum system simulation, particularly Hamiltonian evolution, is widely supported, with many frameworks demonstrating moderate to strong capability in this domain. This reflects the historical importance of quantum simulation as a primary application of quantum computing. Additionally, QML support is concentrated in a smaller set of frameworks that provide differentiable programming interfaces and seamless integration with classical machine learning libraries, whereas traditional simulators typically lack these features.

Finally, the implementation languages highlight a fundamental trade-off between usability and performance. Python is the primary interface in most hybrid frameworks because its simplicity and extensive ecosystem enable rapid prototyping and accessibility. However, performance-critical simulators frequently utilize C++ or GPU-accelerated backends to handle large-scale computations more efficiently, with some frameworks adopting hybrid architectures that combine Python frontends with compiled kernels. Julia-based tools offer an intermediate approach by providing high-level syntax with near-native execution speed. Overall, the comparison underscores that no single framework simultaneously maximizes all capabilities; instead, the choice of simulator depends on whether the priority is realistic noise modeling, hybrid optimization workflows, high-performance, large-scale simulation, or specialized applications such as QML.

\vspace{-10pt}

\subsection{Hardware‑Agnostic Quantum Platform Ecosystems}
Beyond single-vendor quantum ecosystems such as IBM Quantum, D-Wave, and Rigetti, which provide access to proprietary hardware and software stacks, a new class of unified platforms has emerged in recent years, enabling access to multiple quantum processing units (QPUs) through a common interface. These platforms let users run quantum programs across diverse hardware technologies with fewer code changes and backend-specific adaptations, reducing vendor lock-in and improving portability. As a result, hardware-agnosticism in quantum computing has evolved into a multi-layered concept that spans the entire ecosystem, from direct hardware access to middleware orchestration and software-level abstraction. This layered perspective provides a structured understanding of how modern quantum platforms facilitate interoperability, flexibility, and scalability in quantum application development. Hardware-agnosticity is not a binary property but a continuum of abstraction, with different platforms addressing interoperability at the infrastructure, middleware, or software-development level. Based on the degree of abstraction, quantum platforms fall into three categories: hardware-agnostic platforms, hybrid (middle-layer) platforms, and software-agnostic frameworks, as shown in Table \ref{hardware_agnostic_classification}.

Hardware-agnostic platforms, including Strangeworks \cite{strangeworks}, qBraid \cite{Hill_qBraid-SDK_Platform-agnostic_quantum_2026}, AWS Braket \cite{braket2019}, and Azure Quantum \cite{Azure_Quantum}, operate at the infrastructure level and provide direct access to multiple hardware providers through a unified interface. These platforms abstract the heterogeneity of different quantum technologies by exposing standardized APIs for job submission and execution. For example, AWS Braket enables users to run quantum circuits on diverse hardware backends, including IonQ, Rigetti, IQM, QuEra, and OQC, through a common cloud-based execution interface. Similarly, Azure Quantum integrates providers such as IonQ, Quantinuum, Rigetti, and Pasqal, while supporting multiple programming environments including Q\# and Python-based frameworks. qBraid extends this capability by supporting multiple quantum software development kits (SDKs) (e.g., Qiskit, Cirq, and Braket), enabling interoperability across quantum software ecosystems and facilitating cross-platform execution across heterogeneous quantum computing environments. Strangeworks aggregates multiple hardware and classical compute resources, providing a flexible environment for enterprise and research applications. These platforms form the foundational infrastructure layer, providing direct, unified access to heterogeneous QPUs.

At the intermediate level, hybrid (middle-layer) platforms such as Orquestra (Zapata Quantum) \cite{Orquestra_Zapata}, Superstaq (Infleqtion) \cite{Superstaq_Infleqtion}, and Q-CTRL (Fire Opal) \cite{Q-CTRL_Fire_Opal} provide middleware functionality that operates on top of hardware-agnostic infrastructures. These platforms do not directly expose hardware; instead, they focus on workflow management, performance optimization, and execution-level enhancements across multiple backends. Orquestra (Zapata AI) is a workflow orchestration platform that enables the design and execution of hybrid quantum-classical pipelines across cloud-accessible quantum backends such as IBM Quantum, AWS Braket, and Azure Quantum. In contrast, Superstaq is a hardware-agnostic compiler and optimization platform that improves quantum circuit performance through techniques such as noise-aware compilation and qubit mapping across multiple backends, including Infleqtion, IBM, and AQT. Q-CTRL (Fire Opal) complements this by focusing on quantum control, error suppression, and performance optimization, improving the reliability of quantum computations through noise suppression and performance tuning across backends such as IBM, IonQ, and AWS Braket. Together, these platforms form a middleware layer that bridges hardware access and high-level applications by providing orchestration, optimization, and execution-enhancement capabilities.

At the highest level of abstraction, software-agnostic platforms such as PennyLane \cite{pennylane_website}, Classiq \cite{classiq_platform}, and CUDA-Q \cite{cudaq} provide frameworks that decouple quantum program development from specific hardware implementations. These platforms let users design quantum algorithms in a hardware-independent way and run them across different backends through plugin-based or cloud-integrated mechanisms. PennyLane offers a Python-based interface with built-in support for QML and hybrid algorithms, enabling seamless execution on platforms such as IBM Quantum, AWS Braket, Xanadu’s photonic systems, Quantinuum, IonQ, and Rigetti backends. Classiq, by contrast, focuses on model-based quantum program synthesis, automatically generating hardware-compatible quantum circuits from high-level functional specifications, allowing users to define quantum algorithms declaratively and target multiple hardware providers via integrated cloud platforms. CUDA-Q by NVIDIA extends this paradigm by providing a high-performance quantum programming framework that integrates quantum workflows with classical high-performance computing environments, offering C++ and Python interfaces for hybrid quantum-classical execution on simulators and external hardware backends. Unlike infrastructure-level platforms, these frameworks primarily focus on program development and abstraction rather than direct hardware provisioning, relying on backend integrations, plugins, or cloud services for execution.

\begin{table*}[t]
\centering
\caption{Classification of smart grid applications (based on Section \ref{QC Smart Grid Applications}) into computational problem types.}
\label{tab:problem_classification}
\footnotesize
\begin{tabular}{p{4.0cm} p{3.8cm} p{8.2cm}}
\hline
\textbf{Application Category} &
\textbf{Section V Subsection} &
\textbf{Computational Problem Type} \\
\hline

\multirow{4}{*}{Monitoring \& Estimation}
& A. Optimal PMU Placement
& Combinatorial optimization \\
& C. State Estimation
& Linear/nonlinear least-squares optimization \\
& J. Power System Simulation
& Differential-algebraic equations (DAEs), numerical simulation \\
& O1. Power Flow Analysis
& Nonlinear algebraic equations, sparse linear systems \\
\hline

\multirow{4}{*}{System Planning}
& B. Facility Location Allocation
& Combinatorial optimization \\
& P. Unit Commitment
& Mixed-integer combinatorial optimization (MILP/MIQP/MINLP) \\
& Q. Energy Management
& Mixed-integer, convex/nonconvex, and multi-objective optimization \\
& R. Grid Partitioning and Others
& Graph partitioning, combinatorial optimization, network optimization \\
\hline

\multirow{3}{*}{Operation \& Control}
& H. Smart Grid Control
& Dynamic optimization, optimal control \\
& O2. Optimal Power Flow
& Nonlinear optimization, constrained optimization \\
& Q. Economic Dispatch
& Continuous optimization, convex/nonconvex optimization \\
\hline

\multirow{3}{*}{Security, Reliability \& Resilience}
& E. Contingency Analysis
& Combinatorial optimization, linear systems, system evaluation \\
& F. Reliability Analysis
& Probabilistic analysis, stochastic simulation, uncertainty quantification \\
& G. Restoration \& Resiliency
& Stochastic and combinatorial optimization, network reconfiguration \\
\hline

\multirow{1}{*}{Stability \& Assessment}
& D. Stability Assessment
& Nonlinear dynamical systems, differential equations \\
\hline

\multirow{4}{*}{Data-Driven Intelligence}
& K. Anomaly Detection
& Machine learning, classification, clustering \\
& L. Fault Diagnostics
& Pattern recognition, classification, inference \\
& M. Battery Management Systems
& State estimation, control, machine learning \\
& N. Resource Forecasting
& Time-series forecasting, machine learning, probabilistic prediction \\
\hline

\multirow{1}{*}{Digital Twin Systems}
& I. Digital Twin Operation
& Physics-informed modeling, hybrid physics--data systems \\
\hline
\end{tabular}
\vspace{-6pt}
\end{table*}

\section{Quantum Computing for Smart Grids Applications} \label{QC Smart Grid Applications} 
The smart grid applications using quantum computing discussed in this section can be systematically categorized into functional domains, including monitoring and estimation, planning and optimization, operation and control, security and resilience, stability assessment, data-driven intelligence, and digital twin systems. As summarized in Table~\ref{tab:problem_classification}, these domains map naturally to a set of fundamental computational problem classes. Monitoring and estimation tasks are typically formulated as least-squares problems, algebraic systems, and differential–algebraic equation (DAE) models, while planning and optimization problems are predominantly expressed as combinatorial and mixed-integer formulations with convex, nonconvex, and multi-objective characteristics. Operation and control tasks correspond to constrained nonlinear and continuous optimization problems, whereas security and resilience applications involve probabilistic analysis, stochastic optimization, and combinatorial formulations. Stability assessment involves nonlinear dynamical systems, while data-driven intelligence relies on machine learning and time-series modeling. Digital twin systems also integrate physics-informed and data-driven models for real-time system representation and optimization. The following subsections discuss these application domains and current quantum computing trends in detail, and summarize notable contributions across quantum hardware, simulators, and algorithms.

\subsection{Optimal PMU Placement}
Ensuring full network observability is fundamental to reliable monitoring, protection, and control of modern power systems. However, because phasor measurement units (PMUs) are expensive and installation is constrained, determining optimal placement to achieve full observability with the fewest devices remains a challenging combinatorial optimization problem. Quantum optimization is increasingly recognized as a promising approach to addressing this challenge, particularly given the exponential scaling limitations of classical techniques. Early work reformulates the PMU placement problem as a minimum dominating set problem and expresses it using QUBO/Binary Quadratic Model (BQM) representations, which can be solved on D‑Wave annealers; these studies demonstrated encouraging performance on small IEEE test systems, although embedding challenges emerge as system size increases \cite{jones2020computational}. More recent developments explore gate‑based paradigms that combine QAOA with graph learning. Specifically, Graph Convolutional Networks generate informed initial states, accelerating convergence and improving solution quality for PMU placement tasks \cite{jiang2026qaoa, jiang2025accelerating}. These results suggest that enhanced encoding and initialization strategies can play a critical role in improving scalability. Beyond direct PMU placement, quantum optimization settings have also addressed broader observability planning and metering design. QAOA‑driven frameworks, for example, incorporate observability constraints and customized objective functions to better capture operational requirements and often outperform classical heuristics \cite{jiang2025optimal}. In related work, researchers treat general meter‑placement problems as QUBO instances and solve them via annealing or hybrid algorithms, demonstrating feasibility and performance gains as system dimensions and sampling complexity increase \cite{ganeshamurthy2025quantum}.

\subsection{Facility Location Allocation}
Facility location–allocation problems are central to energy infrastructure planning, where decisions must balance cost, accessibility, and system-level constraints. In power systems, this matters for applications such as electric vehicle charging station placement (EVCSP), where optimal siting directly affects grid performance and user convenience. A common methodology reformulates placement decisions as QUBO models, solves them with quantum annealers, and then refines the solutions with classical optimization techniques. For example, combining QA-based initialization with genetic algorithms has been shown to improve spatial alignment between charging stations and points of interest \cite{chandra2022towards}. Decomposition-based strategies have also been proposed, in which quantum annealers handle master allocation problems while classical methods resolve operational constraints \cite{veshchezerova2023hybrid}. Other hybrid designs assign placement tasks to quantum solvers while retaining grid analysis in classical processes, achieving noticeable runtime improvements \cite{rao2023hybrid}, especially for large-scale systems. These early developments highlight the effectiveness of hybrid quantum–classical frameworks.

At a more detailed level, multi-objective EVCSP formulations have also benefited from both quantum and quantum-inspired optimization methods. BQM-based approaches incorporate metrics such as cost, travel time, and traffic conditions to determine suitable station locations \cite{subramanian2023ev}, while QA-based formulations efficiently address competing objectives, including coverage and spacing \cite{sakib2024quantum}. Alternative strategies include Grover-based search techniques that bypass explicit QUBO formulations while offering quadratic speedups \cite{radvand2024quantum}, as well as digital annealers that perform well in large urban deployment scenarios \cite{ou2025solving}. The current literature consistently supports hybrid frameworks as a practical approach to balancing computational efficiency and modeling accuracy. A broader perspective on quantum optimization in energy systems emerges when linking these application-specific studies with foundational problem formulations. For example, facility location–allocation problems map to a quadratic assignment problem and subsequently to Ising/QUBO representations, enabling solutions via both annealing and gate-based methods such as VQE \cite{ajagekar2019quantum}. This approach achieves near-optimal results with fast runtimes for moderate-sized instances. Similarly, generation siting is formulated as a Max-Cut problem and solved using QAOA, demonstrating compatibility with NISQ hardware \cite{hidary2020application}. Building on these ideas, ref. \cite{yin2025optimised} applies QUBO formulations to practical battery placement problems in distribution systems. Extending QUBO-based optimization to energy storage planning, a hybrid quantum–classical framework effectively explores large placement search spaces and generates high-quality siting solutions for distribution networks \cite{hasan2026quantum}.

\subsection{State Estimation}
State estimation is typically formulated as a large-scale nonlinear least-squares optimization problem, aimed at reconstructing system states from redundant and noisy measurements. Accurate state estimation is essential for reliable monitoring, control, and operation, yet the growing size and complexity of modern grids pose significant computational challenges, motivating the exploration of quantum-enhanced techniques. Research on quantum approaches to state estimation has largely concentrated on overcoming NISQ‑era limitations, including restricted precision, circuit depth, and qubit availability. Recent studies adapt quantum linear solvers, such as HHL, to distributed AC state estimation, in which simplified circuit implementations reduce circuit depth while maintaining acceptable accuracy on IEEE test systems \cite{NEW_State_Estimation}. To further address encoding inefficiencies, researchers have introduced iterative refinement strategies, enabling accurate DC state estimation with fewer qubits \cite{tran2025noise}. Despite these advances, comprehensive evaluations suggest that HHL‑ and VQLS‑based approaches face challenges related to state preparation and measurement overhead. In contrast, QAOA and annealing formulations appear better suited to near‑term applications \cite{Stoyanova2025QuantumCM}. Complementary work explores hybrid and variational methodologies to improve robustness and scalability. Quantum implementations of Gauss–Newton and preconditioned quantum linear solvers with the HHL algorithm demonstrate that estimation accuracy comparable to classical techniques can be achieved while improving tolerance to disturbances \cite{feng2022quantum}. Efforts to design NISQ‑compatible circuits have further reduced depth requirements, enabling practical estimation on larger networks \cite{feng2024noisy}. In addition, learning-based quantum approaches, such as reinforcement learning (RL) augmented with variational circuits, have improved convergence speed and dynamic parameter estimation \cite{addo2025deep}. 

\vspace{-6pt}
\subsection{Stability Assessment}
Maintaining system stability following disturbances remains a fundamental requirement for secure power system operation, particularly as increasing penetration of inverter-based resources introduces new dynamics and uncertainties. In this context, stability assessment is commonly treated as a dynamic prediction and classification problem, where system trajectories are analyzed to determine post-disturbance security. Interest in quantum computing for stability analysis has grown, particularly in learning‑based and feature‑extraction approaches. Hybrid and federated learning frameworks incorporating quantum circuits have been shown to enhance scalability and reduce communication overhead while maintaining robustness to noise \cite{yu2025quantum, ren2024enhancing}. Moreover, quantum-enhanced RL further improves adaptability in inverter‑dominated systems \cite{masoumi2025quantum}. For transient stability prediction and classification tasks, techniques such as variational classifiers and quantum kernels embed system states into higher‑dimensional spaces, thereby improving system performance \cite{zhou2022noise, sabadra2024quantum}. In parallel, algorithms such as QFT and quantum principal component analysis (QPCA) accelerate dimensionality reduction and transient feature extraction, thereby offering notable speedups over classical methods \cite{li2023transient, chen2024quantum}.

Recent efforts have increasingly focused on robustness, distributed learning, and privacy preservation. For example, quantum models incorporating adversarial training improve resilience to perturbations \cite{yu2024quantum}, while federated quantum learning enables multi‑region analysis without exchanging raw data \cite{yu2024quantum_2}. Besides, extensions toward short‑term voltage stability introduce explainable and digital‑twin‑enhanced approaches \cite{ren2024esqfl}. More advanced designs combine quantum circuits with Transformer attention architectures, demonstrating potential to improve feature representation and robustness \cite{li2026qstaformer}. Beyond electromechanical stability, quantum learning also addresses data-scarce and techno-economic stability scenarios \cite{ahmad2025qm}, while secure multi-party frameworks combine quantum key distribution with federated learning for protected model aggregation \cite{ren2024qfdsa}.

\vspace{-2pt}
\subsection{Contingency Analysis}
Power system operation requires the rapid identification of critical component outages that could jeopardize system security, especially under credible $N-k$ failure scenarios. This task is intrinsically a combinatorial screening and ranking problem, in which a large number of potential contingencies must be evaluated to efficiently detect the most critical events. The recent quantum contingency analysis framework, for instance, employs VQC and VQLS to evaluate steady‑state conditions, achieving theoretical speedups and favorable qubit scaling with effective noise-mitigation strategies suited to NISQ hardware \cite{feng2025quantum}. Complementing this, the QA–based backward $N-k$ formulation encodes worst-case contingency selection as QUBOs, enabling parallel exploration of multiple fault combinations, achieving significant speed‑ups, and identifying critical outages and stressed lines \cite{cremer2025n}. Beyond annealing, QAOA-based methods incorporate power-flow deviations and outage constraints into the Hamiltonians, delivering near-optimal results for small systems and highlighting trade-offs between circuit depth and sampling \cite{jiang2025quantum}. Recent ideal simulations on the IEEE 9-bus system further show that QAOA can recover optimal line-contingency solutions and near-optimal system-wide contingency solutions, although low feasible-sample rates remain a limitation \cite{lange2026contingency}. Hybrid approaches further integrate quantum RL (QRL), in which variational circuits enhance policy learning and outperform classical baselines on grid‑security decision tasks \cite{peter2025quantum}.

In parallel, broader studies evaluate quantum algorithms and theoretical foundations for contingency analysis. Benchmarking of Grover search, annealing, and photonic methods identifies potential speedups but also highlights current challenges, such as circuit depth, penalty tuning, and hardware limitations \cite{neumann2024quantum}. Early formulations demonstrated exact, parallelizable post‑contingency flow computation on small systems \cite{antoli2023quantum}. Foundational work also highlights that quantum linear‑system solvers, such as HHL, could eventually provide exponential speed-ups for large‑scale $N-k$ analysis by reformulating power‑flow equations into quantum‑compatible forms \cite{eskandarpour2020quantum_security}. These studies show a clear progression toward hybrid quantum–classical frameworks that enhance scalability and speed in security assessment, while current performance remains constrained by NISQ‑era limitations.

\subsection{Reliability Analysis}
Assessing power system reliability requires quantifying the likelihood and impact of component failures under uncertainty, often involving large-scale stochastic simulations. This task is fundamentally a probabilistic inference and risk evaluation problem, where various system performance metrics must be estimated from high-dimensional uncertainty spaces. In reliability assessment, quantum approaches have primarily focused on accelerating probabilistic evaluation by replacing classical sampling with amplitude‑based estimation. Early QAE applications showed that it can compute reliability indices with significantly improved sampling efficiency \cite{nikmehr2022quantum}. Moreover, quantum Bayesian frameworks incorporate dependencies among system components, allowing for a more accurate representation of system‑level reliability behavior \cite{carrascal2023bayesian}. Variants of amplitude estimation further expand their applicability to stochastic operational studies involving uncertainty sources such as wind generation \cite{jong2023quantum}. Complementary efforts integrate Bayesian inference with amplitude amplification to identify rare but impactful events, such as wildfire‑related risks \cite{silva2023quantum}. These early approaches illustrate how quantum probability estimation can fundamentally change reliability computations.

Subsequent developments extend these ideas to large‑scale systems, multi‑hazard modeling, and adequacy assessment. For example, the QAE has been applied to larger distribution networks to compute load‑point and system‑level reliability indices with improved convergence \cite{nikmehr2023quantum}, while QML models capture multi‑hazard risks and spatiotemporal failure propagation \cite{thelasingha2024energy}. Quantum Monte Carlo and Bayesian frameworks further accelerate load‑loss and operational‑reliability estimation under uncertainty \cite{hosseini2024modern}. Moreover, adequacy analysis has been reformulated using superposition‑based state evaluation and QAE to efficiently compute loss of load probability and expected energy not served \cite{yang2024power}, while extensions to cyber–physical reliability integrate attack–defense models into quantum circuits \cite{alvarez2025quantum}. Additional formulations use amplitude amplification for fault‑tree analysis and stochastic quantum power flow for overload‑risk estimation \cite{bara2025quantum, saevarsson2025stochastic}. More recently, QAE has been applied to power-system reliability assessment, offering a theoretical quadratic speed-up over classical Monte Carlo sampling while demonstrating empirical runtime improvements on proof-of-concept test systems \cite{alvarez2026quantum}.

\subsection{Restoration and Resiliency}
\subsubsection{Power Systems Restoration}
Following major disturbances, restoring service in power systems requires rapid decision-making to reconfigure the network, resupply loads, and maintain operational constraints. This process is typically a large-scale mixed-integer optimization problem involving both discrete switching actions and continuous power flow constraints under time-critical conditions. Recent research increasingly positions quantum and hybrid quantum–classical approaches as promising enablers for fast fault isolation and service restoration, in which classical problem formulations are systematically transformed into quantum-compatible representations and solved by integrating quantum routines into decomposed solution architectures. Early investigations primarily focused on embedding restoration problems within distributed optimization frameworks, such as combining the alternating direction method of multipliers (ADMM) with the QAOA, thereby revealing critical trade‑offs between solution accuracy and current hardware limitations \cite{ngo2022evaluate}. Moreover, advances in gate-based quantum methods have further strengthened compatibility with convex subproblems while reducing qubit requirements. Notably, QAOA–QRAO-based ADMM architectures have shown they can preserve classical solution quality while running on near-term quantum hardware, such as IBMQ platforms \cite{ngo2024quantum}. In parallel, QA approaches have shown strong potential to accelerate resilient microgrid formation by reformulating mixed-integer linear programming (MILP) problems as QUBO models. These methods achieve near-optimal restoration performance with reduced computational time and demonstrate favorable scalability on larger IEEE test systems \cite{nikmehr2023quantum_restoration, nikmehr2024quantum}.

Recently, hybrid decomposition strategies have also emerged as an effective way to partition restoration tasks by assigning discrete decision variables to quantum solvers while preserving continuous physical modeling within classical optimization frameworks. Representative approaches include Benders decomposition and ADMM-based architectures that integrate QUBO-formulated QA with classical optimization, thereby improving convergence behavior and scalability. These frameworks have been further extended to support coordinated sensing and network reconfiguration, as well as distributed quantum architectures for multi-island restoration scenarios \cite{fu2023coordinated, qu2024quantum, fu2024quantum, lin2025distributed}. Beyond core restoration, quantum optimization is increasingly applied to adjacent services, including black-start planning and energy storage system reconfiguration, enabling faster decision-making and enhanced system resilience compared with purely classical methods \cite{shao2025quantum}. In parallel, emerging hardware paradigms such as photonic Ising machines provide highly parallel capabilities for QUBO optimization. For instance, surrogate absolute-value Lagrangian-relaxation frameworks for post-event restoration achieve near-optimal solutions with significant computational speedups \cite{fu2026quantum}, underscoring the scalability and practical potential of hybrid quantum–classical architectures for next-gen power system restoration.

\subsubsection{Power Systems Resiliency}
Improving power system resilience involves preparing for, withstanding, and rapidly recovering from high-impact disturbances while maintaining critical functionality. This challenge is fundamentally a multi-stage optimization and coordination problem that requires decisions across prevention, response, and recovery under complex system interdependencies. Quantum computing is emerging as a promising approach for enhancing grid resilience through faster decision‑making, improved coordination, and scalable optimization. For example, a quantum game‑theoretic framework models interconnected microgrids as coalition structures and solves the resulting Ising‑encoded payoff problem using VQE, enabling coordinated resilience strategies that balance local and system‑wide objectives \cite{sanjani2024quantum}. Quantum-enhanced optimization frameworks also address uncertainty-dominated scenarios, such as wildfire-prone systems, by combining QA with quantum-kernel representations to better capture stochastic dependencies and improve critical-load support and system stability \cite{zhao2025enhancing}. %These approaches demonstrate how quantum methods can enhance both coordination and uncertainty modeling, enabling more adaptive and resilient grid operation under high‑impact, low‑probability events.

In addition, integrated and large‑scale resilience frameworks combine quantum optimization, sampling, and hybrid decomposition. Unified QUBO‑based models use QAOA with recursive refinement to jointly optimize prevention, response, and recovery, achieving significant improvements in restoration benefits \cite{zhou5292390scalable}. Hybrid frameworks further exploit quantum acceleration for scenario evaluation via QAE and large‑scale optimization, delivering substantial runtime reductions compared to classical solvers \cite{xie2025quantum}. Complementing these advances, recent work formulates dynamic grid partitioning into self‑reliant communities as a QUBO problem and demonstrates that hybrid quantum–classical approaches can achieve high‑quality solutions within practical time limits despite the problem’s exponential complexity \cite{bucher2024evaluating}. For system-level resilience improvement, quantum-integrated approaches enhance load-flow computation, distributed generation placement, and resilient microgrid formation through compact QUBO encodings and scalable hardware execution, achieving near-optimal performance with fewer qubits \cite{salman2025quircy, lin2025reforming}.

\subsection{Smart Grid Control}
Modern smart grids increasingly rely on real-time control strategies to coordinate distributed energy resources, maintain stability, and optimize system performance under dynamic conditions. These tasks belong to high-dimensional control and sequential decision-making problems, often requiring fast optimization and adaptive learning under uncertainty. Quantum computing is reshaping control and optimization in power systems, with early advances in distributed microgrid control, synchronization, and learning-based regulation that improve scalability and robustness by reducing communication and parameter complexity compared to classical schemes. For example, VQCs replace deep RL actors in distributed secondary frequency control, achieving better disturbance and delay tolerance with fewer parameters \cite{yan2022multiagent}. Additionally, quantum synchronization frameworks based on Lindblad dynamics enable coordinated AC/DC control via quantum-state exchange rather than classical communication \cite{babahajiani2022quantum, babahajiani2022employing}. Beyond coordination, quantum learning also supports real‑time controller design, enabling VQC‑based models to infer optimal PI gains from historical data with fast, accurate responses \cite{jahed2025quantum}. 

From an optimization perspective, quantum solvers address computational bottlenecks in model predictive control (MPC) for renewable energy systems. Both QA and gate‑based approaches accelerate mixed‑integer MPC by reformulating problems into QUBO or linear‑system representations, enabling faster decision‑making while preserving classical solution quality \cite{deng2023quantum, jing2024hhl}. Hybrid QAOA/HHL frameworks further simplify MPC design and improve scalability, while quantum-walk-enhanced approaches strengthen prediction robustness \cite{jing2024integrating, luo2024quantum}. Beyond MPC, quantum‑inspired search improves maximum power point tracking in renewable energy systems, and QAOA‑based optimization supports power‑electronics design \cite{gao2024quantum, paterakis2025quantum}. Constrained QAOA has been used for modular multilevel converter cell selection to balance switching activity and capacitor-voltage regulation \cite{pan2026constrained}, while QA-based real-time control of DC–DC buck converters in EV charging systems reduces switching losses and voltage ripple under varying operating conditions \cite{sharmiladevi2026quantum}. In distribution networks, quantum‑enhanced RL enables efficient Volt–Var control with a smaller model and faster learning \cite{lin2025quantum}. Recent work further demonstrates that QRL can enable cooperative secondary frequency control in microgrids, achieving faster frequency recovery and greater adaptability to disturbances than classical approaches \cite{liu2025cooperative}.

\subsection{Digital Twin Operation}
Integrating digital twins (DTs) in power systems enables continuous monitoring, prediction, and optimization by maintaining real-time virtual replicas of physical assets. These applications involve data-intensive, high-dimensional optimization and dynamic system estimation problems that require efficient processing of streaming data and complex system models. DT technologies are increasingly integrated with quantum computing to address high-dimensional optimization and stability analysis in smart grids. In voltage‑stability applications, hybrid classical–quantum learning frameworks embed DT replicas via secure, explainable quantum federated learning to predict stability margins, reducing communication overhead and enhancing privacy \cite{ren2024esqfl}. Extending to large‑scale infrastructures, DT platforms synchronize real‑time data with cloud‑based models and use hybrid solvers, including QA, to optimize multi‑timescale energy routing and dispatch \cite{li2025study}. More broadly, conceptual DT–quantum architectures leverage QAOA to accelerate complex optimization tasks while enabling predictive, continuously updated virtual grid models for real‑time operation \cite{saber2025potential}.

In parallel, DT–quantum integration is advancing cyber‑physical security, equipment modeling, and system‑level control. Quantum-enhanced DT frameworks, such as Grover‑assisted anomaly detection and hybrid optimization, strengthen resilience against cyberattacks in smart‑city infrastructures \cite{naderi2025securing}. At the equipment level, QAOA‑based DT models optimize physical parameters such as insulation design \cite{lemo2025towards}, while quantum algorithms for differential equations enable accurate real‑time tracking of converter dynamics \cite{ma2026quandt}. At the network level, DT-driven optimization frameworks formulate reactive-power and voltage control as QUBO problems solved via hybrid QAOA approaches, enabling efficient closed-loop control with DT validation \cite{jiang2026digital}. These studies collectively indicate a coherent trend toward quantum-enhanced DT systems that improve predictive accuracy, security, and real‑time decision‑making.

\subsection{Power System Simulation: Dynamic, Steady-State, and Co-Simulation} 
Simulating power system behavior across dynamic, steady-state, and multi-domain environments is essential for planning, operation, and validation of modern grids. These tasks involve solving large-scale systems of linear and nonlinear differential-algebraic equations, often coupled across temporal and spatial scales, which are computationally intensive and difficult to scale. In recent years, interest in quantum‑enhanced power-system simulation has grown significantly, particularly for dynamic analysis under complex operating conditions. Several early contributions focus on electromagnetic transient (EMT) simulation, where nodal equations from EMTP formulations are recast as quantum linear‑system problems and solved using VQLS frameworks compatible with NISQ hardware \cite{zhou2022noisy}. In a related direction, dynamic simulation models based on DAEs are transformed into ordinary differential equation representations, encoded via amplitude encoding, and solved through HHL‑based Hamiltonian evolution \cite{tran2023applying}. Hybrid implementations further combine quantum encoding with classical iterative updates to accurately simulate both single- and multi‑machine system dynamics \cite{tran2024solving}. Alongside these efforts, quantum routines have been integrated into co‑simulation platforms, including Quantum‑in‑the‑Loop environments and real‑time simulators such as RTDS and MOSAIK, enabling tighter interaction between quantum computing modules and practical grid studies \cite{chanda2023architecture, vereno2023quantum, vereno2023exploiting}.

Beyond foundational simulation tasks, recent developments emphasize scalability and broader applicability. Hybrid quantum–classical approaches, such as a combination of the quantum linear systems algorithm and the variational
quantum solvers, incorporate nonlinear updates alongside quantum solvers to improve both accuracy and computational speed in EMT simulation \cite{anjimoon2023quantum}. Efforts to reduce resource requirements have led to low-dimensional, matrix-based qubit-casting methods that decrease qubit and gate complexity in high‑frequency switching scenarios \cite{lou2025matrix}. In parallel, QA has been explored to support tasks such as optimal grid partitioning and hardware‑in‑the‑loop simulation, where partition quality improves despite overhead challenges \cite{hartmann2025quantum, kaseb2026quantum}. Extending beyond deterministic analysis, quantum techniques also contribute to uncertainty modeling: amplitude estimation accelerates Monte Carlo simulations \cite{lange2025quantum}, while quantum neural networks offer accurate approximations of nonlinear transient responses \cite{soltaninia2025quantum}. 

\subsection{Power Systems Anomaly Detection}
Detecting abnormal behavior in power systems is critical for ensuring secure and reliable operation in increasingly complex cyber–physical environments. This task is naturally a high-dimensional pattern recognition and classification problem, in which subtle deviations in spatial and temporal data must be identified amid noise, uncertainty, and potential adversarial interference. Recent work increasingly leverages QML and hybrid quantum–classical models for anomaly detection in cyber–physical power systems, driven by the need for improved separability in the presence of high dimensionality, noise, and adversarial manipulation. A common strategy is hybrid representation learning, in which grid measurements are first compressed into compact embeddings and then mapped into expressive quantum feature spaces, enabling simpler yet more robust decision boundaries \cite{sakhnenko2022hybrid}. QML methods have also been progressively tailored to address critical threat surfaces, most notably false data injection attacks (FDIAs), which are particularly challenging to detect because they overlap with normal operating conditions. In this context, VQCs, quantum kernel methods, and hybrid quantum–classical deep models have demonstrated improved detection accuracy in relays and monitoring systems, while QAOA-based formulations further extend these capabilities to vulnerability identification under stochastic grid environments \cite{hammadia2025quantum, hammadia2026optimal, huang2026aligning, wu2025stochastic, saber2026quantum}. 

Beyond FDIAs, QML has expanded to detect communication-layer intrusions and time-dependent anomalies, where complex, high-dimensional, and dynamic data streams challenge classical methods. In particular, quantum Support Vector Machines (SVMs) and quantum‑enhanced RL improve DDoS detection and adaptability in evolving network conditions \cite{said2023quantum, said2024quantum}. At the same time, federated quantum learning and quantum autoencoding architectures address data privacy, robustness to poisoning attacks, and distributed ownership of measurements \cite{addo2025federated, cirillo2025quantum}. For sequential grid data such as PMU streams, quantum circuits enable temporal modeling through unitary evolution, quantum Long Short-Term Memory (LSTM)-based forecasting, and hybrid generative models, thereby enabling anomaly detection from trajectory deviations and forecast residual structures \cite{jafari2024quantum, nguyen2024integrating, yao2025synchrophasor}. These approaches emphasize the growing role of quantum representations in capturing both spatial and temporal dependencies in grid behavior. In parallel, QML has been applied to physical disturbances and asset-level anomalies, including power‑quality disturbances, battery degradation, and electricity theft. Hybrid quantum‑kernel methods, variational quantum neural networks, and quantum–classical convolutional architectures improve the identification of nonlinear disturbance signatures under noisy and inverter-dominated conditions \cite{phassadawongse2025quantum, li2025quantum}. For energy storage systems, quantum embeddings and clustering approaches enhance the detection of degradation trends and fault patterns in lithium‑ion batteries \cite{mutua2025quantum, sivakumar2025real}. Extending beyond physical faults to behavioral anomalies, hybrid variational circuits address electricity‑theft detection amid class imbalance and noisy consumption profiles \cite{blazakis2025power}. 

Complementing these directions, recent studies further highlight the breadth of quantum-enhanced approaches across power-system analytics. For instance, quantum-inspired signal processing has been explored for power quality monitoring, where a Gram matrix–based approach enables robust voltage sag detection by capturing high-dimensional similarities, improving accuracy, reducing false positives, and enabling faster computation than classical techniques \cite{mohapatra2025voltage}. At the system level, partitioned quantum neural networks show how scalable hybrid architectures can process complex-valued grid measurements and improve anomaly-detection performance while increasing robustness to adversarial perturbations via quantum-induced noise mechanisms \cite{ngo2026qupid}. Similarly, for physical disturbances, hybrid quantum–classical convolutional neural networks (CNNs) have achieved near-perfect accuracy in detecting and classifying power quality disturbances by leveraging quantum feature extraction alongside classical deep learning layers \cite{li2025hybrid}. Along with other work \cite{lakshmi2023quantum, hangun2025classical, ogiesoba2025quantum}, these studies report potential improvements in representation capacity, computational efficiency, and noise resilience under their respective experimental configurations; however, they have not established the generalizability of these improvements.

\vspace{-6pt}
\subsection{Fault Diagnostics}
Accurate fault identification and localization are essential to ensure a fast protection response and minimize service disruptions in power systems. This task can be viewed as a combinatorial inference and pattern classification problem, in which system measurements must be analyzed to determine the fault type, location, and severity under noisy and partially observed conditions. Quantum optimization provides a natural framework for casting fault diagnosis and localization as combinatorial problems suitable for near‑term hardware execution. Early work formulates multi‑fault diagnosis as QUBO mapped onto D‑Wave annealers, solving instances beyond classical engines \cite{perdomo2015quantum}. This is extended to distribution‑scale fault localization using decoupled QUBO/Ising formulations that reduce qubit requirements while maintaining accuracy and runtime efficiency \cite{xie2024fault}. In parallel, gate‑based methods leverage polynomial unconstrained binary optimization‑to‑Ising mappings solved via QAOA, achieving protection‑consistent diagnostics with improved scaling relative to classical higher‑order solvers \cite{fei2024power}. Complementing optimization, quantum‑assisted learning enhances feature extraction, representation, and classification across diverse fault‑diagnosis tasks. Quantum‑trained generative models improve missed‑detection and false‑alarm rates in simulated grids \cite{ajagekar2021fault}, while hybrid quantum–classical neural networks achieve competitive performance in photovoltaic fault classification \cite{uehara2021quantum}. Variational quantum layers further strengthen vibration‑based diagnostics by mitigating overfitting and improving generalization \cite{gbashi2024hybrid}, and quantum shadow learning provides parameter‑efficient, noise‑robust dissolved‑gas analysis for transformer monitoring \cite{he2025power}. These advances consistently highlight the role of quantum feature embeddings and hybrid pipelines in improving separability and robustness under noisy, high‑dimensional conditions.

More recent work extends to time-series monitoring, perception, and system-level integration, bridging quantum learning and real-world grid analytics. Quantum LSTM models and hardware‑validated quantum features improve early‑fault detection and noise resilience in sequential data streams \cite{zhang2025wind, bowden2026machine}, while quantum‑inspired CNN architectures enhance image‑based inspection of assets such as transmission insulators without increasing computational cost \cite{lin2025quantumfault}. At the protection‑logic level, quantum decision primitives enable minimal‑query fault discrimination, illustrating new paradigms for relay design. At the system level, hybrid optimization frameworks integrate prediction, sensing, and decision‑making, for example, in weather‑aware outage prediction and sensor placement under budget constraints \cite{prasad2025weather}. In parallel, quantum federated learning supports privacy-preserving, cross-utility diagnostics with strong accuracy under non-independent and identically distributed data conditions \cite{li2026qftd}. 

\vspace{-6pt}
\subsection{Battery Management Systems}
Effective battery management requires continuous monitoring, prediction, and control of electrochemical states to ensure safety, longevity, and optimal performance in energy systems. These tasks constitute a data-driven estimation and forecasting problem, often involving nonlinear system identification and decision-making under uncertainty. Recent efforts in battery health monitoring increasingly explore quantum and hybrid quantum–classical models to enable next‑generation battery management systems (BMS). Circuit‑based approaches use amplitude and angle encoding with VQCs to capture nonlinear degradation patterns, enabling early‑cycle lifetime classification and accurate capacity‑fade modeling on real datasets \cite{haris2021lifetime, ngo2023quantum}. These methods extend to state‑of‑health (SoH) estimation and safety classification, where Variational Quantum Neural Network (VQNN)‑based and Quantum CNN (QCNN) architectures achieve competitive accuracy with compact parameterization, while Quantum LSTM (QLSTM) models further enhance SoH and remaining useful life prediction in data‑limited environments \cite{mutua2025predicting, liang2025stochastic, wang2025state}. A recent study further reinforces this trend by showing that a hybrid QML framework that combines variational quantum neural networks with agent-based modeling can accurately predict battery SoH and thermal-runaway risks, enabling real-time, context-aware decision-making in electric vehicle systems \cite{mutiso2025quantum}. Complementing these developments, a recent comprehensive review shows that quantum and hybrid models consistently improve prediction accuracy and parameter efficiency for battery health analytics, although their practical deployment remains limited by NISQ-era hardware constraints and reliance on simulation environments \cite{komarsofla2026quantum}. 

In parallel, quantum techniques strengthen robustness and cyber‑resilience in BMS. QA formulations, such as quantum boost, map SoH estimation to QUBO problems and enhance resistance to adversarial sensor perturbations \cite{akash2023quantum}, while quantum‑inspired Ising solvers provide faster and energy‑efficient alternatives \cite{khot2025quantum2}. Hybrid ensemble approaches further improve prediction accuracy by integrating quantum learners with classical models, leveraging quantum feature mappings to better capture nonlinear electro‑thermal behavior and reduce estimation error \cite{wang2024lithium, soon2025hybrid, soon2025quantum2, sheilla2025framework}. These recent developments highlight the role of hybrid architectures in balancing accuracy, robustness, and computational efficiency. Beyond estimation, quantum learning supports predictive maintenance, control, and system integration. As a case in point, variational quantum classifiers improve predictive maintenance and battery health monitoring using sensor data \cite{grandhi2024quantum}, while quantum‑enhanced RL accelerates control convergence and reduces degradation in hybrid energy storage systems \cite{tran2025quantum}. At the system level, quantum‑assisted optimization enables real‑time SoH‑aware energy dispatch, reducing battery wear \cite{wang2025quantum}, and quantum feature‑selection methods extract compact health indicators from high‑dimensional data streams \cite{khot2025quantum}.

\subsection{Resource Forecasting}
Reliable forecasting of renewable generation and electricity demand is essential for efficient grid operation, market participation, and energy planning. These tasks are fundamentally time-series prediction and regression problems that require models capable of capturing nonlinear, stochastic, and temporally correlated patterns across multiple scales.
\subsubsection{Photovoltaic Power}
Quantum and hybrid quantum–classical learning have been increasingly applied to solar irradiance and photovoltaic (PV) power forecasting, where strong nonlinearity motivates expressive feature mappings. Early studies show that hybrid neural networks with parameterized quantum layers achieve parameter-efficient performance comparable to Bidirectional LSTM (BiLSTM), while standalone PQCs are limited by convergence issues and simulation overhead \cite{sushmit2023forecasting}. In temporal modeling, QLSTM architectures replace classical gates with VQCs, improving temporal correlation learning and reducing forecasting errors across benchmarks \cite{yu2023prediction}. Building on this, hybrid recurrent models with quantum embeddings enhance nonlinear feature extraction and efficiency, as demonstrated by hybrid VQE‑LSTM frameworks that integrate quantum‑optimized features into classical LSTMs to improve forecasting accuracy and robustness \cite{phan2025hybrid}. VQC‑based Gated Recurrent Unit (GRU) variants improve accuracy with fewer parameters, and advanced designs such as modified QLSTM reduce latency while outperforming classical and earlier quantum models across multiple PV datasets \cite{jeong2024short, phan2025modified}. 

In parallel, quantum and kernel‑based approaches offer alternative formulations, where hybrid quantum neural networks and quantum sequence models significantly reduce forecasting error and improve data efficiency compared to classical baselines \cite{sagingalieva2025photovoltaic}. VQNNs and QFT‑enhanced quantum kernels improve long‑horizon forecasting and capture periodic patterns more effectively than classical methods \cite{oliveira2024application, mechiche2025quantum}. Advanced QLSTM implementations further achieve higher accuracy and faster convergence than classical LSTMs, though they remain constrained by simulation costs. Furthermore, quantum-enhanced forecasting, combined with optimization frameworks such as a dual attention mechanism-based QLSTM integrated with VQCs, improves prediction accuracy and enables more efficient energy management in PV systems \cite{wang5336951quantum}. Advanced QLSTM implementations also show higher accuracy and faster convergence than classical LSTMs, though they remain constrained by simulation costs \cite{khan2024quantum}.

\subsubsection{Wind Power Generation}
Wind speed and power forecasting is another domain where quantum learning shows strong potential to capture the nonlinear, nonstationary, and long‑range temporal dynamics inherent in wind resources. Early progress is driven by hybrid models: LSTM–QNN architectures with instantaneous quantum polynomial embeddings and entangling layers improve seasonal robustness and outperform classical baselines \cite{hong2023robust}, while QNN‑based feature extractors enhance SVR performance and generalization on long‑term offshore datasets \cite{hangun2024hybrid}. These approaches demonstrate how quantum feature mappings enrich representation capacity for complex spatiotemporal wind patterns. Furthermore, fully quantum neural architectures explore expressive circuit designs for improved forecasting fidelity. Variational QNNs with angle encoding and entanglement achieve accuracy comparable to or exceeding classical multi-layer perceptrons across multiple sites \cite{pires2025quantum}, while systematic studies show that Z‑feature‑map QNNs consistently deliver strong performance and effective trend capture \cite{hangun2025comparative}. 

Further enhancements integrate quantum layers into deeper models, such as QAOA-embedded residual LSTMs, improving prediction accuracy by leveraging richer embeddings, albeit with increased simulation complexity \cite{hong2025implementing}. Additional studies confirm that quantum models perform particularly well in low‑data regimes and maintain high accuracy despite increasing circuit-depth costs \cite{hangun2025quantum}. Recent developments extend toward scalable hybrid architectures and uncertainty‑aware forecasting, linking quantum learning with practical deployment needs. Hybrid feed‑forward and recurrent models with quantum layers improve multi‑step forecasting and seasonal consistency while also enabling better uncertainty quantification in wind predictions \cite{kumar2026wind}.

\subsubsection{Electricity Load and Price}
Quantum methods have expanded beyond renewables to include grid‑level forecasting tasks, such as load prediction, consumption disaggregation, and electricity price prediction. For short‑term load forecasting, quantum neural models achieve notable error reductions over classical approaches on large smart‑meter datasets \cite{kumar2023quantum}. At finer granularity, VQC classifiers improve appliance‑level disaggregation \cite{arvanitidis2023quantum}, while quantum kernel methods, such as quantum SVM, enhance residential consumption forecasting, particularly in low‑data regimes \cite{nutakki2024quantum}. Hybrid ANN–quantum models further maintain competitive accuracy with limited input features \cite{habibi2025electrical}. Recent benchmarking studies also show that hybrid QLSTM models applied to real electrical meter data can achieve predictive accuracy comparable to classical LSTMs while converging faster and using fewer parameters, demonstrating improved learning efficiency in practical load forecasting tasks \cite{lalindeautomatic}. 

In addition, quantum‑enhanced recurrent architectures enable scalable, privacy‑aware forecasting. Hybrid quantum Recurrent Neural Network (RNN) models with annealing‑based tuning improve accuracy and seasonal robustness over classical RNN/LSTM methods \cite{wang5114742power}, while federated QLSTM frameworks reduce communication overhead by sharing only quantum parameters, achieving higher accuracy with fewer model parameters \cite{zhang2025privacy}. These approaches highlight the integration of quantum learning with distributed and privacy‑preserving analytics. Finally, quantum methods extend to market forecasting, where adiabatic quantum models encode system variables into Hamiltonians for electricity price prediction and achieve consistent error reductions over LSTM and GRU across day‑ahead horizons \cite{dash2025dynamic}.

\subsection{Power Flow}
\subsubsection{Power Flow Analysis}\label{Power_Flow}
Evaluating steady-state operating conditions in power systems requires solving nonlinear algebraic equations that describe the balance of power injections and network constraints. This task is fundamentally a large-scale linear and nonlinear equation-solving problem, often approached through iterative numerical methods, which can be computationally burdensome for large networks. Early efforts to accelerate AC power‑flow (PF) with quantum methods focus on reformulating PF equations as linear system problems. The quantum fast‑decoupled PF in \cite{feng2021quantum} maps PF equations to Hermitian systems solvable via enhanced HHL, reducing per‑iteration complexity while preserving classical accuracy. To overcome HHL’s depth limitations, NISQ-compatible approaches employ VQC-based VQLS solvers for shallow-depth PF iterations \cite{feng2023noise}, while hybrid DC-PF schemes reduce qubit requirements using optimized phase-estimation strategies \cite{gao2023solving}. In parallel, annealing-based PF formulations recast PF as QUBO/Ising problems, enabling fast solution times and scalable voltage estimation across small- and medium-size systems with accuracy comparable to the Newton–Raphson (NR) method \cite{kaseb2024adiabatic, kaseb2025combinatorial}. Complementary progress has emerged through quantum-assisted learning for PF estimation, bridging physics-based modeling and data-driven methods. Hybrid quantum physics-informed neural networks embed PF constraints directly into the loss function while leveraging quantum circuits to improve accuracy and robustness \cite{kaseb2024hybrid}. QCNNs and QNNs further reduce training data needs and improve estimation accuracy \cite{kaseb2024quantum}, while Bayesian QNNs enhance generalization under renewable uncertainty \cite{zhu2025bayesian}. Variational quantum frameworks that exploit PF sparsity also enable efficient measurement and scalable prediction \cite{le2025learning}. 

Recent work integrates quantum solvers into NR workflows and system‑level optimization to improve practical scalability, but HHL‑based NR methods reduce iteration counts while facing significant resource constraints \cite{el2024newton, zheng2025early}, motivating hybrid and variational alternatives for Jacobian inversion \cite{liu2024quantum}. Moreover, recent studies highlight that while quantum‑enhanced PF methods, ranging from early NR integrations to hybrid and annealing‑based approaches, demonstrate potential computational advantages, their practical deployment remains constrained by resource requirements, noise sensitivity, and scalability limitations, reinforcing the need for hybrid quantum–classical frameworks in real‑world power system applications \cite{zheng2024early, kaseb2024power, pareek2025limitations}. More scalable approaches, such as quantum singular value transformation‑based quantum Newton methods, show accurate PF solutions on larger systems and stochastic PF tasks \cite{fan2026quantum}. Additionally, quantum‑enhanced RL improves NR initialization by optimizing starting points through QUBO‑based environments solved on annealers, reducing iterations and improving convergence \cite{kaseb2025quantumRL}. 

\subsubsection{Optimal Power Flow}
Determining optimal operating points in power systems, such as optimal power flow (OPF), requires minimizing generation cost or losses while satisfying network constraints and system limits. This task is innately a large-scale, constrained nonlinear optimization problem that combines continuous variables with discrete operational decisions under physical and security constraints. Quantum computing for OPF has progressed from early annealing‑based formulations to a broader set of hybrid and quantum‑accelerated optimization methods. Initial studies reformulated distribution‑level OPF tasks, such as EV charging, PV sizing, and network planning, into QUBO models solvable on quantum annealers \cite{morstyn2022annealing}. Moreover, linear-solver-based approaches embed HHL within Newton–Raphson iterations to accelerate DC-OPF, with quantum signal-processing techniques improving gate efficiency and indicating potential gains in iteration time \cite{amani2023quantum}. Additionally, variational and learning‑based methods employ parameterized quantum circuits as differentiable approximators for OPF mappings, incorporating privacy mechanisms and physics‑informed constraints to enhance feasibility, robustness, and generalization under uncertainty \cite{cao2024differentially, hu2024advancing}. 

A parallel research direction integrates quantum solvers into interior‑point and decomposition‑based OPF methods to improve scalability. A quantum-enhanced primal–dual interior-point method leverages HHL and VQLS for Newton updates, while noise‑tolerant strategies, preconditioning, and hybrid switching improve convergence under NISQ limitations \cite{amani2024quantum, hafshejani2025quantum}. Recent studies also show that noise‑resilient and variationally enhanced quantum interior‑point methods, incorporating HHL and coherent VQLS, can achieve accurate DC‑ and AC‑OPF solutions with improved convergence and reduced computational complexity, although scalability and performance remain constrained by quantum noise and current hardware limitations \cite{amani2025optimal, amani2025quantum}. Beyond linear solvers, hybrid formulations such as ADMM‑based decompositions and adiabatic QUBO models assign discrete subproblems to quantum processors and continuous components to classical solvers \cite{magar2025quantum, carrillo2025quantum}. More recent work emphasizes constraint‑preserving and scalable variational designs, including QAOA architectures that enforce physical laws during optimization and primal–dual quantum formulations that reduce measurement overhead in large‑scale OPF \cite{prasad2025constraint, le2026solving}. 

\subsection{Unit Commitment}
Unit commitment (UC) problems determine the optimal on/off scheduling of generation units over time to satisfy demand while minimizing operational costs and respecting system constraints. It is a mixed-integer optimization problem with strong temporal coupling and uncertainty, making it computationally challenging for real-world systems, especially in large networks. Early quantum approaches to UC focus on hybrid formulations that separate binary commitment decisions from continuous dispatch, in which variational and QAOA‑based methods efficiently recover commitment decisions at shallow circuit depths, while classical solvers handle dispatch \cite{koretsky2021adapting}. Subsequent advances improve solution quality and convergence through Grover‑based search, warm‑started QAOA, and variational filtering, alongside more efficient encodings, such as logarithmic discretization, that significantly reduce qubit requirements \cite{zheng2024fast, salgado2024hybrid, christeson2024quantum, aboumrad2025new}. To enhance scalability, distributed frameworks, including quantum ADMM and distributed VQE, decompose UC into smaller, time‑coupled subproblems, enabling larger systems through coordinated QUBO formulations \cite{yang2025scalable, hasanzadeh2025d2, nikmehr2022quantumUC}. These developments are expected to establish a practical foundation for NISQ‑compatible UC optimization.

In stochastic and security-constrained UC, recent work integrates decomposition with learning-based and qubit-efficient strategies to address uncertainty and system complexity. For example, annealing‑based pipelines combined with augmented Lagrangian and ADMM techniques maintain fixed‑size QUBO representations while improving convergence and runtime performance \cite{hong2025qubit}. In parallel, learning‑assisted Benders frameworks reduce problem size by leveraging LSTM‑based recourse estimation and cut selection prior to QAOA optimization \cite{mahroo2026machine}. In security-constrained settings, qubit-aware encodings, such as semidefinite programming-based quadratic approximations, preserve solution fidelity while reducing the number of binary variables, and distributed quantum frameworks using annealers or photonic coherent Ising machines improve convergence and cost efficiency on benchmark systems \cite{zheng2025toward, huang2025consensus, gao2025distributed, ling2026hybrid}. These approaches illustrate how quantum UC formulations increasingly adapt to uncertainty, network constraints, and large‑scale system requirements.

Recent research further advances robust and large-scale UC through tightly integrated hybrid, learning-enhanced, and decomposition-driven frameworks. Quantum‑embedded column‑and‑constraint generation and ADMM methods solve discrete UC subproblems using QAOA or annealing while retaining classical treatment of continuous variables, with performance varying across hardware platforms \cite{fu2025quantum}. Extensions incorporating surrogate Lagrangian relaxation, learning‑guided ADMM, and quantum‑assisted cut selection reduce iteration counts and improve scalability \cite{feng2022novel, mahroo2023learning, paterakis2023hybrid, christeson2026hybrid, wei2026quantum}. Integration with OPF and broader system scheduling tasks highlights the feasibility of quantum-assisted co-optimization, while annealing-based studies reveal trade-offs among encoding strategies, feasibility, and optimality \cite{magar2024dc, ling2025hybrid, muller2025quantum}. The literature points to a clear transition toward hybrid quantum–classical UC frameworks that improve convergence, solution quality, and scalability, yet remain constrained by current NISQ hardware limitations and embedding challenges.

\subsection{Energy Management}
\subsubsection{Demand Response}
Demand response (DR) is a stochastic, behavior-driven scheduling problem involving demand flexibility, uncertainty, and temporal coupling across multiple users, aiming to adjust consumer electricity usage in response to price signals or grid conditions to improve system efficiency and reliability. The growing maturity of quantum optimization methods is driving DR approaches, especially for large‑scale price‑driven scheduling. For instance, hybrid annealing frameworks reformulate discount‑based DR programs as QUBO problems, enabling scalable optimization on platforms such as D‑Wave, where classical solvers struggle with high‑dimensional combinatorial problems \cite{bucher2023dynamic, bucher2024incentivizing}. These approaches show that quantum and hybrid pipelines can efficiently handle population‑level pricing while maintaining solution quality. Beyond pricing, quantum optimization has also been applied to microgrid scheduling and distribution planning, where QUBO-based models solved on annealing and photonic hardware deliver significant runtime reductions and improved convergence in demand-side management and DR-integrated energy scheduling \cite{wan2024optimization, xin2025quantum, minghong2025behavior}.

In parallel, QML and RL enhance DR control and behavior modeling under uncertainty. Hybrid quantum–classical RL frameworks use VQCs to approximate value functions, thereby reducing energy consumption and emissions compared to MPC and classical RL baselines \cite{ajagekar2024demand, ajagekar2024variational}. Quantum learning models also support DR prediction and coordination, using compact QNNs and quantum‑enhanced sequence architectures to capture consumer behavior and price‑response dynamics while improving risk evaluation efficiency \cite{safari2024neuroquman, zhuang2025quantum}. Advanced controllers such as quantum-enhanced proximal policy optimization further enable real-time DR operation with faster convergence, better supply–demand balancing, and improved stability in renewable-rich grids \cite{yeboah2025quantum}. 

\subsubsection{Energy Trading}
Electricity markets are increasingly evolving toward decentralized, data-driven platforms where autonomous agents exchange energy based on dynamic pricing and local constraints. In this setting, energy trading becomes a distributed decision-making and optimization problem, involving strategic interactions among participants under uncertainty and market rules. As quantum computing becomes increasingly integrated with energy markets, recent research has shifted toward decentralized trading architectures supported by optimization‑driven and learning‑based decision frameworks. In this context, peer‑to‑peer (P2P) energy systems have emerged as a key application area where blockchain‑enabled QRL combines secure transaction infrastructure with adaptive market intelligence. In particular, QRL‑based double‑auction mechanisms frame trading as a Markov decision process and use quantum amplitude amplification to accelerate learning and improve market efficiency \cite{kumar2023blockchain}. This paradigm naturally extends to local energy markets, where VQC policies optimize decentralized trading strategies, while blockchain mechanisms ensure data privacy and reliable settlement \cite{moniruzzaman2025peer}. 

Complementing learning‑based designs, researchers have applied quantum and quantum‑inspired optimization to trading, grid interaction, and allocation problems. For instance, QUBO‑based formulations solved via QAOA and annealing methods enable prosumer scheduling and P2P matching while maintaining physical consistency \cite{mastroianni2023assessing, blenninger2024q}. However, benchmarking studies show that classical solvers can still outperform QA in dense, fully connected market models due to embedding overheads \cite{bucher2025grid}. In parallel, quantum‑enhanced policy learning enables market intelligence at larger scales, where VQCs approximate trading strategies for energy arbitrage with faster convergence and fewer parameters than classical models \cite{huang2026quantum}.

\subsubsection{EV Energy Management}
The rapid integration of electric vehicles introduces significant operational challenges in coordinating charging demand with grid constraints and renewable generation. In this context, EV energy management is a large-scale, discrete–continuous optimization and control problem involving scheduling, uncertainty, and user behavior across interconnected energy systems. Quantum computing is emerging as a promising tool for EV energy‑management optimization, particularly for complex charging coordination under network constraints. Early work focuses on combinatorial scheduling, where EV charging decisions are formulated as discrete optimization problems. For instance, QAOA has been applied to knapsack‑based charging selection to achieve high‑quality, constraint‑aware scheduling \cite{kea2023leveraging}, while QA frameworks embed pricing, grid constraints, and user behavior into QUBO models to enable near‑real‑time coordination and load shifting in EV networks \cite{wang2025employing}. Along these lines, QA has also been used for coordinated EV charging in distribution feeders and residential settings, reducing power losses, voltage deviations, peak demand, and overall operational costs through scalable binary optimization formulations \cite{rashnu2025optimization, deng2025quantum}.
Beyond combinatorial optimization, quantum learning and control methods are increasingly applied to EV‑integrated energy systems. For example, QNNs capture nonlinear interactions among PV, storage, and vehicle-to-grid‑enabled EVs, improving cost efficiency and constraint satisfaction in microgrid management \cite{khatiri2025quantum}. For real-time decision-making, QRL frameworks use variational circuits to improve policy approximation, achieving faster convergence and better charging performance than classical DRL \cite{xu2025quantum}. At the grid level, quantum‑enhanced control improves voltage regulation and reduces losses in EV‑dominated feeders \cite{yeboah2025quantumEV}.

\subsubsection{Economic Dispatch and Optimal Planning}
Balancing generation schedules with system constraints while minimizing operational and investment costs is central to efficient power system operation and planning. These tasks span short-term dispatch and long-term planning horizons and are commonly formulated as large-scale, mixed-integer, stochastic optimization problems involving tightly coupled energy resources and network constraints. To address the scale and combinatorial complexity of these problems, recent research in economic dispatch and multi‑energy optimization has increasingly embraced hybrid quantum–classical approaches. Early studies focus on integrating QA into Benders decomposition, in which binary master problems are reformulated as QUBOs and continuous subproblems are solved using classical methods, demonstrating feasibility, albeit with limited speed‑ups compared to conventional MILP solvers \cite{leenders2024integrating}. Building on this foundation, subsequent advancements, including quantum-driven multi-cut Benders methods, improve convergence by reducing iteration counts, while quantum-inspired and annealing-based techniques further enhance scalability and convergence performance in renewable-rich, multi-objective energy systems \cite{zhao2023optimal, awan2024quantum}. Furthermore, hybrid frameworks that incorporate VQCs and PQCs into control and optimization processes have been shown to improve energy efficiency and accuracy and reduce carbon emissions, while annealing-based scheduling methods for hybrid energy systems demonstrate notable gains in computational speed and cost-effectiveness over classical approaches \cite{ajagekar2024variationalEMS, shi2025quantum}.

Beyond these initial formulations, recent efforts target large-scale, uncertainty-aware, and integrated energy system optimization problems, including planning, stochastic dispatch, and microgrid operation. For instance, hybrid quantum–classical strategies accelerate generation‑transmission expansion and multi‑energy co‑optimization through techniques such as warm‑starting and combinatorial Benders decomposition, thereby reducing computational time and improving scalability \cite{xu2025joint, li2025quantum_EMS, wu2025hybrid}. At the same time, quantum‑enhanced stochastic dispatch leverages amplitude estimation and QAOA to reduce scenario complexity, while QAOA‑ and annealing‑based microgrid optimization approaches improve computational efficiency, cost performance, and renewable integration \cite{han2025quantum, liu2025quantum_sus}. Additional developments, including physics‑informed variational models and quantum‑accelerated distributionally robust optimization, further improve convergence and robustness and enable secure cloud‑based implementations through blind quantum computing \cite{zhang2026physics, han2026quasi}. Other studies demonstrate feasibility for multistack fuel-cell vehicles, microgrid scheduling, and energy-water nexus optimization \cite{shi2025optimal, wang2025optimal, qian2025integrated}.

\subsection{Grid Partitioning and Others}

Partitioning and clustering techniques play an important role in power system analysis by enabling network decomposition, improved scalability, and efficient control of large-scale grids. These problems are commonly formulated as graph-based combinatorial optimization and unsupervised learning tasks, where system components are grouped based on electrical, topological, or operational similarities. Quantum machine‑learning and optimization techniques have shown early promise in addressing power‑system analysis problems such as clustering and partitioning, where high‑dimensional data and combinatorial structures pose significant challenges for classical approaches. Quantum $K$‑means clustering, which leverages swap‑test‑based distance estimation, achieves practical accuracy when combined with subspace parallelization, although its performance remains sensitive to noise on NISQ hardware \cite{diadamo2022practical}. Similarly, in network partitioning applications, electrical‑grid segmentation can be formulated as QUBO problems and solved on quantum annealers, enabling scalable and feasible solutions for IEEE benchmark systems \cite{wang2022quantum}. To further improve scalability and computational efficiency, enhanced formulations such as integer‑slack QUBO models eliminate the need for iterative constraint handling and provide measurable runtime gains as system size increases \cite{wang2023quantum}, while hybrid quantum–classical solvers enhance partition quality in community detection tasks, often outperforming classical heuristics under fixed computational budgets \cite{bucher2025evaluating}. 

Beyond partitioning, these approaches extend to broader grid-operation tasks that involve combinatorial decision-making and system-level coordination. For example, distribution‑network reconfiguration and surplus partitioning can be efficiently modeled as QUBO problems and solved on annealing platforms, yielding near‑optimal solutions with favorable quality‑to‑time trade‑offs despite embedding limitations \cite{silva2023quantumothers, colucci2023power}. In parallel, QAOA-based optimization addresses temporal aggregation problems, such as representative‑day selection, enabling significant data reduction while maintaining planning accuracy \cite{singha2025application}. Similarly, kernel-based clustering of renewable-energy profiles has been reformulated as a QUBO problem and solved on a coherent Ising machine, demonstrating near-constant computation time and alleviating the curse of dimensionality in large-scale profile aggregation tasks \cite{liu2025alleviating}. At the operational level, QRL enhances transmission‑switching decisions by improving convergence behavior and reducing model complexity\cite{lin2025two}, while QAOA‑based search and variational optimization further support renewable‑energy control and critical‑network identification tasks \cite{zhang2025quantum}. Recent quantum-assisted microgrid energy-management frameworks also formulate dispatch and coalition-formation decisions as QUBO problems and employ QAOA within deadline-aware hybrid quantum–classical agent architectures for real-time operational control \cite{ioannou2026quantum}. Similarly, recent QAOA‑based islanding frameworks \cite{jiang2026regrid} achieve classical‑optimal solutions with improved quantum resource efficiency, reinforcing the practicality of hybrid quantum optimization for grid-operation applications.

\subsection{Key Insights and Observations}
The literature reviewed in this section demonstrates, as summarized in Table~6, that quantum computing has evolved from a largely conceptual research topic into an emerging computational framework for addressing a broad range of smart grid challenges. Existing studies span nearly all major domains of power system planning and operation, including monitoring, optimization, control, security, reliability assessment, forecasting, and digital twin applications. However, the maturity, demonstrated benefits, and technological readiness of these applications vary considerably. A clear trend across the literature is the dominance of hybrid quantum--classical frameworks. Because of NISQ hardware limitations, most implementations combine classical optimization, simulation, or machine learning with quantum subroutines rather than relying on fully quantum workflows. Consequently, hybrid approaches currently represent the most practical pathway for integrating quantum computing into smart grid applications.

From an application perspective, combinatorial optimization problems appear particularly well aligned with current quantum capabilities. Unit commitment, optimal PMU placement, facility location allocation, network reconfiguration, restoration, and energy management can often be reformulated as QUBO or Ising models and addressed using quantum annealing or variational optimization techniques. Among currently available technologies, quantum annealing has shown strong practical applicability because of its mature hardware implementations and comparatively large qubit counts. In contrast, gate-based algorithms such as QAOA and VQE provide greater modeling flexibility and strong theoretical foundations but remain constrained by circuit depth, hardware noise, and parameter-optimization challenges.
The review also reveals an important distinction between optimization- and machine-learning-oriented applications. Quantum optimization studies primarily seek to improve combinatorial decision-making, solution quality, or computational efficiency. In contrast, quantum machine learning applications generally derive their reported benefits from quantum feature mappings, parameterized circuit representations, or hybrid learning architectures. Consequently, improvements in forecasting, anomaly detection, fault diagnosis, stability assessment, and battery management are more commonly reported in terms of predictive accuracy, model expressiveness, or robustness than demonstrated computational speedup. Such application-specific improvements should not be interpreted as evidence of general quantum advantage.

Across the reviewed application areas, several recurring limitations are evident. Most studies rely on small test systems, classical simulation, or narrowly selected datasets, while relatively few have been evaluated on physical quantum hardware at practically relevant scales. Classical comparisons are often limited, and end-to-end costs associated with data encoding, problem embedding, quantum execution, repeated measurements, parameter optimization, communication, and solution decoding are not consistently included. Performance is also frequently evaluated using a single hardware backend, noise model, dataset, or problem configuration, with limited sensitivity analysis regarding system size, circuit depth, penalty parameters, sampling budget, and hardware noise. In addition, incomplete reporting of circuit architectures, optimizer settings, embedding procedures, and computational resources restricts reproducibility and prevents independent verification of some reported improvements.

Accordingly, evidence of practical quantum advantage in smart grid applications remains limited. Improvements reported for a particular dataset, test system, algorithm configuration, or hardware platform should not be interpreted as broadly applicable results. Their generalizability across network sizes, operating conditions, datasets, noise levels, quantum backends, and classical benchmarks has generally not been established. In several application areas, the claimed benefits also remain theoretical and depend on assumptions such as efficient state preparation, favorable problem structure, efficient oracle construction, error-free computation, or the availability of future fault-tolerant quantum computers. Moreover, faster convergence, improved predictive accuracy, or better solution quality does not necessarily demonstrate an end-to-end computational speedup.
Nevertheless, the reviewed literature indicates substantial long-term potential for quantum computing in next-generation smart grids, particularly in large-scale combinatorial optimization, uncertainty quantification, and data-driven decision-making. Advances in hardware scalability, fault tolerance, problem encoding, reproducible benchmarking, and application-specific algorithm design are still required before widespread operational deployment becomes feasible. In the near term, the greatest impact will likely come from hybrid quantum--classical frameworks that apply quantum resources selectively while retaining the maturity, scalability, and reliability of classical computing infrastructure.

\footnotesize
\onecolumn
\label{Literature_Summary}
\begin{longtable}%{p{2.4cm} p{0.65cm} p{0.90cm} p{1.2cm} p{1.30cm} p{2.3cm} p{1.5cm} p{4.7cm}}
{p{1.2cm} p{0.7cm} p{0.70cm} p{1.3cm} p{1.10cm} p{2.7cm} p{1.5cm} p{4.9cm}}
\caption{Literature summary in smart grid applications using quantum computing.}\\ 
\hline
\textbf{Application} & \textbf{Ref.} & \textbf{Year} & \textbf{Real HW} & \textbf{Simulator} & \textbf{Platform} & \textbf{Method} & \textbf{Reported Evidence-calibrated Findings} \\
\hline
\endfirsthead

\multicolumn{8}{c}%
{{ \tablename\ \thetable{} -- continued from previous page}} \\
\hline
\textbf{Application} & \textbf{Ref.} & \textbf{Year} & \textbf{Real HW} & \textbf{Simulator} & \textbf{Platform} & \textbf{Method} & \textbf{Reported Evidence-calibrated Findings} \\
\hline
\endhead

\hline \multicolumn{8}{r}{{Continued on next page}} \\ \hline
\endfoot

\hline
\endlastfoot

\multirow{3}{*}{Optimal}   & \cite{jones2020computational}       & 2020      &  \checkmark     &   \checkmark     &  D-Wave      & QA  & Outperforms classical counterpart \\
\multirow{3}{*}{PMU}     & \cite{jiang2026qaoa}      &  2026     &  \ding{55}     &  \checkmark      & IBM   & QAOA  & Reported runtime improvement \\ 
\multirow{3}{*}{Placement}   & \cite{jiang2025accelerating}      &  2025     &    \ding{55}    &    \checkmark     & Classical Emulation   &  QAOA   & Reported computational speed up \\ 
                                     &\cite{jiang2025optimal}      & 2025      &  \checkmark      &  \checkmark      &  IBM      &  QAOA &  Up to  Order of runtime reduction   \\  
                                     & \cite{ganeshamurthy2025quantum}      & 2025      & \checkmark      &  \checkmark      &  D-Wave      &  QA &  Improved solution over SA at high reads \\ \hline
\multirow{9}{*}{Facility}  &  \cite{chandra2022towards}            &  2022     &  \checkmark      &   \checkmark     &  D-Wave      &  QA &  Scale of runtime improvement  \\ 
\multirow{9}{*}{Location}   & \cite{veshchezerova2023hybrid}            &  2023     &  \checkmark      &  \ding{55}      &  D-Wave      &  QA &  Supports better for scalability  \\
\multirow{9}{*}{Allocation}     & \cite{rao2023hybrid}   &  2023  &  \checkmark       &  \checkmark       &  D-Wave, Google      &  QA, SA & Order of runtime/scalability gain  \\ 
                                     & \cite{subramanian2023ev} &  2023     &  \checkmark      & \ding{55}        &  D-Wave      & QA  &  Not reported \\ 
                                    & \cite{sakib2024quantum}  &   2024    &  \checkmark      &  \ding{55}      &  D-Wave      & QA  &  Outperforms classical counterparts  \\ 
                                    & \cite{radvand2024quantum}  & 2024      &  \ding{55}      &  \checkmark      & IBM       &  GAS, QPE &  Quadratic speedup over classical methods  \\ 
                                     & \cite{ou2025solving} &   2025    &   \ding{55}      &     \checkmark   & Fujitsu       &  DA  & Practical runtime speed‑up  \\
                                    & \cite{ajagekar2019quantum}       & 2019      &   \checkmark    & \ding{55}    &  IBM, D-Wave    & VQE, QA  & Scale of runtime advantage \\
                                    & \cite{hidary2020application}             &  2020     &   \ding{55}     & \checkmark       &  IBM      &  QAOA &  Theoretical runtime advantage  \\ 
                                     & \cite{yin2025optimised}      & 2025      &  \ding{55}     &  \checkmark     &  D-Wave      &  QA &  Demonstrates superior scalability \\ 
                                     & \cite{hasan2026quantum}  &   2026    &   \checkmark      &   \ding{55}     & IonQ    &  VQA  & No quantum speed-up demonstrated \\ \hline
\multirow{4}{*}{State}   & \cite{NEW_State_Estimation}   &  2026     &   \ding{55}     &   \checkmark     &    IBM    & HHL   &  Reduced gate count and depth  \\
\multirow{4}{*}{Estimation}                                    &  \cite{tran2025noise}      &   2025     &   \checkmark     &  \ding{55}      &   D-Wave      & QA  &  Improves accuracy with fewer qubits \\ 
                                    &  \cite{feng2022quantum}            & 2022      &   \ding{55}      &  \checkmark      &  IBM      &  HHL & Potential of exponential speedups   \\ 
                                    & \cite{feng2024noisy} &   2024    &   \checkmark     & \checkmark       &  IBM      & VQC, VQLS  &  Reduces circuit depth vs HHL \\  
                                    & \cite{addo2025deep} &    2025   &   \ding{55}     &   \checkmark     &  IBM      & VQC  & Enhanced accuracy and convergence   \\  \hline
\multirow{12}{*}{Stability} &  \cite{yu2025quantum}            &  2025     &  \ding{55}      &    \checkmark    &     IBM, PennyLane   & QNN  &    Improves noise robustness \\ 
\multirow{12}{*}{Analysis}  & \cite{ren2024enhancing}  & 2026 & \ding{55} & \checkmark & Paddle Quantum & PQC & High accuracy with fewer parameters\\ 
                                &  \cite{masoumi2025quantum}           &  2025     &  \ding{55}      &    \checkmark    &     PennyLane   & PQC  &    Classical-comparable performance \\ 
                                & \cite{zhou2022noise} & 2023 & \checkmark & \checkmark & IBM, PennyLane & VQC & Theoretical speedup, not practical \\
                                & \cite{sabadra2024quantum} & 2024 & \ding{55} & \checkmark & IBM & VQC & Shows high accuracy and noise resilience \\
                                & \cite{li2023transient}  & 2023 & \ding{55} & \checkmark & Classical Emulation & QFT & Enabling theoretical exponential speedup \\
                                & \cite{chen2024quantum} & 2024 & \ding{55} & \checkmark & Classical Emulation & VQC & Theoretical dimensionality reduction\\
                                & \cite{yu2024quantum} & 2024 & \ding{55} & \checkmark & Classical Emulation & VQC & Reported accuracy improvement \\  
                                & \cite{yu2024quantum_2} & 2024 & \checkmark & \checkmark & IBM, PennyLane & VQC & High accuracy with fewer parameters\\ 
                                & \cite{ren2024esqfl} & 2024 & \ding{55} & \checkmark & Classical Emulation & QML & Higher accuracy than classical methods\\ 
                                & \cite{li2026qstaformer} & 2026 & \ding{55} & \checkmark & PennyLane & PQC & Higher accuracy than classical methods\\
                                & \cite{ahmad2025qm} & 2025 & \ding{55} & \checkmark & IBM & QNN & Comparable accuracy to classical models \\ 
                                & \cite{ren2024qfdsa} & 2024 & \ding{55} & \ding{55} & Classical & QKD & Obtain high accuracy solution \\ \hline  
\multirow{7}{*}{Contingency} &  \cite{feng2025quantum}           &  2025     &  \checkmark      &    \checkmark    &     IBM   & VQC, VQLS &   Theoretical exponential speed up \\
\multirow{7}{*}{Analysis}                                 & \cite{cremer2025n} & 2026 & \checkmark & \ding{55} & D-Wave & QA & Order of speed up vs exhaustive search\\ 
                                & \cite{jiang2025quantum} & 2025 & \ding{55} & \checkmark & IBM & QAOA & Comparable to classical methods \\  
                                & \cite{lange2026contingency} & 2026 & \ding{55} & $\checkmark$ & IBM & QAOA & Provides near-optimal solutions \\
                                & \cite{peter2025quantum} & 2025 & \checkmark & \checkmark & IBM & PQC & Improved learning performance \\   
                                & \cite{neumann2024quantum} & 2024  & \ding{55} & \checkmark & IBM, D-Wave & GS, QA & Theoretical quadratic speed up\\    
                                & \cite{antoli2023quantum} & 2023 & \ding{55} & \checkmark & IBM & HHL & Compatible accuracy vs  classical models\\    
                                & \cite{eskandarpour2020quantum_security} & 2020 & \ding{55} & \checkmark & Not Reported & HHL & Exponential speed up vs classical models \\  \hline
\multirow{10}{*}{Reliability} &  \cite{nikmehr2022quantum}           &  2022     &  \ding{55}      &    \checkmark    &     IBM   & QAE &   Quadratic convergence speed up \\  
\multirow{10}{*}{Assessment}                                & \cite{carrascal2023bayesian} & 2023 & \checkmark & \checkmark & IBM & RQBN & Matches classical Monte Carlo accuracy\\ 
                                & \cite{jong2023quantum} & 2023 & \checkmark & \checkmark & IBM & QAE & Quadratic speed up \\ 
                                & \cite{silva2023quantum} & 2023 & \ding{55} & \checkmark & Classical Emulation & AA & Theoretical quadratic speed up \\ 
                                & \cite{nikmehr2023quantum} & 2023 & \ding{55} & \checkmark & IBM & QAE & Quadratic speed‑up \\  
                                & \cite{thelasingha2024energy} & 2024 & \checkmark & \checkmark & IBM & VQC & Demonstrates potential for risk modeling \\  
                                & \cite{hosseini2024modern} & 2024 & \ding{55} & \ding{55} & Not Reported & QFT & Theoretical quadratic speed up \\   
                                & \cite{yang2024power} & 2024 & \ding{55} & \checkmark & Classical Emulation & QAE & Theoretical quadratic speed up \\    
                                & \cite{alvarez2025quantum} & 2025 & \ding{55} & \checkmark & IBM & QAE & Theoretical quadratic speed up \\     
                                & \cite{bara2025quantum} & 2025 & \ding{55} & \checkmark & IBM & QAA & Theoretical quadratic speed up \\      
                                & \cite{saevarsson2025stochastic} & 2025 &  \ding{55} & \checkmark & IBM  & QAE & Theoretical quadratic speed up \\ 
                                & \cite{alvarez2026quantum} & 2026 &  \checkmark & \checkmark & IBM  & QAE &  Empirical speed-up \\ \hline                         
\multirow{9}{*}{System} &  \cite{ngo2022evaluate}           &  2022     &   \checkmark     &    \checkmark    &     IBM   & QAOA &   Potential computational speed up \\                                  
\multirow{9}{*}{Restoration} & \cite{ngo2024quantum} & 2024 & \checkmark & \checkmark & IBM  & QAOA & No speed up reported \\   
                                & \cite{nikmehr2023quantum_restoration} & 2023 & \ding{55} & \checkmark & D-Wave & QA  & Order of speed up vs classical approach\\  
                                & \cite{nikmehr2024quantum} & 2025 & \checkmark & \ding{55} & D-Wave & QA & Faster convergence than classical method \\  
                                & \cite{fu2023coordinated} & 2023 & \ding{55} & \checkmark & D-Wave & QA & Order of computation speed up \\
                                & \cite{qu2024quantum} & 2024 & \ding{55} & \checkmark & Not Reported & QA &  Runtime reduction vs classical solvers \\     
                                & \cite{fu2024quantum} & 2024 & \checkmark & \checkmark & IBM, D-Wave & QAOA, QA & Reported runtime reduction \\   
                                & \cite{lin2025distributed} & 2026 & \checkmark & \checkmark & IBM, D-Wave & QAOA, QA & Order speed-up vs classical \\     

                                & \cite{shao2025quantum} & 2025 & \checkmark & \ding{55} & D-Wave & QA & Reported decision-time reduction \\  
                                & \cite{fu2026quantum} & 2026 & \checkmark & \checkmark & QBosoN & Q‑SAVLR & Over 30\%–efficiency improvement\\ \hline
\multirow{6}{*}{Resiliency} &  \cite{sanjani2024quantum}           &  2024     &   \ding{55}     &    \checkmark    &     IBM   & VQE &   Competitive to classical solutions \\                                 
\multirow{6}{*}{Improvement}  & \cite{zhao2025enhancing} & 2025 & \checkmark & \ding{55} & D-Wave & QA & Qualitative performance improvement \\       %
                                & \cite{zhou5292390scalable} & 2025 & \ding{55} & \checkmark & IBM & QAOA & Limited runtime and performance gains\\      %
                                & \cite{xie2025quantum} & 2025 & \checkmark & \checkmark & QBoson & QAE & 40\%–75\% runtime reduction vs classical\\      
                                & \cite{bucher2024evaluating} & 2025 & \checkmark & \ding{55} & D-Wave & QA & Competitive quantum-classic solution  \\                               
                                & \cite{salman2025quircy} & 2025 & \ding{55} & \checkmark & IBM & HHL, QAOA & Order of computation time reduction \\ 
                                & \cite{lin2025reforming} & 2025 & \checkmark & \ding{55} & D-Wave & QA & Reducing qubit count significantly\\ \hline 
 &  \cite{yan2022multiagent}   &  2022     &   \ding{55}     &    \checkmark    &     PennyLane   & VQC &   Competitive or improved performance \\
\multirow{10}{*}{Smart Grid}   & \cite{babahajiani2022quantum} & 2022 & \ding{55} &  \checkmark & QuTiP & QDC & Control‑layer convergence benefits \\ 
\multirow{10}{*}{Control} & \cite{babahajiani2022employing} & 2023 & \ding{55} &  \checkmark & QuTiP & QDC & Provable exponential synchronization \\  
                                & \cite{jahed2025quantum} & 2025 & \ding{55} & \checkmark & IBM & VQC & Improvements in feature extraction\\  
                                & \cite{deng2023quantum} & 2023 & \checkmark & \ding{55} & D‑Wave & QA & Order of computational time reduction\\   
                                & \cite{jing2024hhl}  & 2024 & \ding{55} & \checkmark & Classical Emulation & HHL & Achieves polynomial speed-up \\ 
                                & \cite{jing2024integrating} & 2024 & \ding{55} & \checkmark & Classical Emulation & VQA, HHL & Polynomial complexity improvement \\  
                                & \cite{luo2024quantum} & 2024 & \ding{55} & \checkmark & Classical Emulation & QW & Improves prediction accuracy \\  
                                & \cite{gao2024quantum} & 2024 & \ding{55} & \checkmark & Classical Emulation & GS & Faster performance vs classical method \\  
                                & \cite{paterakis2025quantum} & 2026 & \checkmark & \checkmark & IBM, PennyLane &  QAOA & Proof-of-concept quantum optimization \\ 
                                & \cite{pan2026constrained} & 2026 & \ding{55} & $\checkmark$ &
                                Not reported & QAOA  & Demonstrates competitive performance. \\
                                & \cite{sharmiladevi2026quantum}  & 2026 & \ding{55} & \checkmark & Classical Emulation & QA & Control-performance improvement \\  
                                & \cite{lin2025quantum} & 2025 & \ding{55} & \checkmark & PennyLane & VQC & Ensure parameter-efficient learning \\  
                                & \cite{liu2025cooperative} & 2025 & \ding{55} & \checkmark & Classical Emulation & GS & Heuristic learning improvement \\    \hline
\multirow{5}{*}{Digital Twin} &  \cite{li2025study}   &  2025     &   \checkmark &   \ding{55}     &     D-Wave   & QA &   Heuristic speedup \\
                                & \cite{naderi2025securing}  & 2025 & \ding{55} & \checkmark & IBM & GA & Quadratic theoretical speed‑up \\ 
                                & \cite{lemo2025towards} & 2025 & \ding{55} & \checkmark & IBM & QAOA & Improved modeling accuracy \\ 
                                & \cite{ma2026quandt} & 2026 & \ding{55} & \checkmark & PennyLane & HHL, VQLS & Theoretical exponential HHL speed up \\  
                                & \cite{jiang2026digital}  & 2026 & \ding{55} & \checkmark & IBM, PennyLane & QAOA & Heuristic improvement in convergence \\  \hline  
\multirow{8}{*}{Power Grid} &  \cite{zhou2022noisy}   &  2023      &   \checkmark   &   \checkmark  &    IBM   & VQLS &   NISQ-compatible quantum performance \\
\multirow{8}{*}{Simulation} & \cite{tran2023applying} & 2023 & \ding{55} & \checkmark & Yao & HHL & Theoretical polynomial–log complexity \\
                                & \cite{tran2024solving} & 2024 & \ding{55} & \checkmark & Yao & HHL & Theoretical speedup and memory savings \\
                                &  \cite{vereno2023quantum}  & 2023 & \checkmark & \checkmark & IBM & HHL & PoC power flow co-simulation \\
                                & \cite{lou2025matrix} & 2025 & \ding{55} & \checkmark & PyQpanda & VQLS &  Exponential speed‑up in matrix mapping \\ 
                                & \cite{hartmann2025quantum} & 2025 & \checkmark & \ding{55} & D-Wave & QA & Outshine classical models at small scale \\ 
                                & \cite{kaseb2026quantum} & 2025 & \checkmark & \checkmark & D‑Wave,  Fujitsu & QA & Classical comparable performance \\
                               & \cite{lange2025quantum} & 2025 & \ding{55}  & \checkmark & Classical Emulation & QAE & Provides quadratic speedup over classical \\ 
                               & \cite{soltaninia2025quantum}  & 2025 & \ding{55}  & \checkmark  & PennyLane & VQC & Accurate solution with fewer parameters\\ \hline
\multirow{23}{*}{Anomaly} &  \cite{sakhnenko2022hybrid}   &  2022      &   \ding{55}   &   \checkmark  &    IBM   & PQC &  Empirical performance gains  \\ 
\multirow{23}{*}{Detection} & \cite{hammadia2025quantum}  & 2025 & \ding{55}  &   \checkmark & PennyLane & VQC & Outperforms classical approaches \\
                               & \cite{hammadia2026optimal} & 2026 & \ding{55}  &   \checkmark & PennyLane & VQC & Outperforms classical approaches\\ 
                               & \cite{huang2026aligning} & 2026 & \ding{55}  &   \checkmark   & PennyLane & PQC & Order of run-time reduction\\
                               & \cite{wu2025stochastic} & 2025 & \ding{55}  &   \checkmark  & IBM & QAE, QAOA & Potential quadratic speed up \\
                               & \cite{saber2026quantum} & 2026 & \ding{55}  &   \checkmark  & PennyLane & VQC & Improves detection accuracy\\
                               & \cite{said2023quantum} & 2022 & \ding{55}  &   \checkmark  &  IBM & PQC & Improved accuracy and faster execution \\ 
                               & \cite{said2024quantum} & 2024 & \ding{55}  &   \checkmark  &  IBM  & HHL & Faster convergence and accuracy gain \\
                               & \cite{addo2025federated} & 2025 & \ding{55}  &   \checkmark & IBM  & PQC &  Reduces communication overhead \\
                               & \cite{cirillo2025quantum} & 2025 & \ding{55}  &   \checkmark  & Not Reported & VQC & Classical comparable performance \\
                               & \cite{jafari2024quantum} & 2024 & \ding{55}  &   \checkmark & IBM & VQC & Faster convergence and accuracy gain\\  
                               & \cite{nguyen2024integrating} & 2024 & \ding{55}  &   \checkmark & PennyLane & VQC & Faster learning vs. classical \\ 
                               & \cite{yao2025synchrophasor} & 2025 & \ding{55}  &   \checkmark  & IBM  & VQC & Reduces training complexity \\ 
                               & \cite{phassadawongse2025quantum} & 2025 & \ding{55}  &   \checkmark  & IBM & PQC & Improves classification accuracy \\
                               & \cite{li2025quantum} & 2025 & \ding{55}  & \checkmark & MindQuantum & VQC  & Theoretical complexity improvement \\ 
                               & \cite{mutua2025quantum} & 2025 & \ding{55}  &   \checkmark & PennyLane & QNN, VQC & Achieves superior detection performance\\ 
                               & \cite{sivakumar2025real} & 2025 & \ding{55}  &   \checkmark  & IBM & Q K-Means& Improves computational efficiency  \\  
                               & \cite{blazakis2025power} & 2025 & \ding{55}  &   \checkmark & PennyLane & VQC & Improved classification performance \\
                               & \cite{mohapatra2025voltage} & 2025 & \ding{55}  &   \checkmark  & IBM & - & Improved detection and runtime \\ 
                               & \cite{ngo2026qupid} & 2026 & \ding{55}  &   \checkmark & PennyLane & PQC & Improves robustness and scalability\\   
                               & \cite{li2025hybrid}  & 2025 & \ding{55}  &   \checkmark & PennyLane & PQC & Polynomial time and space complexity \\  
                               & \cite{lakshmi2023quantum} & 2023 & \ding{55}  &   \checkmark   & Quirk & QML &  Conceptual modeling and validation \\ 
                               & \cite{hangun2025classical} & 2025 & \ding{55}  &   \checkmark  & PennyLane & VQC & Classical comparable performance \\ 
                               & \cite{ogiesoba2025quantum} & 2025 & \ding{55}  &   \checkmark & IBM & VQC & Hybrid model outdo classical models \\ \hline
\multirow{11}{*}{Fault} &  \cite{perdomo2015quantum}   &  2015      &   \checkmark   &   \ding{55}  &    D‑Wave   & QA &  Near‑optimal diagnoses \\
\multirow{11}{*}{Diagnostics} & \cite{xie2024fault} & 2024 & \checkmark   &   \ding{55}  &    D‑Wave   & QA  & Improved runtime for large networks \\
                               & \cite{fei2024power}  & 2024 & \ding{55} & \checkmark & PennyLane  & QAOA & Efficient multi-qubit gate decomposition \\
                               & \cite{ajagekar2021fault} & 2021 &  \checkmark   &   \ding{55}   & D‑Wave & QA & Improved training quality \\ 
                               & \cite{uehara2021quantum} & 2021 & \ding{55} & \checkmark & IBM & VQC & Demonstrated quantum feature mapping\\   
                               & \cite{gbashi2024hybrid} & 2024 & \ding{55} & \checkmark & IBM  & VQC, PQC & Improved training efficiency \\ 
                               & \cite{he2025power} & 2025 & \ding{55} & \checkmark & Baidu Paddle & PQC & Improved accuracy and robustness\\  
                               & \cite{zhang2025wind} & 2025 & \ding{55} & \checkmark & IBM & PQC & Enhanced temporal modeling accuracy \\ 
                               & \cite{bowden2026machine} & 2026 & \checkmark & \checkmark & IBM & PQC & Enhanced accuracy with noisy data\\  
                               & \cite{lin2025quantumfault} & 2025 & \ding{55} & \checkmark & Classical Emulation &  QCNN & Improved accuracy  \\ 
                               & \cite{prasad2025weather} & 2025 & \checkmark & \ding{55} & IBM, D-Wave & QAOA, QA & Classical competitive solution \\ 
                               & \cite{li2026qftd} & 2026 & \ding{55} & \checkmark & PennyLane & VQC & Theoretical complexity improvement \\ \hline 
\multirow{6}{*}{Battery} &  \cite{haris2021lifetime}   &  2021      &   \checkmark   &   \ding{55}  &    IBM   & VQC &  Improve classification performance \\                               
\multirow{6}{*}{Management} & \cite{ngo2023quantum} & 2023 & \ding{55} & \checkmark & Qulacs & PQC, VQC & Exponential expressivity advantage\\   
                               & \cite{mutua2025predicting} & 2025 & \ding{55} & \checkmark & IBM & PQC & Potential exponential state representation \\
                               & \cite{liang2025stochastic} & 2025 & \ding{55} & \checkmark & PennyLane & VQC & High accuracy with low complexity\\
                               & \cite{wang2025state} & 2025 & \ding{55} & \checkmark & PennyLane & VQC & Improved prediction accuracy \\
                               & \cite{mutiso2025quantum} & 2025 & \ding{55} & \checkmark & Not Reported & PQC & Outperforms classical models in accuracy\\
                               & \cite{komarsofla2026quantum} & 2025 &  \ding{55} & \checkmark & PennyLane & VQC & Accuracy gains vs classical models \\  
                               & \cite{akash2023quantum} & 2023 & \checkmark   &   \ding{55} & D-Wave & QA & Exhibits robustness to adversarial attacks\\
                               & \cite{khot2025quantum2} & 2025 & \checkmark & \ding{55} & D-Wave & QA & Magnitude of speed‑up in training \\    
                               & \cite{wang2024lithium}  & 2024 & \ding{55} & \checkmark & IBM & VQC & Improved estimation accuracy\\    
                               & \cite{soon2025hybrid} & 2025 & \ding{55} & \checkmark & IBM & VQC & Enhanched accuracy\\    
                               & \cite{soon2025quantum2} & 2025 & \ding{55} & \checkmark & IBM & PQC &  Nonlinear modeling and accuracy gain \\    
                               & \cite{sheilla2025framework} & 2025 & \ding{55} & \checkmark & IBM & VQC & Enhanced accuracy and feature mapping\\%                               
                               & \cite{grandhi2024quantum} & 2024 & \ding{55} & \checkmark & IBM & VQC & High‑dimensional feature mapping \\
                               & \cite{tran2025quantum} & 2025 & \ding{55} & \checkmark & PennyLane & VQC & Efficiency gain with fewer parameters \\  
                               & \cite{wang2025quantum} & 2025 & \checkmark & \ding{55} & D-Wave & QA &  Faster convergence and efficiency gain \\  
                               & \cite{khot2025quantum} & 2025 & \checkmark & \ding{55} & D-Wave & QA  & Classical comparable performance \\ \hline
\multirow{23}{*}{Resource} &  \cite{sushmit2023forecasting}   &  2023      &  \ding{55}   &  \checkmark    &    PennyLane   & PQC &  Improve classification performance \\                               
\multirow{23}{*}{Forecasting} & \cite{yu2023prediction} & 2023      &  \ding{55}   &  \checkmark    &    PennyLane   & VQC & Gains consistent accuracy improvement \\                                 %         
                               & \cite{phan2025hybrid} & 2025 & \ding{55}   &  \checkmark & IBM & VQE & Improved nonlinear feature extraction\\
                               & \cite{jeong2024short} & 2024 & \checkmark & \checkmark & IBM, PennyLane & VQC & Higher accuracy with fewer parameters \\ 
                               & \cite{phan2025modified} & 2025 & \ding{55} &  \checkmark    &  PennyLane   & VQC & Higher accuracy and lower latency \\
                               & \cite{sagingalieva2025photovoltaic} & 2025 & \ding{55} &  \checkmark    &  PennyLane   & VQC & Magnitudes of performance improvement \\ 
                               & \cite{oliveira2024application} & 2024 & \ding{55}   &  \checkmark & IBM & VQC & Superior robustness at longer horizons\\
                               & \cite{mechiche2025quantum} & 2025 & \ding{55} &  \checkmark    &  PennyLane & QFT &  Superior capture of periodic patterns \\   
                               & \cite{wang5336951quantum} & 2025 & \checkmark & \ding{55} & Origin Wukong & VQC & Improved accuracy and reduced cost \\  
                               & \cite{khan2024quantum} & 2024 & \ding{55}  &  \checkmark    &  PennyLane   & VQC  & Significant accuracy gains over classical \\   \cline{2-8}
                               % Wind Power Forecasting
                               & \cite{hong2023robust} & 2023 & \ding{55} &  \checkmark    &  PennyLane & PQC & Enhanced nonlinear feature learning \\ 
                               & \cite{hangun2024hybrid} & 2024 &  \ding{55} &  \checkmark    &  PennyLane & PQC & Improved learning efficiency vs classical\\ 
                               & \cite{pires2025quantum} & 2025 & \ding{55} &  \checkmark  & Kuatomu  & PQC & Classical competitive performance \\ 
                               & \cite{hangun2025comparative} & 2025 & \ding{55} &  \checkmark   & IBM & VQC & Architecture-dependent accuracy gains \\  
                               & \cite{hong2025implementing} & 2025 & \ding{55} &  \checkmark    &  PennyLane & PQC, QAOA  & Outperformed classical counterpart\\ 
                               & \cite{hangun2025quantum} & 2025 & \ding{55} &  \checkmark   & IBM & VQC & Classical competitive performance \\ 
                               & \cite{kumar2026wind} & 2026 & \ding{55} &  \checkmark   & PennyLane & VQC & Significantly improved accuracy \\ \cline{2-8}
                               % Electricity Load and Price
                               %
                               & \cite{kumar2023quantum} & 2023 & \ding{55} &  \checkmark & Classical Emulation  & QCNN & Improved forecasting accuracy \\     
                               & \cite{arvanitidis2023quantum} & 2023 & \ding{55} &  \checkmark   & IBM & VQC & Enhanced  classification accuracy  \\ 
                               & \cite{nutakki2024quantum} & 2024 & \ding{55} &  \checkmark & Classical Emulation & PQC & Higher accuracy vs classical models\\ 
                               & \cite{habibi2025electrical} & 2025 & \ding{55} &  \checkmark & Classical Emulation & PQC &  Enhanced uncertainty handling \\ 
                               & \cite{lalindeautomatic} & 2024 & \ding{55} &  \checkmark   & PennyLane & VQC & Demonstrated faster convergence\\
                               & \cite{wang5114742power} & 2025 & \ding{55} &  \checkmark & Classical Emulation & QA & Faster convergence\\
                               & \cite{zhang2025privacy}  & 2026 & \ding{55} &  \checkmark & Classical Emulation & VQC & Reduced communication and parameters\\ 
                               & \cite{dash2025dynamic} & 2025 & \checkmark & \ding{55} & D-Wave & QA & Outperformed classical counterpart\\  \hline
\multirow{17}{*}{Power Flow} &  \cite{feng2021quantum}   &  2021      &  \ding{55}   &  \checkmark    &    IBM   & HHL &  Theoretical exponential speed‑up \\                               
\multirow{17}{*}{Analysis} & \cite{feng2023noise} & 2023 & \checkmark & \checkmark & IBM & VQLS & Competitive performance vs. classical \\ 
                               & \cite{gao2023solving} & 2023 & \ding{55}   &  \checkmark & PyQPanda & HHL & Theoretical exponential speed‑up \\ 
                               & \cite{kaseb2024adiabatic} & 2024 & \checkmark & \checkmark & D‑Wave, Fujitsu & QA & Faster than classical iterations \\ 
                               & \cite{kaseb2025combinatorial} & 2025 & \checkmark & \checkmark & D‑Wave, Fujitsu & QA & Superior performance over classical ones \\
                               & \cite{kaseb2024hybrid} & 2024 & \ding{55} & \checkmark & IBM & PQC & Improved robustness and lower error \\ 
                               & \cite{kaseb2024quantum} & 2024 & \ding{55} & \checkmark &  IBM & PQC  & Higher robustness to noise\\
                               & \cite{zhu2025bayesian}  & 2025 & \ding{55} & \checkmark & PennyLane & VQC & Acheived faster convergence \\ 
                               & \cite{le2025learning} & 2025 & \ding{55} & \checkmark & PennyLane & VQC & Achieved better accuracy \\
                               & \cite{el2024newton} & 2024 &  \ding{55} & \checkmark &  IBM  & HHL & Faster convergence with high fidelity \\ 
                               & \cite{zheng2025early} & 2025 & \ding{55} & \checkmark & NWQSim & HHL & Potential exponential speed up \\
                               & \cite{liu2024quantum} & 2024 & \checkmark & \checkmark & IBM & HHL, VQLS &  Theoretical exponential speed up\\  
                               & \cite{zheng2024early} & 2024 & \ding{55} & \checkmark & NWQSim & HHL & Up to 8$\times$ speed up over IBM Qiskit \\ 
                               & \cite{kaseb2024power} & 2024 & \checkmark & \checkmark & D‑Wave, Fujitsu & QA & Classical comparable performance \\
                               & \cite{pareek2025limitations} & 2025 & \ding{55} & \checkmark & Not Reported & HHL & Exhibited no practical quantum gains \\
                               & \cite{fan2026quantum}  & 2026 & \ding{55} & \checkmark & IBM & QSVT & Obtained accurate and faster convergence \\  
                               & \cite{kaseb2025quantumRL} & 2025 & \checkmark & \checkmark & D‑Wave, Fujitsu & QA & Achieved faster convergence\\  \hline
\multirow{10}{*}{Optimal} &  \cite{morstyn2022annealing}   &  2023       &  \checkmark &  \ding{55}     &   D-Wave   & QA & Order of runtime speed up \\                               
\multirow{10}{*}{Power} & \cite{amani2023quantum}  & 2023 & \ding{55} & \checkmark & IBM & HHL & Showed potential exponential speed up\\
\multirow{10}{*}{Flow} & \cite{cao2024differentially} & 2024 & \ding{55} & \checkmark & PennyLane & VQC & Magnitude of runtime  speed up\\ 
                               & \cite{hu2024advancing} & 2025 & \ding{55} & \checkmark & IBM & PQC & Improved convergence and generalization \\
                               & \cite{amani2024quantum} & 2024 & \ding{55} & \checkmark & IBM & HHL &  Theoretical exponential speed-up \\
                               & \cite{hafshejani2025quantum} & 2025 & \ding{55} & \checkmark & Not Reported & HHL, VQLS & Classical comparable performance \\
                               & \cite{amani2025optimal} & 2025 & \ding{55} & \checkmark & IBM & HHL  & Theoretical speed‑up \\
                               & \cite{amani2025quantum} & 2025 & \checkmark & \checkmark & IBM, PennyLane & HHL, VQLS & Classical comparable performance\\
                               & \cite{magar2025quantum} & 2025 & \ding{55} & \checkmark  & IBM & QAOA & Classical comparable accurate solution\\
                               & \cite{carrillo2025quantum} & 2025 & \ding{55} & \checkmark & Qibo & AQC & Achieved global optimum solution \\ 
                               & \cite{prasad2025constraint} & 2025 & \checkmark & \checkmark & IBM, IonQ, D‑Wave & QAOA, QA & Achieved cost reduction\\
                               & \cite{le2026solving} & 2026 & \ding{55} & \checkmark & PennyLane & PQC &  Potential exponential representation gain \\ \hline
\multirow{6}{*}{Unit} &  \cite{koretsky2021adapting}  &  2021      &  \ding{55}   &  \checkmark    &  IBM   & QAOA &  Potential scalability gain over classical \\  
\multirow{6}{*}{commitment} & \cite{zheng2024fast} & 2024 & \checkmark & \checkmark & IBM & VQC & Achieved quadratic search speed up \\%
                               & \cite{salgado2024hybrid} & 2024 & \ding{55}   &  \checkmark    &  IBM   & QAOA & Theoretical polynomial time complexity \\
                               & \cite{christeson2024quantum}  & 2024 & \checkmark & \ding{55} & D-Wave & QA & Achieved linear computation scaling \\   
                               & \cite{aboumrad2025new}  & 2025 & \checkmark & \checkmark & IonQ & VQA & Achieved low approximation error  \\   
                               & \cite{yang2025scalable} & 2025 & \ding{55} & \checkmark & IBM & QAOA & Reduced qubit requirement\\
                               & \cite{hasanzadeh2025d2} & 2025 & \ding{55} & \checkmark & IBM & VQE & Achieved faster convergence\\   
                               & \cite{nikmehr2022quantumUC} & 2022 & \ding{55}   &  \checkmark    &  IBM   & QAOA & Improved computational efficiency\\ 
                               & \cite{hong2025qubit} & 2025 & \checkmark & \ding{55} & D-Wave & QA & Runtime gain over classical solvers \\
                               & \cite{mahroo2026machine} & 2026 & \ding{55} & \checkmark & IBM & QAOA & Reduced quantum resources up to $\sim$70\%\\
                               & \cite{huang2025consensus}  & 2025 & \checkmark & \ding{55} & D‑Wave, QBoson & QA & Achieved faster convergence vs classical\\
                               & \cite{gao2025distributed} & 2025 & \checkmark & \checkmark & D‑Wave & QA & Runtime advantage vs classical solvers \\
                               & \cite{ling2026hybrid} & 2026 & \checkmark & \checkmark & QBoson & Ising & Order of runtime speed up over classical \\  
                               & \cite{fu2025quantum} & 2025 & \checkmark & \checkmark & IBM, D-Wave, QBoson  & QAOA, QA & Faster runtime over classical solvers \\
                               & \cite{feng2022novel} & 2023 & \checkmark & \checkmark & IBM & QAOA & Faster convergence and scalability gains \\
                               & \cite{mahroo2023learning} & 2023 & \checkmark & \checkmark & IBM & QAOA & Achieved faster solution \\
                               & \cite{paterakis2023hybrid} & 2023 & \checkmark & \ding{55} & D-Wave & QA &  Practical runtime reductions \\   
                               & \cite{christeson2026hybrid} & 2026 & \checkmark & \ding{55} & D-Wave & QA & Empirical speed-up on large systems \\    
                               & \cite{wei2026quantum} & 2026 & \ding{55} & \checkmark & Not Reported & PQC & Faster convergence over classical model \\  
                               & \cite{magar2024dc} & 2024 & \ding{55}   &  \checkmark    &  IBM   & QAOA  & Achieved comparable optimal solution \\
                               & \cite{ling2025hybrid}  & 2025 & \checkmark & \checkmark & D‑Wave & QA & Real hardware speedup over simulator \\   
                               & \cite{muller2025quantum} & 2025 & \checkmark & \ding{55} & D-Wave & QA & Reduced qubit requirement \\ \hline
\multirow{9}{*}{Demand} &  \cite{bucher2023dynamic}  &  2023        &  \checkmark  &  \checkmark   &  D‑Wave   & QA &  Empirical speed‑up over classical \\  
\multirow{9}{*}{Response} & \cite{bucher2024incentivizing} & 2024 & \checkmark  &  \checkmark   &  D‑Wave   & QA & Enhanced scalability over classical model \\ 
                               & \cite{wan2024optimization} & 2024 & \checkmark & \ding{55} & QBoson & QUBO & Order of speed up over classical solver\\
                               & \cite{xin2025quantum}  & 2025 & \checkmark & \checkmark & QBosoN & Ising  & Runtime reduction over classical solver  \\
                               & \cite{minghong2025behavior} & 2025 & \ding{55} & \checkmark & D-Wave & SA, QA &  Improved scalability and convergence\\
                               & \cite{ajagekar2024demand} & 2024 & \ding{55} & \checkmark & Not Reported & VQC & Enhanced parameter efficiency\\
                               & \cite{ajagekar2024variational}  & 2024 & \checkmark & \ding{55} & IBM & VQC & Achieved logarithmic scaling with qubits\\ 
                               & \cite{safari2024neuroquman}  & 2024 & \ding{55} & \checkmark & IBM, PennyLane & VQC & Achieved higher accuracy \\    
                               & \cite{zhuang2025quantum} & 2025 & \ding{55} & \checkmark & IBM, PennyLane & VQC, QAE & Theoretical quadratic speed‑up \\   
                               & \cite{yeboah2025quantum} & 2025 & \ding{55}  & \checkmark & IBM & PQC & Improved convergence speed \\  \hline
\multirow{5}{*}{Energy} &  \cite{kumar2023blockchain}   &  2023        &   \ding{55}  & \checkmark     &  Rigetti, PyQuil   & GS, GAA  &  Empirical convergence speed-up  \\  
\multirow{5}{*}{Trading} & \cite{moniruzzaman2025peer} & 2025 &  \ding{55}  & \checkmark & IBM & VQC & Empirical speed‑up vs classical model \\ 
                               & \cite{mastroianni2023assessing} &  2023 & \checkmark & \checkmark & IBM & QAOA & Potential asymptotic quantum advantage \\  
                               & \cite{blenninger2024q} & 2024 & \checkmark & \checkmark & D‑Wave & QA & Improved runtime scaling vs classical \\ 
                               & \cite{bucher2025grid} & 2025 & \checkmark & \ding{55} &  D‑Wave & QA  & QA performs worse than classical solvers \\
                               & \cite{huang2026quantum}  & 2026 & \checkmark & \checkmark & IBM & VQC & Empirical speed-up:  faster convergence \\ \hline
\multirow{6}{*}{EV Energy} &  \cite{kea2023leveraging}   &  2023        &   \checkmark  & \checkmark     &  IBM   & QAOA  &  Classic-comparable simulator results  \\  
\multirow{6}{*}{Management} & \cite{wang2025employing} & 2025 &  \checkmark & \ding{55} & D-Wave & QA & Empirical speed‑up vs classical MILP \\
                               & \cite{rashnu2025optimization}  & 2025 & \ding{55} & \checkmark & Classical Emulation & QA & Provided practical convergence speed‑up \\  
                               & \cite{deng2025quantum} & 2025 & \checkmark & \ding{55} & D-Wave & QA & Order of faster than classical solvers\\  
                               & \cite{khatiri2025quantum}  &  2025 & \ding{55} & \checkmark & IBM & QNN & Model‑level performance improvement\\  
                               & \cite{xu2025quantum}  &  2025 & \ding{55} & \checkmark & TensorFlow Quantum & VQC & Faster convergence vs classical \\  
                               & \cite{yeboah2025quantumEV}  &  2025 & \ding{55} & \checkmark & IBM & VQC & Improved convergence speed-up \\  \hline
\multirow{14}{*}{Economic} &  \cite{leenders2024integrating}   &  2024        &   \checkmark  &  \ding{55}    &  D-Wave   & QA   &  Demonstrates feasibility, but slow  \\  
\multirow{14}{*}{Dispatch} & \cite{zhao2023optimal}  & 2024 & \checkmark  &  \ding{55}  & D-Wave & QA & Faster convergence in complex cases\\  
                               & \cite{awan2024quantum}  &  2024 & \ding{55} & \checkmark & QuantumSim & QA  & Improved system-level performance \\   
                               & \cite{ajagekar2024variationalEMS}  &  2024  & \ding{55} & \checkmark & Not Reported & VQC & Model‑complexity speed‑up\\    
                               & \cite{shi2025quantum}  &  2025 & \checkmark  &  \ding{55}  & D-Wave & QA & Computation speed-up vs classical\\ 
                               & \cite{xu2025joint}   &  2025 & \checkmark  &  \ding{55}  & D-Wave & QA  & Oder of reduction in computation time \\
                               & \cite{li2025quantum_EMS} & 2025 & \checkmark  &  \ding{55}  & D-Wave & QA  & Slower solution time scaling vs classical \\
                               & \cite{wu2025hybrid}   &  2025 & \ding{55} & \checkmark & Classical Emulation & PQC & Faster training and improved accuracy  \\
                               & \cite{han2025quantum} & 2025 & \ding{55} & \checkmark & IBM & QAOA, QAE & Quadratic speedup in scenario generation \\
                               & \cite{liu2025quantum_sus}   &  2025 & \checkmark & \checkmark & D‑Wave, PyQuil   & QA, QAOA & Order of speed up vs classical MILP \\
                               & \cite{zhang2026physics}    &  2026 & \ding{55} & \checkmark & IBM & VQA & Faster convergence vs classical model \\       
                               & \cite{han2026quasi}    &  2026 & \checkmark  &  \ding{55}  & D-Wave & QA & Improved computational efficiency \\  
                               & \cite{shi2025optimal}   &  2025 & \ding{55} & \checkmark & Classical Emulation & PQC  & Fewer parameters vs classical model\\    
                               & \cite{wang2025optimal} & 2025 & \ding{55} & \checkmark & Classical Emulation & QA & System-level performance improvement \\   
                               & \cite{qian2025integrated}  & 2025 & \checkmark  &  \ding{55}  & D-Wave & QA   & Faster convergence under uncertainty \\     \hline  
\multirow{10}{*}{Grid} &  \cite{diadamo2022practical}  &  2022        &   \checkmark  &  \checkmark    &  IBM   & AE   &  Improved accuracy \\  
\multirow{10}{*}{Partitioning}                               & \cite{wang2022quantum}  &  2022 & \ding{55} & \checkmark & Classical Emulation & QA & Enabled quantum-compatible solution\\ 
\multirow{10}{*}{and}                               & \cite{wang2023quantum}  &  2023 &  \checkmark  &  \ding{55}  & D-Wave & QA & Achieved faster solution vs classical \\
\multirow{10}{*}{Others}     & \cite{bucher2025evaluating}  &  2025 & \checkmark  &  \ding{55}  & D-Wave & QA  & Improved solution quality vs classical\\                                                   %
                               & \cite{silva2023quantumothers}  &  2023  & \checkmark & \ding{55} & D‑Wave & QA & Improved solution and decision efficiency \\ 
                               & \cite{colucci2023power}  &  2023 & \checkmark  &  \ding{55}  & D-Wave & QA & Empirical improvements in optimality  \\
                               & \cite{singha2025application}  &  2025 & \ding{55} & \checkmark & IBM & QAOA & Acheived faster solution vs classical \\
                               & \cite{liu2025alleviating} & 2025 & \checkmark & \ding{55} & QBoson & Ising & Curse of dimensionality free speed‑up \\
                               & \cite{lin2025two}  &  2025 & \ding{55} & \checkmark & PennyLane   & VQC & Improved training efficiency\\  
                               & \cite{zhang2025quantum}  &  2025 & \ding{55} & \checkmark & Origin Quantum & QAOA & Better accuracy than classical methods\\
                               & \cite{ioannou2026quantum} &  2026 & \ding{55}  & \checkmark & Classical Emulation & QAOA & Achieved optimal solutions in simulation \\
                               & \cite{jiang2026regrid} &  2026 & \checkmark  & \ding{55} & IBM & QAOA & Quant. resource-efficient optimal solution \\
                               & \cite{sinha2025quantum}  &  2025 & \ding{55} & \checkmark & IBM & QAOA, VQE & Convergence speed and accuracy gain \\ \hline
\end{longtable} 
\twocolumn
\normalsize

\section{Challenges and Limitations} \label{Challenges and Limitations}

\subsection{Hardware Limitations}
Current quantum hardware remains constrained by several fundamental limitations that affect reliability, accuracy, and practical deployment. High error rates caused by decoherence, gate imperfections, crosstalk, and measurement noise accumulate with circuit depth and reduce computational fidelity \cite{preskill2018quantum,kandala2017hardware}. Short coherence times further restrict the depth of executable circuits, limiting the performance of iterative and variational algorithms \cite{kjaergaard2020superconducting}. Restricted qubit connectivity introduces additional challenges because nonlocal interactions often require routing operations such as SWAP gates, increasing circuit depth and error accumulation \cite{sheldon2016procedure}. Moreover, quantum measurements are inherently probabilistic and require repeated sampling to estimate expectation values, introducing statistical uncertainty and computational overhead \cite{temme2017error}. Practical operation also requires frequent calibration to compensate for environmental fluctuations and parameter drift \cite{gambetta2017building}. Furthermore, scaling quantum processors while maintaining coherence, controllability, and low error rates remains a major open challenge \cite{arute2019quantum}. The need for specialized infrastructure, including cryogenic cooling and precise environmental control, further increases operational complexity and cost \cite{clarke2008superconducting}. These limitations constrain the reliable execution of large-scale smart grid applications and motivate continued advances in fault-tolerant hardware and error-mitigation techniques.

\subsection{Problem Encoding}
Efficiently representing power system problems in quantum-compatible forms remains a major challenge. Most power system models involve continuous variables, whereas many quantum optimization frameworks, particularly quantum annealing, require discrete QUBO or Ising formulations. Converting continuous variables into binary representations introduces approximation errors and can substantially increase qubit requirements, especially in high-precision applications \cite{djidjev2025extending,arai2023effectiveness}. Constraint handling presents an additional difficulty. Equality and inequality constraints are commonly incorporated through penalty functions, but inappropriate penalty selection may either violate constraints or distort the optimization landscape. Furthermore, auxiliary variables introduced for constraint enforcement can significantly increase problem size and resource requirements \cite{kanatbekova2025qubit,gabbassov2025lagrangian}. Many power system models also exhibit nonlinear and nonconvex characteristics. For example, AC power-flow equations often require approximations before they can be mapped onto quantum optimization frameworks, potentially reducing model fidelity and generating dense interaction structures that are difficult to embed on current hardware \cite{kaseb2024power}. In gate-based quantum computing, additional challenges arise from quantum state preparation, Hamiltonian construction, and observable estimation, all of which complicate the efficient representation of large-scale power system models \cite{arseniev2024tridiagonal,volpe2024towards,ikhtiarudin2025shot,scriva2024challenges}.

\subsection{Algorithmic Limitations}
%Despite substantial progress in recent years, many quantum algorithms remain insufficiently mature for large-scale smart grid deployment. Most current applications rely on general-purpose algorithms such as QAOA and VQE, which are not specifically designed to exploit the structural characteristics of power system problems. Consequently, their performance often remains comparable to, rather than consistently superior to, advanced classical optimization methods. Many contemporary approaches also depend heavily on hybrid quantum-classical frameworks, introducing additional optimization overhead and convergence challenges, particularly in variational settings \cite{cerezo2021variational,mcclean2018barren}. Furthermore, many proposed methods lack rigorous benchmarking, scalability analysis, and systematic comparisons against state-of-the-art classical techniques. Combined with sensitivity to hardware noise and implementation constraints, these limitations highlight the need for robust, scalable, and domain-specific quantum algorithms tailored to power and energy applications.
Despite substantial progress in recent years, many quantum algorithms remain insufficiently mature for large-scale smart grid deployment. Most current applications rely on general-purpose algorithms such as QA, QAOA, and VQE, which are not specifically designed to exploit the structural characteristics of power system problems. Consequently, their performance often remains comparable to, rather than consistently superior to, advanced classical optimization methods. Many contemporary approaches also depend heavily on hybrid quantum-classical frameworks, introducing additional optimization overhead and convergence challenges, particularly in variational settings \cite{cerezo2021variational,mcclean2018barren}. Furthermore, many proposed methods lack rigorous benchmarking, scalability analysis, and systematic comparisons against state-of-the-art classical techniques. Combined with sensitivity to hardware noise and implementation constraints, these limitations highlight the need for robust, scalable, and domain-specific quantum algorithms tailored to power and energy applications. Another challenge arises from the limited transferability of many quantum algorithms across different power system applications. Methods that perform well for a particular optimization or learning task may require substantial reformulation when applied to different network topologies, operational objectives, or uncertainty models. In addition, the performance of many variational algorithms depends strongly on circuit architecture, parameter initialization, and optimizer selection, leading to highly application-dependent outcomes. The absence of standardized evaluation methodologies further complicates objective performance assessment and reproducibility. %Therefore, developing adaptive, transferable, and systematically benchmarked quantum algorithms remains an important research direction for smart grid applications.

\subsection{Scalability Issues}
%Scalability remains one of the most significant barriers to practical quantum computing in smart grids. Large-scale power system applications involve high-dimensional optimization, simulation, and decision-making problems that demand substantial quantum resources. As problem size increases, requirements for qubits, circuit depth, variational parameters, and measurements often grow rapidly, frequently exceeding the capabilities of current NISQ devices \cite{preskill2018quantum}. Variational algorithms are particularly affected because expanding parameter spaces increases optimization complexity and training cost. Similarly, measurement overhead grows with system size because accurate expectation-value estimation requires numerous circuit executions \cite{cerezo2021variational,mcclean2018barren}. Consequently, many realistic transmission and distribution system applications remain beyond the reach of existing quantum platforms. Overcoming these limitations will require resource-efficient algorithms, problem decomposition strategies, distributed quantum architectures, and ultimately fault-tolerant quantum computing.
Scalability remains one of the most significant barriers to practical quantum computing in smart grids. Large-scale power system applications involve high-dimensional optimization, simulation, and decision-making problems that demand substantial quantum resources. As problem size increases, requirements for qubits, circuit depth, variational parameters, and measurements often grow rapidly, frequently exceeding the capabilities of current NISQ devices \cite{preskill2018quantum}. Variational algorithms are particularly affected because expanding parameter spaces increases optimization complexity and training cost. Similarly, measurement overhead grows with system size because accurate expectation-value estimation requires numerous circuit executions \cite{cerezo2021variational,mcclean2018barren}. Consequently, many realistic transmission and distribution system applications remain beyond the reach of existing quantum platforms. Scalability challenges are further intensified by the growing complexity of modern smart grids, which increasingly incorporate distributed energy resources, electric vehicles, responsive loads, microgrids, and advanced communication networks. These systems generate large volumes of dynamic and geographically distributed data that must be processed within operational time constraints. Moreover, decomposing large optimization or simulation problems into smaller quantum-manageable subproblems often introduces coordination overhead between subsystems, reducing potential computational gains. Achieving practical scalability therefore requires not only larger quantum processors but also efficient decomposition strategies, distributed quantum architectures, and hybrid computational frameworks capable of coordinating large-scale tasks.

\subsection{Limited Qubit Precision}
%Quantum algorithms often provide limited numerical precision when representing continuous-valued engineering quantities, particularly under finite sampling and finite-resolution encodings \cite{preskill2018quantum,cerezo2021variational}. This limitation is especially important for power system applications requiring high numerical accuracy, including state estimation, power flow analysis, optimal power flow, and stability assessment. Small numerical inaccuracies may propagate through computations and influence operational decisions, potentially resulting in infeasible or suboptimal solutions. In addition, discretization of continuous variables introduces approximation errors and increases resource requirements. Although hybrid quantum-classical techniques can partially alleviate these effects, limited numerical precision remains a significant challenge for high-fidelity smart grid applications.
Quantum algorithms often provide limited numerical precision when representing continuous-valued engineering quantities, particularly under finite sampling and finite-resolution encodings \cite{preskill2018quantum,cerezo2021variational}. This limitation is especially important for power system applications requiring high numerical accuracy, including state estimation, power flow analysis, optimal power flow, and stability assessment. Small numerical inaccuracies may propagate through computations and influence operational decisions, potentially resulting in infeasible or suboptimal solutions. In addition, discretization of continuous variables introduces approximation errors and increases resource requirements. Although hybrid quantum-classical techniques can partially alleviate these effects, limited numerical precision remains a significant challenge for high-fidelity smart grid applications. Precision constraints can also affect uncertainty quantification and probabilistic decision-making processes that depend on accurate estimation of low-probability events. Applications such as reliability assessment, resiliency analysis, renewable energy forecasting, and risk-aware operational planning often require precise representation of probability distributions and statistical metrics. Even minor numerical deviations may alter the assessment of critical operating conditions and influence risk-sensitive decisions. Consequently, advances in encoding techniques, error mitigation, and hardware stability will be essential for supporting precision-intensive power system analyses.

\vspace{-2pt}
\subsection{Backend Latency}
Practical deployment is further constrained by backend latency. Most quantum processors are presently accessed through cloud-based platforms, requiring communication between classical systems and remote quantum hardware \cite{preskill2018quantum}. This issue is particularly pronounced in hybrid quantum-classical workflows, where repeated cycles of job submission, execution, and result retrieval can introduce substantial cumulative delays. Backend latency limits the applicability of quantum computing to time-sensitive smart grid functions such as optimal power flow, economic dispatch, contingency analysis, fault diagnosis, and real-time control. Although emerging solutions, including edge-based quantum services and tighter hardware-software integration, may alleviate communication overhead, latency remains a significant obstacle to operational deployment. Additional challenges arise from resource sharing in cloud-based quantum environments. Queue waiting times, variable execution schedules, and competing user workloads can introduce unpredictable delays that affect computational consistency and responsiveness. Such variability is particularly problematic for operational applications that require strict computational deadlines and rapid decision-making. Reducing latency through improved scheduling mechanisms, optimized communication protocols, and geographically distributed quantum services will be important for supporting future real-time smart grid applications.

\vspace{-6pt}
\subsection{Data and Integration Issues}
Integrating quantum computing into smart grid infrastructures introduces challenges related to data management, interoperability, and system integration. Power system data originate from heterogeneous sources, including SCADA systems, PMUs, smart meters, and IoT devices, and are often affected by noise, uncertainty, and missing values \cite{montanari2012graphical,zhu2012sparse}. These characteristics complicate preprocessing and quantum-state preparation. Furthermore, the distributed architecture of modern power systems contrasts with the centralized nature of current quantum computing resources. Effective deployment therefore requires new middleware, communication protocols, and hybrid computing architectures capable of integrating quantum and classical platforms. Data privacy, cybersecurity, and regulatory compliance impose additional constraints that further complicate adoption in operational environments. Another important challenge involves maintaining data consistency across interconnected computational environments. Smart grid applications frequently rely on continuously updated measurements and near-real-time information exchange. Synchronizing these data streams between classical infrastructures and quantum workflows introduces additional architectural complexity and communication overhead. Moreover, deploying quantum-enabled applications within existing utility infrastructures may require modifications to legacy software systems, operational procedures, and maintenance practices. Addressing these integration challenges will be crucial for achieving reliable and scalable deployment in real-world power system environments.

\subsection{Ethical Issues}
Beyond technical barriers, quantum computing raises important ethical and societal concerns related to cybersecurity, privacy, transparency, and equitable access. A major issue is the potential impact of large-scale quantum computers on current cryptographic systems. Algorithms such as Shor's algorithm could theoretically compromise widely used public-key cryptography, including RSA and elliptic-curve schemes, threatening the security of critical power system infrastructure \cite{shor1999polynomial,mosca2018cybersecurity}. The transition to post-quantum cryptography introduces additional challenges, including increased computational overhead, larger key sizes, and infrastructure compatibility issues \cite{bernstein2025post}. Blockchain-enabled smart grid applications may face similar risks because many existing digital signature mechanisms are not quantum-resistant \cite{al2025post,parida2023post}. In addition, the limited interpretability of many hybrid and variational quantum algorithms can reduce transparency and trust in operational decision-making processes \cite{cerezo2021variational}. Finally, the high cost and limited availability of quantum resources raise concerns regarding equitable access and the potential widening of technological disparities among stakeholders. Addressing these concerns will require quantum-secure infrastructures, transparent algorithms, and inclusive technology-development strategies.

\subsection{Key Takeaways} 
%The challenges discussed above can be broadly categorized into four interconnected barriers: (i) hardware constraints, (ii) problem representation and encoding complexity, (iii) algorithmic maturity and scalability limitations, and (iv) deployment, integration, and governance concerns. Among these, hardware fidelity, efficient problem encoding, and scalable algorithm design currently represent the most significant technical obstacles to achieving practical quantum advantage in smart grid applications. While future advances in fault-tolerant quantum computing are expected to alleviate many hardware-related limitations, widespread adoption will also require domain-specific algorithms, resource-efficient hybrid quantum-classical frameworks, interoperable software ecosystems, and quantum-secure grid infrastructures.
The challenges discussed above can be broadly categorized into four interconnected barriers: (i) hardware constraints, (ii) problem representation and encoding complexity, (iii) algorithmic maturity and scalability limitations, and (iv) deployment, integration, and governance concerns. Among these, hardware fidelity, efficient problem encoding, and scalable algorithm design currently represent the most significant technical obstacles to achieving practical quantum advantage in smart grid applications. Importantly, these challenges are highly interdependent rather than isolated. Hardware limitations directly influence algorithm design, while encoding complexity often determines scalability and computational feasibility. Similarly, integration challenges associated with latency, interoperability, cybersecurity, and data management become increasingly significant as quantum technologies move beyond laboratory demonstrations toward operational deployment. While future advances in fault-tolerant quantum computing are expected to alleviate many hardware-related limitations, widespread adoption will also require domain-specific algorithms, resource-efficient hybrid quantum-classical frameworks, interoperable software ecosystems, standardized benchmarking practices, and quantum-secure grid infrastructures. Successfully addressing these challenges will require coordinated progress across hardware, software, algorithms, and system integration to fully realize the potential of quantum computing in future smart grid operations.

\section{Future Directions and Roadmaps} \label{Future Directions} \subsection{Democratization of Quantum Technologies} 
The long-term impact of quantum computing on smart grid applications will depend heavily on broad and equitable access to quantum resources. Cloud-based quantum services, open-source software ecosystems, and hardware-agnostic development platforms are expected to lower technical and economic barriers, enabling utilities, researchers, and industry practitioners to experiment with quantum-enabled solutions without direct access to specialized hardware \cite{seskir2023democratization}. Future efforts should prioritize accessible programming environments, standardized development toolkits, open benchmarks, and publicly available test systems that facilitate reproducible research and technology transfer. Such developments can accelerate innovation and support the widespread adoption of quantum computing across the power and energy sector. An equally important objective is reducing knowledge barriers that limit participation in quantum research and development. Community-driven software projects, openly accessible educational resources, and collaborative research initiatives can help broaden engagement across institutions with varying levels of technical expertise and financial resources. Expanding access to practical training environments and cloud-based experimentation platforms will further support innovation by enabling a larger and more diverse research community to contribute to quantum-enabled energy solutions.

\subsection{Advancements in Fault-Tolerant Architectures} 
Achieving practical quantum advantage ultimately requires fault-tolerant quantum computing. Although current NISQ devices have enabled important proof-of-concept demonstrations, their computational capability remains constrained by noise and limited circuit depth. Fault-tolerant architectures based on quantum error correction are expected to enable large-scale and reliable computation by protecting logical qubits against physical errors \cite{shor1995scheme,fowler2012surface}. Future research should focus on reducing the resource overhead of error correction, improving qubit coherence, enhancing gate fidelity, and developing efficient decoding algorithms. Continued progress in these areas will be essential for supporting complex optimization, simulation, and machine-learning tasks relevant to next-generation smart grids. Future developments are also expected to emphasize hardware-software co-design approaches that jointly optimize quantum architectures and application requirements. Such strategies can improve resource utilization, reduce implementation overhead, and enhance computational performance. In addition, advances in logical qubit construction, error-detection mechanisms, and adaptive correction strategies may significantly improve the practicality of fault-tolerant systems and accelerate their transition from research laboratories to operational environments. 

\subsection{Hybrid Quantum-Classical Frameworks} 
Hybrid quantum-classical computing is likely to remain the dominant computational paradigm in the near and medium term. By delegating data processing, constraint management, and optimization orchestration to classical processors while exploiting quantum hardware for computationally demanding subproblems, hybrid frameworks can effectively utilize limited quantum resources. Variational approaches such as VQE and QAOA exemplify this strategy and have demonstrated considerable promise across smart grid applications \cite{cerezo2021variational}. Future work should focus on improving convergence reliability, reducing communication overhead, and developing application-specific hybrid architectures that balance computational efficiency with hardware constraints. Another promising direction involves adaptive workload allocation between quantum and classical resources. Intelligent orchestration mechanisms capable of dynamically selecting the most suitable computational platform for specific tasks could improve overall efficiency and system responsiveness. Such adaptive architectures may become particularly valuable for large-scale power system applications involving heterogeneous datasets, uncertainty, and continuously changing operating conditions.

\subsection{Application-Specific Algorithm Development} 
The development of domain-specific quantum algorithms represents a critical step toward practical deployment. Generic algorithms provide valuable foundations but often fail to fully exploit the structural characteristics of power system problems. Future research should emphasize algorithms tailored to optimization, simulation, uncertainty quantification, forecasting, cybersecurity, and control applications. By leveraging problem-specific sparsity, locality, and physical constraints, such approaches can improve efficiency, reduce resource requirements, and enhance robustness to noise \cite{montanaro2016quantum,bharti2022noisy}. Identifying applications where meaningful quantum advantage can be achieved remains a key research priority. Future investigations should also focus on integrating physical system knowledge directly into algorithmic design. Incorporating network topology, operational constraints, and domain-specific heuristics may enable more compact quantum representations and improved solution quality. Furthermore, benchmark-driven algorithm development can help identify application categories that provide the greatest practical benefits under realistic hardware limitations.

\subsection{Scalable Hardware and System Integration} 
Large-scale deployment will require significant advances in both hardware capability and system-level integration. Future quantum platforms must support larger qubit counts while maintaining acceptable levels of fidelity, coherence, and controllability \cite{kjaergaard2020superconducting,arute2019quantum}. Promising directions include modular and distributed quantum architectures in which multiple processors operate cooperatively. Equally important is the development of seamless interfaces between quantum and classical computing environments. Efficient co-processing frameworks, communication infrastructures, and standardized integration methodologies will be essential for incorporating quantum technologies into real-world operational workflows. Research is also expected to explore geographically distributed quantum computing infrastructures that can interact with distributed smart grid architectures. Such systems may enable collaborative processing across multiple quantum resources while improving reliability and resource utilization. Advances in communication technologies and quantum networking could further support the development of scalable and interconnected computing ecosystems suitable for future power system operations.

\subsection{Resource-Efficient Quantum Computing} 
Because quantum resources remain scarce and expensive, improving computational efficiency is a fundamental research objective. Resource-efficient quantum computing seeks to minimize qubit requirements, circuit depth, gate operations, and measurement overhead while preserving solution quality. Potential approaches include compact problem encodings, circuit compression, qubit reuse, sparse representations, and hardware-aware circuit design \cite{cerezo2021variational,bharti2022noisy}. Advances in resource-efficient computing will be particularly important for large-scale power system applications, where practical deployment depends on balancing computational performance with hardware limitations. Additional opportunities exist in adaptive approximation methods that strategically balance accuracy and computational cost. Developing algorithms that intelligently allocate resources based on problem complexity can improve scalability while reducing execution requirements. Such techniques may enable larger smart grid applications to be addressed with available quantum hardware before fully fault-tolerant systems become practical.

\subsection{Interoperable Quantum Software Ecosystems} 
As the quantum computing ecosystem expands, interoperability will become increasingly important. Current hardware platforms often employ different programming languages, software stacks, and execution environments, creating barriers to portability and long-term adoption. Future progress will require hardware-agnostic software frameworks, standardized interfaces, unified programming abstractions, and portable intermediate representations that support execution across multiple platforms \cite{cross2017open,javadi2024quantum}. In parallel, advances in compiler technology, transpilation, circuit optimization, and noise-aware scheduling will improve hardware utilization and execution fidelity. Collectively, these developments will simplify application deployment, enhance reproducibility, and reduce vendor dependence. Beyond technical compatibility, interoperable ecosystems can facilitate collaborative research and accelerate technology transfer between academia, industry, and software developers. Standardized benchmarking methodologies, common data formats, and portable workflow pipelines will improve reproducibility and support more objective comparisons across competing hardware and software platforms.

\subsection{Workforce Development and Interdisciplinary Research} 
The successful integration of quantum computing into smart grids depends on the availability of a highly skilled and interdisciplinary workforce. Future progress will require expertise spanning quantum information science, electrical engineering, computer science, applied mathematics, and energy systems. Educational programs, specialized training initiatives, and collaborative research centers can help cultivate the necessary talent base. Strengthening partnerships among academia, industry, utilities, and government agencies will further accelerate knowledge transfer and facilitate the conversion of scientific advances into practical applications. Long-term workforce development efforts should also emphasize interdisciplinary problem-solving and application-oriented training. Encouraging collaboration between quantum scientists and power engineers can help bridge disciplinary boundaries and promote solutions that address realistic operational challenges. Such collaborative ecosystems will be essential for translating theoretical advances into deployable smart grid technologies.

\subsection{Roadmap Summary} 
The future development of quantum computing for smart grid applications can be viewed through four complementary pathways. First, advances in fault-tolerant architectures, scalable hardware, and resource-efficient computing will improve computational capability and reliability. Second, application-specific algorithms and hybrid quantum-classical frameworks will enhance the practical utility of near-term quantum devices. Third, interoperable software ecosystems will promote portability, accessibility, and large-scale deployment across heterogeneous platforms. Finally, democratization initiatives and workforce development efforts will strengthen the research community and accelerate technology adoption. Importantly, these pathways are highly interconnected. Progress in hardware capabilities will influence algorithm development, while improvements in software ecosystems can enhance accessibility and adoption. Likewise, workforce development and democratization initiatives provide the foundation necessary to support innovation across all technological layers. These directions collectively provide a roadmap toward practical, scalable, secure, and trustworthy quantum-enabled smart grid systems capable of addressing the increasing complexity of future energy infrastructures.

\section{Conclusions}\label{Conclusion}
This paper presented a comprehensive review of quantum computing and its emerging applications in next-gen smart grid operations. The review covered the fundamental principles of quantum computing, key quantum algorithms, and the evolving hardware and software ecosystem that supports quantum application development. Furthermore, it provided a structured assessment of existing research on quantum-enabled optimization, control, forecasting, reliability assessment, cybersecurity, energy management, and other critical smart grid functions.
The literature reviewed in this work demonstrates that quantum computing has significant potential to address the growing computational complexity of modern power and energy systems. In particular, hybrid quantum-classical frameworks have emerged as a practical and promising approach for leveraging the capabilities of current NISQ-era devices, offering opportunities to improve computational efficiency, scalability, and solution quality for large-scale smart grid problems. Although many reported studies remain at the proof-of-concept stage, the results indicate that quantum methods can provide promising application-specific improvements for selected optimization, learning, and decision-making tasks, although broadly generalizable and end-to-end quantum advantage has not yet been established.

However, realizing the full potential of quantum computing in real-world smart grid applications will require overcoming several important challenges. Current limitations in quantum hardware, including noise, qubit count, coherence time, and error rates, continue to constrain algorithm performance and scalability. Additional challenges include efficient problem encoding, resource management, benchmark standardization, and integrating quantum solutions with existing power system infrastructure and operational practices. Addressing these issues will require continued advances in fault-tolerant quantum hardware, resource-efficient and application-specific algorithms, scalable hybrid quantum–classical architectures, and comprehensive validation using realistic power system datasets and operating conditions.
Looking forward, the convergence of quantum computing, artificial intelligence, advanced communication networks, and digitalized energy infrastructures presents a promising pathway toward next-gen smart grids. Continued research, technological advances, and strong interdisciplinary collaboration among quantum scientists, computer scientists, and power engineers will be essential to accelerate this transition. Although quantum computing is still in its early stages, its long-term potential to transform next-gen smart grid operations positions it as a potentially important enabling technology for future intelligent, resilient, sustainable, and highly efficient power and energy systems.

\section*{Acknowledgment}
While preparing this manuscript, the authors used the generative artificial intelligence tool Microsoft Copilot to improve grammar, clarity, and sentence structure. The authors carefully reviewed and edited all AI-assisted content and take full responsibility for the manuscript’s accuracy, originality, and integrity.

\vspace{-1pt}
\bibliographystyle{IEEEtran}
\bibliography{references}

@article{cavus2025advancing,
  title={Advancing power systems with renewable energy and intelligent technologies: A comprehensive review on grid transformation and integration},
  author={Cavus, Muhammed},
  journal={Electronics},
  volume={14},
  number={6},
  pages={1159},
  year={2025},
  publisher={MDPI}
}

@article{kurella2025comprehensive,
  title={A comprehensive review of DERMS for smart and secure energy networks: standards, protocols, and security considerations},
  author={Kurella, Pavan Kumar and Mikkili, Suresh},
  journal={International Journal of Ambient Energy},
  volume={46},
  number={1},
  pages={2506703},
  year={2025},
  publisher={Taylor \& Francis}
}

@book{liu2025smart,
  title={Smart grid handbook, 3 volume set},
  author={Liu, Chen-Ching and McArthur, Stephen and Lee, Seung-Jae},
  volume={3},
  year={2025},
  publisher={John Wiley \& Sons}
}

@article{khan2025comprehensive,
  title={Comprehensive review of hybrid energy systems: challenges, applications, and optimization strategies},
  author={Khan, Aqib and Bressel, Mathieu and Davigny, Arnaud and Abbes, Dhaker and Ould Bouamama, Belkacem},
  journal={Energies},
  volume={18},
  number={10},
  pages={2612},
  year={2025},
  publisher={MDPI}
}

@incollection{feynman2018simulating,
  title={Simulating physics with computers},
  author={Feynman, Richard P},
  booktitle={Feynman and computation},
  pages={133--153},
  year={2018},
  publisher={cRc Press}
}

@article{deutsch1985quantum,
  title={Quantum theory, the Church--Turing principle and the universal quantum computer},
  author={Deutsch, David},
  journal={Proceedings of the royal society of London. A. mathematical and physical sciences},
  volume={400},
  number={1818},
  pages={97--117},
  year={1985},
  publisher={The Royal Society London}
}

@inproceedings{shor1994algorithms,
  title={Algorithms for quantum computation: discrete logarithms and factoring},
  author={Shor, Peter W},
  booktitle={Proceedings 35th annual symposium on foundations of computer science},
  pages={124--134},
  year={1994},
  organization={Ieee}
}

@inproceedings{grover1996fast,
  title={A fast quantum mechanical algorithm for database search},
  author={Grover, Lov K},
  booktitle={Proceedings of the twenty-eighth annual ACM symposium on Theory of computing},
  pages={212--219},
  year={1996}
}

@misc{nstc2018quantum,
  title={National Strategic Overview for Quantum Information Science},
  author={{National Science and Technology Council}},
  year={2018},
  note={Executive Office of the President of the United States}
}

@misc{eu2018flagship,
  title={Quantum Technologies Flagship},
  author={{European Commission}},
  year={2018},
  note={EU research and innovation initiative}
}

@misc{mckinsey2026quantum,
  title={Quantum Technology Monitor 2026: A commercial tipping point},
  author={{McKinsey \& Company}},
  year={2026},
  note={Industry report}
}

@article{ullah2022quantum,
  title={Quantum computing for smart grid applications},
  author={Ullah, Md Habib and Eskandarpour, Rozhin and Zheng, Honghao and Khodaei, Amin},
  journal={IET Generation, Transmission \& Distribution},
  volume={16},
  number={21},
  pages={4239--4257},
  year={2022},
  publisher={Wiley Online Library}
}

@article{gyongyosi2019survey,
  title={A survey on quantum computing technology},
  author={Gyongyosi, Laszlo and Imre, Sandor},
  journal={Computer Science Review},
  volume={31},
  pages={51--71},
  year={2019},
  publisher={Elsevier}
}

@article{shafique2024quantum,
  title={Quantum computing: Circuits, algorithms, and applications},
  author={Shafique, Muhammad Ali and Munir, Arslan and Latif, Imran},
  journal={IEEE Access},
  volume={12},
  pages={22296--22314},
  year={2024},
  publisher={IEEE}
}

@article{oukaira2025quantum,
  title={Quantum hardware devices (QHDs): opportunities and challenges},
  author={Oukaira, Aziz},
  journal={IEEE Access},
  year={2025},
  publisher={IEEE}
}

@article{assad2022smart,
  title={Smart grid, demand response and optimization: A critical review of computational methods},
  author={Assad, Ussama and Hassan, Muhammad Arshad Shehzad and Farooq, Umar and Kabir, Asif and Khan, Muhammad Zeeshan and Bukhari, S Sabahat H and Jaffri, Zain ul Abidin and Olah, Judit and Popp, Jozsef},
  journal={Energies},
  volume={15},
  number={6},
  pages={2003},
  year={2022},
  publisher={MDPI}
}

@article{patari2021distributed,
  title={Distributed optimization in distribution systems: Use cases, limitations, and research needs},
  author={Patari, Niloy and Venkataramanan, Venkatesh and Srivastava, Anurag and Molzahn, Daniel K and Li, Na and Annaswamy, Anuradha},
  journal={IEEE Transactions on Power Systems},
  volume={37},
  number={5},
  pages={3469--3481},
  year={2021},
  publisher={IEEE}
}

@article{pavon2023review,
  title={A review of modern computational techniques and their role in power system stability and control},
  author={Pavon, Wilson and Jaramillo, Manuel and Vasquez, Juan C},
  journal={Energies},
  volume={17},
  number={1},
  pages={177},
  year={2023},
  publisher={MDPI}
}

@software{Precedence_Research,
  author       = {{Precedence Research}},
  title        = {Quantum Computing Market Size, Share, and Trends 2026 to 2035},
  year         = {2026},
  url          = {https://www.precedenceresearch.com/quantum-computing-market}
}

@article{golestan2023quantum,
  title={Quantum computation in power systems: An overview of recent advances},
  author={Golestan, Saeed and Habibi, MR and Mousavi, SY Mousazadeh and Guerrero, Josep M and Vasquez, Juan C},
  journal={Energy Reports},
  volume={9},
  pages={584--596},
  year={2023},
  publisher={Elsevier}
}

@article{li2025application,
  title={Application of quantum computing for power systems},
  author={Li, Yan and Venayagamoorthy, Ganesh K and Du, Liang},
  journal={Smart Cyber-Physical Power Systems: Solutions from Emerging Technologies},
  volume={2},
  pages={313--322},
  year={2025},
  publisher={Wiley Online Library}
}

@article{ganeshamurthy2024next,
  title={Next generation power system planning and operation with quantum computation},
  author={Ganeshamurthy, Priyanka Arkalgud and Ghosh, Kumar and O'Meara, Corey and Cortiana, Giorgio and Schiefelbein-Lach, Jan and Monti, Antonello},
  journal={IEEE Access},
  volume={12},
  pages={182673--182692},
  year={2024},
  publisher={IEEE}
}

@article{morstyn2024opportunities,
  title={Opportunities for quantum computing within net-zero power system optimization},
  author={Morstyn, Thomas and Wang, Xiangyue},
  journal={Joule},
  volume={8},
  number={6},
  pages={1619--1640},
  year={2024},
  publisher={Elsevier}
}

@article{zhou2022quantum,
  title={Quantum computing in power systems},
  author={Zhou, Yifan and Tang, Zefan and Nikmehr, Nima and Babahajiani, Pouya and Feng, Fei and Wei, Tzu-Chieh and Zheng, Honghao and Zhang, Peng},
  journal={IEnergy},
  volume={1},
  number={2},
  pages={170--187},
  year={2022},
  publisher={TUP}
}

@article{eskandarpour2020quantum,
  title={Quantum-enhanced grid of the future: A primer},
  author={Eskandarpour, Rozhin and Ghosh, Kumar Jang Bahadur and Khodaei, Amin and Paaso, Aleksi and Zhang, Liuxi},
  journal={IEEE Access},
  volume={8},
  pages={188993--189002},
  year={2020},
  publisher={IEEE}
}

@article{giani2021quantum,
  title={Quantum computing opportunities in renewable energy},
  author={Giani, Annarita and Eldredge, Zachary},
  journal={SN Computer Science},
  volume={2},
  number={5},
  pages={393},
  year={2021},
  publisher={Springer}
}

@article{zhang2024review,
  title={A Review of Quantum Computing Applications in Optimal Power Flow Calculations in Power Systems},
  author={Zhang, Runhao},
  journal={Authorea Preprints},
  year={2024},
  publisher={Authorea}
}

@article{chen2025review,
  title={A review of quantum computing technologies in power system optimization},
  author={Chen, Yousu and Vu, Thanh Long},
  year={2025},
  publisher={Pacific Northwest National Laboratory (PNNL), Richland, WA (United States)}
}

@article{jang2024review,
  title={Review of applications of quantum computing in power flow calculation},
  author={Jang, Ye-Eun and Kim, Na-Yeon and Kim, Young-Jin},
  journal={Journal of Electrical Engineering \& Technology},
  volume={19},
  number={2},
  pages={877--886},
  year={2024},
  publisher={Springer}
}

@incollection{tightiz2026next,
  title={Next-gen smart grids: A quantum approach},
  author={Tightiz, Lilia and Yoo, Joon and Oh, Eunsung and Bang, Seunghyun and Kim, Shiho},
  booktitle={Advances in Computers},
  volume={140},
  pages={173--214},
  year={2026},
  publisher={Elsevier}
}

@article{strata2025quantum,
  title={Quantum Machine Learning early opportunities for the energy industry: A scoping review},
  author={Strata, Francesco and Migliori, Luca and Gebran, Nour and Guarino, Nicolina and Colombo, Giacomo Carlo and Pezzuolo, Sara and Luzietti, Emiliano},
  journal={Frontiers in Quantum Science and Technology},
  volume={4},
  pages={1653104},
  year={2025},
  publisher={Frontiers Media SA}
}

@article{corli2025quantum,
  title={Quantum machine learning algorithms for anomaly detection: A review},
  author={Corli, Sebastiano and Moro, Lorenzo and Dragoni, Daniele and Dispenza, Massimiliano and Prati, Enrico},
  journal={Future Generation Computer Systems},
  volume={166},
  pages={107632},
  year={2025},
  publisher={Elsevier}
}

@book{mermin2007quantum,
  title={Quantum computer science: an introduction},
  author={Mermin, N David},
  year={2007},
  publisher={Cambridge University Press}
}

@article{javadi2024quantum,
  title={Quantum computing with Qiskit},
  author={Javadi-Abhari, Ali and Treinish, Matthew and Krsulich, Kevin and Wood, Christopher J and Lishman, Jake and Gacon, Julien and Martiel, Simon and Nation, Paul D and Bishop, Lev S and Cross, Andrew W and others},
  journal={arXiv preprint arXiv:2405.08810},
  year={2024}
}

@article{isakov2021simulations,
  title={Simulations of quantum circuits with approximate noise using qsim and cirq},
  author={Isakov, Sergei V and Kafri, Dvir and Martin, Orion and Heidweiller, Catherine Vollgraff and Mruczkiewicz, Wojciech and Harrigan, Matthew P and Rubin, Nicholas C and Thomson, Ross and Broughton, Michael and Kissell, Kevin and others},
  journal={arXiv preprint arXiv:2111.02396},
  year={2021}
}

@article{suzuki2021qulacs,
  title={Qulacs: a fast and versatile quantum circuit simulator for research purpose},
  author={Suzuki, Yasunari and Kawase, Yoshiaki and Masumura, Yuya and Hiraga, Yuria and Nakadai, Masahiro and Chen, Jiabao and Nakanishi, Ken M and Mitarai, Kosuke and Imai, Ryosuke and Tamiya, Shiro and others},
  journal={Quantum},
  volume={5},
  pages={559},
  year={2021},
  publisher={Verein zur F{\"o}rderung des Open Access Publizierens in den Quantenwissenschaften}
}

@article{steiger2018projectq,
  title={ProjectQ: an open source software framework for quantum computing},
  author={Steiger, Damian S and H{\"a}ner, Thomas and Troyer, Matthias},
  journal={Quantum},
  volume={2},
  pages={49},
  year={2018},
  publisher={Verein zur F{\"o}rderung des Open Access Publizierens in den Quantenwissenschaften}
}

@article{lambert2026qutip,
  title={QuTiP 5: The quantum toolbox in Python},
  author={Lambert, Neill and Gigu{\`e}re, Eric and Menczel, Paul and Li, Boxi and Hopf, Patrick and Su{\'a}rez, Gerardo and Gali, Marc and Lishman, Jake and Gadhvi, Rushiraj and Agarwal, Rochisha and others},
  journal={Physics Reports},
  volume={1153},
  pages={1--62},
  year={2026},
  publisher={Elsevier}
}

@article{smith2016practical,
  title={A practical quantum instruction set architecture},
  author={Smith, Robert S and Curtis, Michael J and Zeng, William J},
  journal={arXiv preprint arXiv:1608.03355},
  year={2016}
}

@article{karalekas2020quantum,
  title={A quantum-classical cloud platform optimized for variational hybrid algorithms},
  author={Karalekas, Peter J and Tezak, Nikolas A and Peterson, Eric C and Ryan, Colm A and Da Silva, Marcus P and Smith, Robert S},
  journal={Quantum Science and Technology},
  volume={5},
  number={2},
  pages={024003},
  year={2020},
  publisher={IOP Publishing}
}

@article{bergholm2018pennylane,
  title={Pennylane: Automatic differentiation of hybrid quantum-classical computations},
  author={Bergholm, Ville and Izaac, Josh and Schuld, Maria and Gogolin, Christian and Ahmed, Shahnawaz and Ajith, Vishnu and Alam, M Sohaib and Alonso-Linaje, Guillermo and AkashNarayanan, Bharath and Asadi, Ali and others},
  journal={arXiv preprint arXiv:1811.04968},
  year={2018}
}

@article{broughton2020tensorflow,
  title={Tensorflow quantum: A software framework for quantum machine learning},
  author={Broughton, Michael and Verdon, Guillaume and McCourt, Trevor and Martinez, Antonio J and Yoo, Jae Hyeon and Isakov, Sergei V and Massey, Philip and Halavati, Ramin and Niu, Murphy Yuezhen and Zlokapa, Alexander and others},
  journal={arXiv preprint arXiv:2003.02989},
  year={2020}
}

@inproceedings{svore2018q,
  title={Q\# enabling scalable quantum computing and development with a high-level dsl},
  author={Svore, Krysta and Geller, Alan and Troyer, Matthias and Azariah, John and Granade, Christopher and Heim, Bettina and Kliuchnikov, Vadym and Mykhailova, Mariia and Paz, Andres and Roetteler, Martin},
  booktitle={Proceedings of the real world domain specific languages workshop 2018},
  pages={1--10},
  year={2018}
}

@article{jones2019quest,
  title={QuEST and high performance simulation of quantum computers},
  author={Jones, Tyson and Brown, Anna and Bush, Ian and Benjamin, Simon C},
  journal={Scientific reports},
  volume={9},
  number={1},
  pages={10736},
  year={2019},
  publisher={Nature Publishing Group UK London}
}

@article{guerreschi2020intel,
  title={Intel Quantum Simulator: A cloud-ready high-performance simulator of quantum circuits},
  author={Guerreschi, Gian Giacomo and Hogaboam, Justin and Baruffa, Fabio and Sawaya, Nicolas PD},
  journal={Quantum Science \& Technology},
  volume={5},
  number={3},
  pages={034007},
  year={2020},
  publisher={IOP Publishing}
}

@inproceedings{khammassi2017qx,
  title={QX: A high-performance quantum computer simulation platform},
  author={Khammassi, Nader and Ashraf, Imran and Fu, Xiang and Almudever, Carmen G and Bertels, Koen},
  booktitle={Design, Automation \& Test in Europe Conference \& Exhibition (DATE), 2017},
  pages={464--469},
  year={2017},
  organization={IEEE}
}

@article{luo2020yao,
  title={Yao. jl: Extensible, efficient framework for quantum algorithm design},
  author={Luo, Xiu-Zhe and Liu, Jin-Guo and Zhang, Pan and Wang, Lei},
  journal={Quantum},
  volume={4},
  pages={341},
  year={2020},
  publisher={Verein zur F{\"o}rderung des Open Access Publizierens in den Quantenwissenschaften}
}

@article{gheorghiu2018quantum++,
  title={Quantum++: A modern C++ quantum computing library},
  author={Gheorghiu, Vlad},
  journal={PloS one},
  volume={13},
  number={12},
  pages={e0208073},
  year={2018},
  publisher={Public Library of Science San Francisco, CA USA}
}

@article{kelly2018simulating,
  title={Simulating quantum computers using OpenCL},
  author={Kelly, Adam},
  journal={arXiv preprint arXiv:1805.00988},
  year={2018}
}

@article{killoran2019strawberry,
  title={Strawberry fields: A software platform for photonic quantum computing},
  author={Killoran, Nathan and Izaac, Josh and Quesada, Nicol{\'a}s and Bergholm, Ville and Amy, Matthew and Weedbrook, Christian},
  journal={Quantum},
  volume={3},
  pages={129},
  year={2019},
  publisher={Verein zur F{\"o}rderung des Open Access Publizierens in den Quantenwissenschaften}
}

@article{kottmann2021tequila,
  title={Tequila: A platform for rapid development of quantum algorithms},
  author={Kottmann, Jakob S and Alperin-Lea, Sumner and Tamayo-Mendoza, Teresa and Cervera-Lierta, Alba and Lavigne, Cyrille and Yen, Tzu-Ching and Verteletskyi, Vladyslav and Schleich, Philipp and Anand, Abhinav and Degroote, Matthias and others},
  journal={Quantum Science \& Technology},
  volume={6},
  number={2},
  pages={024009},
  year={2021},
  publisher={IOP Publishing}
}

@article{efthymiou2022qibo,
  title={Qibo: a framework for quantum simulation with hardware acceleration},
  author={Efthymiou, Stavros and Ramos-Calderer, Sergi and Bravo-Prieto, Carlos and P{\'e}rez-Salinas, Adri{\'a}n and Garc{\'\i}a-Mart{\'\i}n, Diego and Garcia-Saez, Artur and Latorre, Jos{\'e} Ignacio and Carrazza, Stefano},
  journal={Quantum Science \& Technology},
  volume={7},
  number={1},
  pages={015018},
  year={2022},
  publisher={IOP Publishing}
}

@article{lyakh2022exatn,
  title={ExaTN: Scalable GPU-accelerated high-performance processing of general tensor networks at exascale},
  author={Lyakh, Dmitry I and Nguyen, Thien and Claudino, Daniel and Dumitrescu, Eugene and McCaskey, Alexander J},
  journal={Frontiers in Applied Mathematics and Statistics},
  volume={8},
  pages={838601},
  year={2022},
  publisher={Frontiers Media SA}
}

@article{fishman2022itensor,
  title={The ITensor software library for tensor network calculations},
  author={Fishman, Matthew and White, Steven and Stoudenmire, Edwin Miles},
  journal={SciPost Physics Codebases},
  pages={004},
  year={2022}
}

@article{villalonga2019flexible,
  title={A flexible high-performance simulator for verifying and benchmarking quantum circuits implemented on real hardware},
  author={Villalonga, Benjamin and Boixo, Sergio and Nelson, Bron and Henze, Christopher and Rieffel, Eleanor and Biswas, Rupak and Mandr{\`a}, Salvatore},
  journal={npj Quantum Information},
  volume={5},
  number={1},
  pages={86},
  year={2019},
  publisher={Nature Publishing Group UK London}
}

@article{gray2018quimb,
  title={quimb: A python package for quantum information and many-body calculations},
  author={Gray, Johnnie},
  journal={Journal of Open Source Software},
  volume={3},
  number={29},
  pages={819},
  year={2018}
}

@software{cudaq,
  author       = {{NVIDIA Corporation}},
  title        = {{CUDA-Q}},
  year         = {2025},
  url          = {https://nvidia.github.io/cuda-quantum/latest/index.html},
  publisher    = {NVIDIA Corporation}
}

@inproceedings{strano2023exact,
  title={Exact and approximate simulation of large quantum circuits on a single GPU},
  author={Strano, Daniel and Bollay, Benn and Blaauw, Aryan and Shammah, Nathan and Zeng, William J and Mari, Andrea},
  booktitle={2023 IEEE International Conference on Quantum Computing and Engineering (QCE)},
  volume={1},
  pages={949--958},
  year={2023},
  organization={IEEE}
}

@inproceedings{wang2022quantumnas,
  title={Quantumnas: Noise-adaptive search for robust quantum circuits},
  author={Wang, Hanrui and Ding, Yongshan and Gu, Jiaqi and Lin, Yujun and Pan, David Z and Chong, Frederic T and Han, Song},
  booktitle={2022 IEEE International Symposium on High-Performance Computer Architecture (HPCA)},
  pages={692--708},
  year={2022},
  organization={IEEE}
}

@article{carleo2019netket,
  title={NetKet: A machine learning toolkit for many-body quantum systems},
  author={Carleo, Giuseppe and Choo, Kenny and Hofmann, Damian and Smith, James ET and Westerhout, Tom and Alet, Fabien and Davis, Emily J and Efthymiou, Stavros and Glasser, Ivan and Lin, Sheng-Hsuan and others},
  journal={SoftwareX},
  volume={10},
  pages={100311},
  year={2019},
  publisher={Elsevier}
}

@article{seitz2025tool,
  title={Tool: Qadence: A Differentiable Interface for Digital and Analog Programs},
  author={Seitz, Dominik and Heim, Niklas and Moutinho, Jo{\~a}o P and Guichard, Roland and Abramavicius, Vytautas and Wennersteen, Aleksander and Both, Gert-Jan and Quelle, Anton and de Groot, Caroline and Velikova, Gergana V and others},
  journal={IEEE Software},
  volume={42},
  number={6},
  pages={44--54},
  year={2025},
  publisher={IEEE}
}

@article{seidel2024qrisp,
  title={Qrisp: A framework for compilable high-level programming of gate-based quantum computers},
  author={Seidel, Raphael and Bock, Sebastian and Zander, Ren{\'e} and Petri{\v{c}}, Matic and Steinmann, Niklas and Tcholtchev, Nikolay and Hauswirth, Manfred},
  journal={arXiv preprint arXiv:2406.14792},
  year={2024}
}

@article{amy2020staq,
  title={staq—a full-stack quantum processing toolkit},
  author={Amy, Matthew and Gheorghiu, Vlad},
  journal={Quantum Science \& Technology},
  volume={5},
  number={3},
  pages={034016},
  year={2020},
  publisher={IOP Publishing}
}

@article{kramer2018quantumoptics,
  title={QuantumOptics. jl: A Julia framework for simulating open quantum systems},
  author={Kr{\"a}mer, Sebastian and Plankensteiner, David and Ostermann, Laurin and Ritsch, Helmut},
  journal={Computer Physics Communications},
  volume={227},
  pages={109--116},
  year={2018},
  publisher={Elsevier}
}

@article{gawron2018quantuminformation,
  title={QuantumInformation. jl—a julia package for numerical computation in quantum information theory},
  author={Gawron, Piotr and Kurzyk, Dariusz and Pawela, {\L}ukasz},
  journal={PLoS One},
  volume={13},
  number={12},
  pages={e0209358},
  year={2018},
  publisher={Public Library of Science San Francisco, CA USA}
}

@inproceedings{lykov2021performance,
  title={Performance evaluation and acceleration of the QTensor quantum circuit simulator on GPUs},
  author={Lykov, Danylo and Chen, Angela and Chen, Huaxuan and Keipert, Kristopher and Zhang, Zheng and Gibbs, Tom and Alexeev, Yuri},
  booktitle={2021 IEEE/ACM Second International Workshop on Quantum Computing Software (QCS)},
  pages={27--34},
  year={2021},
  organization={IEEE}
}

@misc{qtensorai_github,
  author       = {Minzhao Liu},
  title        = {{QTensorAI}},
  year         = {2026},
  howpublished = {Available: \url{https://github.com/sss441803/QTensorAI}},
  note         = {{A}ccessed June 13, 2026}
}

@software{dwave_ocean_sdk,
  author  = {{D-Wave Systems Inc.}},
  title   = {Ocean {SDK}},
  year    = {2026},
  url     = {https://docs.dwavequantum.com/en/latest/ocean/},
  note    = {{A}ccessed: 2026-06-13}
}

@software{Strangeworks,
  author  = {{Strangeworks}},
  title   = {Strangeworks Documentations},
  url     = {https://docs.strangeworks.com/},
  note    = {{A}ccessed: 2026-06-13}
}

@software{Hill_qBraid-SDK_Platform-agnostic_quantum_2026,
author = {Hill, Ryan James and Gupta, Harshit and Young, Ricky and Setia, Kanav},
doi = {10.5281/zenodo.12627596},
license = {Apache-2.0},
month = may,
title = {{qBraid-SDK: Platform-agnostic quantum runtime framework.}},
url = {https://github.com/qBraid/qBraid},
version = {0.12.1},
year = {2026}
}

@misc{braket2019,
      title  = {{Amazon Braket: a fully managed quantum computing service provided by AWS}},
      author = {{Braket Developers}},
      year   = {2019},
      url    = {https://aws.amazon.com/braket/}, 
}

@misc{Azure_Quantum,
      title  = {{Azure Quantum}},
      author = {{Microsoft}},
      url    = {https://azure.microsoft.com/en-us/solutions/quantum-computing}, 
      note    = {{A}ccessed: 2026-06-13}
}

@misc{Orquestra_Zapata,
      title  = {{Orquestra}},
      author = {{Zapata Quantum}},
      url    = {https://zapataquantum.com/solutions\#platform}, 
      note    = {{A}ccessed: 2026-06-13}
}

@misc{Superstaq_Infleqtion,
      title  = {{Superstaq Documentation}},
      author = {{Infleqtion}},
      url    = {https://superstaq.readthedocs.io/en/latest/resources/contact.html}, 
      note    = {{A}ccessed: 2026-06-13}
}

@misc{Q-CTRL_Fire_Opal,
      title  = {{Q-CTRL Fire Opal}},
      author = {{Q-CTRL}},
      url    = {https://q-ctrl.com/fire-opal}, 
      note    = {{A}ccessed: 2026-06-13}
}

@misc{pennylane_website,
  author       = {{PennyLane AI}},
  title        = {PennyLane: Quantum Programming Software},
  year         = {2026},
  howpublished = {\url{https://pennylane.ai/}},
  note         = {Accessed: 2026-06-13}
}

@misc{classiq_platform,
  author       = {{Classiq Technologies Ltd.}},
  title        = {Classiq Quantum Development Platform},
  year         = {2026},
  howpublished = {\url{https://classiq.io/}},
  note         = {Accessed: 2026-06-13}
}

@article{farhi2014quantum,
  title={A quantum approximate optimization algorithm},
  author={Farhi, Edward and Goldstone, Jeffrey and Gutmann, Sam},
  journal={arXiv preprint arXiv:1411.4028},
  year={2014}
}

@article{zhou2020quantum,
  title={Quantum approximate optimization algorithm: Performance, mechanism, and implementation on near-term devices},
  author={Zhou, Leo and Wang, Sheng-Tao and Choi, Soonwon and Pichler, Hannes and Lukin, Mikhail D},
  journal={Physical Review X},
  volume={10},
  number={2},
  pages={021067},
  year={2020},
  publisher={APS}
}

@article{pellow2024effect,
  title={The effect of classical optimizers and Ansatz depth on QAOA performance in noisy devices},
  author={Pellow-Jarman, Aidan and McFarthing, Shane and Sinayskiy, Ilya and Park, Daniel K and Pillay, Anban and Petruccione, Francesco},
  journal={Scientific reports},
  volume={14},
  number={1},
  pages={16011},
  year={2024},
  publisher={Nature Publishing Group UK London}
}

@inproceedings{kashapogu2024exploring,
  title={Exploring the versatility of QAOA: A comprehensive review},
  author={Kashapogu, Rajavardhan and Hasib, Saima and Rasool, Akhtar},
  booktitle={2024 15th International Conference on Computing Communication and Networking Technologies (ICCCNT)},
  pages={1--8},
  year={2024},
  organization={IEEE}
}

@article{crooks2020gates,
  title={Gates, states, and circuits},
  author={Crooks, Gavin E},
  journal={Gates states and circuits},
  year={2020}
}

@article{jiang2026qaoa,
  title={{QAOA}-Driven PMU placement optimization with graph learning-based parameter initialization refinement},
  author={Jiang, Yuqi and Wang, Xiangyue and Liang, Zhiding and Li, Yan and Morstyn, Thomas and Lopes, Pedro LS and Brandsema, Matthew J},
  journal={Quantum Machine Intelligence},
  volume={8},
  number={1},
  pages={32},
  year={2026},
  publisher={Springer}
}

@inproceedings{jiang2025accelerating,
  title={Accelerating Quantum Optimization with Graph Learning for Optimal PMU Placement},
  author={Jiang, Yuqi and Wang, Xiangyue and Liang, Zhiding and Li, Yan and Morstyn, Thomas and Du, Liang},
  booktitle={2025 IEEE/AIAA Transportation Electrification Conference and Electric Aircraft Technologies Symposium (ITEC+ EATS)},
  pages={1--5},
  year={2025},
  organization={IEEE}
}

@article{jiang2025optimal,
  title={Optimal {PMU} Placement via Quantum Optimization},
  author={Jiang, Yuqi and Liang, Zhiding and Li, Yan and Morstyn, Thomas},
  journal={IEEE Transactions on Smart Grid},
  volume={16},
  number={4},
  pages={3125--3141},
  year={2025},
  publisher={IEEE}
}

@inproceedings{ganeshamurthy2025quantum,
  title={Quantum Annealing for Strategic Meter Placement in Power Distribution Networks},
  author={Ganeshamurthy, Priyanka Arkalgud and Ponci, Ferdinanda and Monti, Antonello},
  booktitle={2025 International Conference on Quantum Communications, Networking, and Computing (QCNC)},
  pages={532--538},
  year={2025},
  organization={IEEE}
}

@inproceedings{jones2020computational,
  title={On the computational viability of quantum optimization for PMU placement},
  author={Jones, Eric B and Kapit, Eliot and Chang, Chin-Yao and Biagioni, David and Vaidhynathan, Deepthi and Graf, Peter and Jones, Wesley},
  booktitle={2020 IEEE Power \& Energy Society General Meeting (PESGM)},
  pages={1--5},
  year={2020},
  organization={IEEE}
}

@article{yin2025optimised,
  title={Optimised battery placement in distribution grids using quantum and quantum inspired QUBO solvers},
  author={Yin, Yingyi and Gentile, Fabio and Cortines, Aser and Mugel, Sam and Or{\'u}s, Rom{\'a}n and Asensio, Miguel Rodr{\'\i}guez and Castro, Cristina Barrag{\'a}n and Vel{\'a}zquez, Carlos S{\'a}nchez},
  journal={Authorea Preprints},
  year={2025},
  publisher={Authorea}
}

@article{hasan2026quantum,
  title={A quantum-classical hybrid framework for optimal energy storage systems planning},
  author={Hasan, Md Shamim and Aboumrad, Willie and Marthi, Phani RV and Epifanovsky, Evgeny and Siopsis, George and Roetteler, Martin and Debnath, Suman},
  journal={Scientific Reports},
  year={2026},
  publisher={Nature Publishing Group UK London}
}

@article{ajagekar2019quantum,
  title={Quantum computing for energy systems optimization: Challenges and opportunities},
  author={Ajagekar, Akshay and You, Fengqi},
  journal={Energy},
  volume={179},
  pages={76--89},
  year={2019},
  publisher={Elsevier}
}

@techreport{hidary2020application,
  title={The application of QAOA on a cloud-based quantum computer for clean energy grid optimization},
  author={Hidary, David R and Libenson, Samuel},
  year={2020},
  institution={Tech. Rep., 2020.[Online]. Available: https://www. academia. edu/44397754~…}
}

@inproceedings{chandra2022towards,
  title={Towards an optimal hybrid algorithm for ev charging stations placement using quantum annealing and genetic algorithms},
  author={Chandra, Aman and Lalwani, Jitesh and Jajodia, Babita},
  booktitle={2022 International Conference on Trends in Quantum Computing and Emerging Business Technologies (TQCEBT)},
  pages={1--6},
  year={2022},
  organization={IEEE}
}

@inproceedings{veshchezerova2023hybrid,
  title={A hybrid quantum-classical approach to the electric mobility problem},
  author={Veshchezerova, Margarita and Somov, Mikhail and Bertsche, David and Limmer, Steffen and Schmitt, Sebastian and Perelshtein, Michael and Tripathi, Ayush Joshi},
  booktitle={2023 IEEE International Conference on Quantum Computing and Engineering (QCE)},
  volume={1},
  pages={636--641},
  year={2023},
  organization={IEEE}
}

@article{rao2023hybrid,
  title={Hybrid quantum-classical solution for electric vehicle charger placement problem.},
  author={Rao, Poojith U and Sodhi, Balwinder},
  journal={Soft Computing-A Fusion of Foundations, Methodologies \& Applications},
  volume={27},
  number={18},
  year={2023}
}

@article{sakib2024quantum,
  title={Quantum-Powered Optimization for Electric Vehicle Charging Infrastructure Deployment},
  author={Sakib, Nazmush and Chen, Xin},
  journal={arXiv preprint arXiv:2411.06684},
  year={2024}
}

@article{radvand2024quantum,
  title={A Quantum Optimization Algorithm for Optimal Electric Vehicle Charging Station Placement for Intercity Trips},
  author={Radvand, Tina and Talebpour, Alireza and Khosravian, Homa},
  journal={arXiv preprint arXiv:2410.16231},
  year={2024}
}

@article{ou2025solving,
  title={Solving optimal electric vehicle charging station placement problem using digital quantum annealing},
  author={Ou, Chia-Ho and Cheng, Chung-Chieh and Chen, Chih-Yu and Bhumkittipich, Krischonme and Romphochai, Sillawat},
  journal={Journal of Communications and Networks},
  volume={27},
  number={4},
  pages={252--263},
  year={2025},
  publisher={KICS}
}

@inproceedings{lange2026contingency,
  title={Contingency Analysis via QAOA in the NISQ Era},
  author={Lange, Joseph Maxwell and Li, Yan},
  booktitle={2026 IEEE Transportation Electrification Conference \& Expo (ITEC) \& Electric Aircraft Technologies Symposium (EATS)(ITEC+ EATS)},
  pages={1--5},
  year={2026},
  organization={IEEE}
}

@inproceedings{peter2025quantum,
  title={Quantum-Enhanced Reinforcement Learning for Power Grid Security Assessment},
  author={Peter, Benjamin M and Korkali, Mert},
  booktitle={2025 57th North American Power Symposium (NAPS)},
  pages={1--6},
  year={2025},
  organization={IEEE}
}

@article{feng2025quantum,
  title={Quantum contingency analysis for power system steady-state security identification},
  author={Feng, Fei and Zhou, Yifan and Bragin, Mikhail A and Shamash, Yacov A and Zhang, Peng},
  journal={Scientific Reports},
  volume={15},
  number={1},
  pages={15148},
  year={2025},
  publisher={Nature Publishing Group UK London}
}

@article{cremer2025n,
  title={N-$ k $ Security Assessment with Quantum Annealing},
  author={Cremer, Jochen Lorenz},
  journal={IEEE Transactions on Power Systems},
  volume={41},
  number={2},
  year={2026},
  pages={1484 -- 1497},
  publisher={IEEE}
}

@article{eskandarpour2020quantum_security,
  title={Quantum computing for enhancing grid security},
  author={Eskandarpour, Rozhin and Gokhale, Pranav and Khodaei, Amin and Chong, Frederic T and Passo, Aleksi and Bahramirad, Shay},
  journal={IEEE Transactions on Power Systems},
  volume={35},
  number={2},
  pages={4135--4137},
  year={2020},
  publisher={IEEE}
}

@inproceedings{jiang2025quantum,
  title={Quantum Computing-Enabled Contingency Analysis for Power Systems},
  author={Jiang, Yuqi and Liang, Zhiding and Li, Yan and Guan, Qiang and Venayagamoorthy, Ganesh Kumar},
  booktitle={2025 IEEE International Conference on Quantum Computing and Engineering (QCE)},
  volume={1},
  pages={1938--1944},
  year={2025},
  organization={IEEE}
}

@mastersthesis{antoli2023quantum,
  title={Quantum computations for N-1 secure power systems},
  author={Antol{\'\i} Gil, Eduard},
  year={2023},
  school={Universitat Polit{\`e}cnica de Catalunya}
}

@article{neumann2024quantum,
  title={Quantum algorithms for N-1 security in power grids},
  author={Neumann, Niels MP and van der Linde, Stan and de Kok, Willem and Leijnse, Koen and Boschero, Juan and Aguilera, Esteban and Berg, Peter Elias-van den and Koppen, Vincent and Jaspers, Nikki and Zwetsloot, Jelte},
  journal={arXiv preprint arXiv:2405.00434},
  year={2024}
}

@article{feng2022quantum,
  title={Quantum microgrid state estimation},
  author={Feng, Fei and Zhang, Peng and Zhou, Yifan and Tang, Zefan},
  journal={Electric Power Systems Research},
  volume={212},
  pages={108386},
  year={2022},
  publisher={Elsevier}
}

@inproceedings{tran2025noise,
  title={Noise-Aware Quantum Annealing for State Estimation in Power Systems},
  author={Tran, Quoc-Bao and Dinh, Thang N},
  booktitle={2025 IEEE International Conference on Quantum Computing and Engineering (QCE)},
  volume={1},
  pages={1945--1954},
  year={2025},
  organization={IEEE}
}

@article{feng2024noisy,
  title={Noisy-intermediate-scale quantum power system state estimation},
  author={Feng, Fei and Zhang, Peng and Zhou, Yifan and Shamash, Yacov A},
  journal={iEnergy},
  volume={3},
  number={3},
  pages={135--141},
  year={2024},
  publisher={TUP}
}

@article{Stoyanova2025QuantumCM,
  title={Quantum Computing Methods for Dynamic State Estimation in Power Systems},
  author={Ivelina Stoyanova and Orkun Şensebat and Yanjun Ji and Priyanka Arkalgud Ganeshamurthy and Sonja Kajganic and Dennis Willsch and Antonello Monti},
  journal={2025 IEEE PES Innovative Smart Grid Technologies Conference Europe (ISGT Europe)},
  year={2025},
  pages={1-5},
  organization={IEEE}
}

@inproceedings{addo2025deep,
  title={Deep Reinforcement Learning With Quantum Enhancements For Power System Inertia Estimation},
  author={Addo, Kwabena and Moloi, Katleho and Kabeya, Musasa and Ojo, Evans Eshiemogie},
  booktitle={2025 33rd Southern African Universities Power Engineering Conference (SAUPEC)},
  pages={1--6},
  year={2025},
  organization={IEEE}
}

@article{NEW_State_Estimation,
author = {Huang, Shyh-Jier and Yang, Po-Wei and
Yen, Shen-Kun and 
Li, Yong-Tai},
title = {Enhancing Distributed State Estimation of Power Grid With a Simplified Quantum Algorithm},
journal = {IET Generation, Transmission \& Distribution},
pages = {1-1},
year={2026},
publisher={Wiley Online Library}
}

@article{masoumi2025quantum,
  title={Quantum-Embedded Dynamic Security Control using Hybrid Deep Reinforcement Learning},
  author={Masoumi, Amin and Korkali, Mert},
  journal={arXiv preprint arXiv:2512.04095},
  year={2025}
}

@article{ren2024enhancing,
  title={Enhancing dynamic security assessment in smart grids through quantum federated learning},
  author={Ren, Chao and Dong, Zhao Yang and Skoglund, Mikael and Gao, Yulan and Wang, Tianjing and Zhang, Rui},
  journal={IEEE Transactions on Automation Science and Engineering},
  volume={23},
  pages={3255--3267},
  year={2026},
  publisher={IEEE}
}

@article{zhou2022noise,
  title={Noise-resilient quantum machine learning for stability assessment of power systems},
  author={Zhou, Yifan and Zhang, Peng},
  journal={IEEE Transactions on Power Systems},
  volume={38},
  number={1},
  pages={475--487},
  year={2023},
  publisher={IEEE}
}

@inproceedings{li2023transient,
  title={Transient Stability Assessment of Islanded Microgrids Through Quantum Computing},
  author={Li, Yan and Du, Liang},
  booktitle={2023 IEEE Transportation Electrification Conference \& Expo (ITEC)},
  pages={1--5},
  year={2023},
  organization={IEEE}
}

@inproceedings{chen2024quantum,
  title={Quantum-Accelerated Transient Stability Assessment for Power Systems},
  author={Chen, Jianing and Li, Yan},
  booktitle={2024 IEEE Computer Society Annual Symposium on VLSI (ISVLSI)},
  pages={593--594},
  year={2024},
  organization={IEEE}
}

@inproceedings{yu2024quantum,
  title={Quantum adversarial machine learning for robust power system stability assessment},
  author={Yu, Sijia and Zhou, Yifan},
  booktitle={2024 IEEE Power \& Energy Society General Meeting (PESGM)},
  pages={1--5},
  year={2024},
  organization={IEEE}
}

@inproceedings{sabadra2024quantum,
  title={Quantum Kernel Based Transient Stability Assessment of Power Systems and Its Implementation in NISQ Environment},
  author={Sabadra, Trisha and Yu, Sijia and Zhou, Yifan},
  booktitle={2024 IEEE International Conference on Big Data (BigData)},
  pages={7402--7406},
  year={2024},
  organization={IEEE}
}

@inproceedings{yu2024quantum_2,
  title={Quantum-Enabled Distributed Transient Stability Assessment of Power Systems},
  author={Yu, Sijia and Zhou, Yifan and Wang, Lizhi},
  booktitle={2024 IEEE International Conference on Quantum Computing and Engineering (QCE)},
  volume={1},
  pages={593--599},
  year={2024},
  organization={IEEE}
}

@article{li2026qstaformer,
  title={QSTAformer: A quantum-enhanced Transformer for robust short-term voltage stability assessment against adversarial attacks},
  author={Li, Yang and Ma, Chong and Li, Yuanzheng and Li, Sen and Chen, Yanbo and Dong, Zhaoyang},
  journal={Applied Energy},
  volume={405},
  pages={127196},
  year={2026},
  publisher={Elsevier}
}

@inproceedings{ahmad2025qm,
  title={QM-OCL: Quantum Many-Shot Out-Of-Context Learning for Metamodeling Techno-Economic Stability in Microgrids and Smart Grids},
  author={Ahmad, Syed Farhan and Gillani, Hassan and Billah, Al Raji and Wesolowski, Adam and Byrd, Gregory T and Namoun, Abdallah},
  booktitle={2025 IEEE PES Conference on Innovative Smart Grid Technologies-Middle East (ISGT Middle East)},
  pages={1--5},
  year={2025},
  organization={IEEE}
}

@article{ren2024qfdsa,
  title={{QFDSA}: A quantum-secured federated learning system for smart grid dynamic security assessment},
  author={Ren, Chao and Yan, Rudai and Xu, Minrui and Yu, Han and Xu, Yan and Niyato, Dusit and Dong, Zhao Yang},
  journal={IEEE Internet of Things Journal},
  volume={11},
  number={5},
  pages={8414--8426},
  year={2024},
  publisher={IEEE}
}

@inproceedings{yu2025quantum,
  title={Quantum Federated Learning Based Power System Stability Assessment: A Noise Robust Approach},
  author={Yu, Sijia and Zhou, Yifan},
  booktitle={2025 57th North American Power Symposium (NAPS)},
  pages={1--6},
  year={2025},
  organization={IEEE}
}

@inproceedings{nikmehr2022quantum,
  title={Quantum distribution system reliability assessment},
  author={Nikmehr, Nima and Zhang, Peng},
  booktitle={2022 IEEE Power \& Energy Society General Meeting (PESGM)},
  pages={1--5},
  year={2022},
  organization={IEEE}
}

@article{carrascal2023bayesian,
  title={A Bayesian-network-based quantum procedure for failure risk analysis},
  author={Carrascal, Gines and Botella, Guillermo and del Barrio, Alberto and Kremer, David},
  journal={EPJ Quantum Technology},
  volume={10},
  number={1},
  pages={13},
  year={2023},
  publisher={Springer Berlin Heidelberg}
}

@article{jong2023quantum,
  title={Quantum amplitude estimation for probabilistic methods in power systems},
  author={Jong, Emilie and S{\ae}varsson, Brynjar and J{\'o}hannsson, Hj{\"o}rtur and Chatzivasileiadis, Spyros},
  journal={arXiv preprint arXiv:2309.17299},
  year={2023}
}

@inproceedings{silva2023quantum,
  title={Quantum-enhanced reliability assessment of power networks in response to wildfire events},
  author={Silva, Gabriel San Mart{\'\i}n and Parhizkar, Tarannom and Nguyen, Hieu T and Droguett, Enrique L{\'o}pez},
  booktitle={2023 Annual Reliability and Maintainability Symposium (RAMS)},
  pages={1--7},
  year={2023},
  organization={IEEE}
}

@article{nikmehr2023quantum,
  title={Quantum-inspired power system reliability assessment},
  author={Nikmehr, Nima and Zhang, Peng},
  journal={IEEE Transactions on Power Systems},
  volume={38},
  number={4},
  pages={3476--3490},
  year={2023},
  publisher={IEEE}
}

@inproceedings{thelasingha2024energy,
  title={Energy Infrastructure Risk Modeling using Quantum and Classical Machine Learning},
  author={Thelasingha, Neelanga and Munasinghe, Thilanka},
  booktitle={2024 IEEE International Conference on Big Data (BigData)},
  pages={4611--4620},
  year={2024},
  organization={IEEE}
}

@inproceedings{hosseini2024modern,
  title={Modern power system risk assessment using quantum computing},
  author={Hosseini, Seyed Amir and Monti, Antonello and Peyghami, Saeed},
  booktitle={2024 18th International Conference on Probabilistic Methods Applied to Power Systems (PMAPS)},
  pages={1--6},
  year={2024},
  organization={IEEE}
}

@article{yang2024power,
  title={Power system reliability assessment technique and modeling approach based on quantum computing theory},
  author={Yang, Hejun and Liu, Yue and Yue, Yangxu and Zhang, Dabo and Ma, Yinghao},
  journal={Electric Power Systems Research},
  volume={236},
  pages={110957},
  year={2024},
  publisher={Elsevier}
}

@article{alvarez2025quantum,
  title={Quantum Computing for Reliability Assessment of SCADA with Continuous Penetration Testing and High Penetration of Cyber-Physical Attacks},
  author={Alvarez-Alvarado, Manuel S},
  journal={IEEE Transactions on Smart Grid},
  volume={16},
  number={6},
  pages={5392-- 5403},
  year={2025},
  publisher={IEEE}
}

@inproceedings{bara2025quantum,
  title={Quantum Circuit Design for Failure Risk Analysis: An Application to Power Systems in Autonomous Ships},
  author={Bara, Sonia and Dhakad, Gunjana and Salfale, Siddhesh and TK, Abhinav Krishnan and Hazra, Indranil},
  booktitle={International Mechanical Engineering Congress and Exposition-India},
  volume={89152},
  pages={V003T06A008},
  year={2025},
  organization={American Society of Mechanical Engineers}
}

@article{saevarsson2025stochastic,
  title={Stochastic quantum power flow for risk assessment in power systems},
  author={S{\ae}varsson, Brynjar and J{\'o}hannsson, Hj{\"o}rtur and Chatzivasileiadis, Spyros},
  journal={Electric Power Systems Research},
  volume={241},
  pages={111409},
  year={2025},
  publisher={Elsevier}
}

@article{alvarez2026quantum,
  title={Quantum computing for power system reliability assessment with distributed energy resources},
  author={Alvarez-Alvarado, Manuel S and Ramirez-Prado, Maria J and Pico, Nabih and Montero, Estrella and Rodr{\'\i}guez-Gallegos, Carlos D and Velasquez, Washington},
  journal={Electric Power Systems Research},
  volume={259},
  pages={113245},
  year={2026},
  publisher={Elsevier}
}

@article{fu2023coordinated,
  title={Coordinated post-disaster restoration for resilient urban distribution systems: A hybrid quantum-classical approach},
  author={Fu, Wei and Xie, Haipeng and Zhu, Hao and Wang, Hefeng and Jiang, Lizhou and Chen, Chen and Bie, Zhaohong},
  journal={Energy},
  volume={284},
  pages={129314},
  year={2023},
  publisher={Elsevier}
}

@inproceedings{nikmehr2023quantum_restoration,
  title={Quantum annealing for distribution system restoration via resilient microgrids formation},
  author={Nikmehr, Nima and Zhang, Peng and Zheng, Honghao and Shamash, Yacov A},
  booktitle={2023 IEEE Power \& Energy Society General Meeting (PESGM)},
  pages={1--5},
  year={2023},
  organization={IEEE}
}

@inproceedings{ngo2024quantum,
  title={Quantum combinatorial optimization algorithms for network reconfiguration: Qrao vs. qaoa},
  author={Ngo, Anh Phuong and Nguyen, Hieu T},
  booktitle={2024 56th North American Power Symposium (NAPS)},
  pages={1--6},
  year={2024},
  organization={IEEE}
}

@inproceedings{fu2024quantum,
  title={Quantum-inspired resilient restoration approach considering uav-assisted fault location and network reconfiguration},
  author={Fu, Wei and Xie, Haipeng and Wang, Hefeng and Chen, Chen},
  booktitle={2024 IEEE Power \& Energy Society General Meeting (PESGM)},
  pages={1--5},
  year={2024},
  organization={IEEE}
}

@article{lin2025distributed,
  title={Distributed Quantum Computing for Fast Restoration of Power Distribution Systems with Grid-Forming {IBR}s},
  author={Lin, Chaofan and Zhang, Peng and Shamash, Yacov A},
  journal={IEEE Transactions on Industry Applications},
  year={2026},
  volume={62},
  number={2},
  pages={3623--3634},
  publisher={IEEE}
}

@article{nikmehr2024quantum,
  title={Quantum annealing-infused microgrids formation: Distribution system restoration and resilience enhancement},
  author={Nikmehr, Nima and Zhang, Peng and Zheng, Honghao and Wei, Tzu-Chieh and He, Gang and Shamash, Yacov A},
  journal={IEEE Transactions on Power Systems},
  volume={40},
  number={1},
  pages={463--475},
  year={2025},
  publisher={IEEE}
}

@article{shao2025quantum,
  title={Quantum-enabled topological optimization of distributed energy storage for resilient black-start operations},
  author={Shao, Yinchi and Gong, Yu and Wan, Xiaoyu and Huang, Xianmiao and Luo, Shanna and Cao, Yuntao and Zhang, Tao},
  journal={Scientific Reports},
  volume={15},
  number={1},
  pages={18034},
  year={2025},
  publisher={Nature Publishing Group UK London}
}

@article{fu2026quantum,
  title={Quantum-accelerated post-event restoration through quantum surrogate absolute-value Lagrangian relaxation},
  author={Fu, Wei and Xie, Haipeng and Xin, Yu},
  journal={Reliability Engineering \& System Safety},
  pages={111656},
  year={2026},
  publisher={Elsevier}
}

@article{bucher2024evaluating,
  title={Evaluating quantum optimization for dynamic self-reliant community detection},
  author={Bucher, David and Porawski, Daniel and Wimmer, Benedikt and N{\"u}{\ss}lein, Jonas and O’Meara, Corey and Mohseni, Naeimeh and Cortiana, Giorgio and Linnhoff-Popien, Claudia},
  journal={IEEE Transactions on Smart Grid},
  volume={40},
  number={1},
  pages={463--475},
  year={2025},
  publisher={IEEE}
}

@inproceedings{sanjani2024quantum,
  title={A quantum cooperative game approach to resilience-oriented microgrids operation},
  author={Sanjani, Khezr and Zhang, Peng and Nikmehr, Nima and Shamash, Yacov A},
  booktitle={2024 IEEE International Conference on Quantum Computing and Engineering (QCE)},
  volume={1},
  pages={719--725},
  year={2024},
  organization={IEEE}
}

@article{zhao2025enhancing,
  title={Enhancing resilience in Los Angeles energy systems},
  author={Zhao, Alexis Pengfei and Gu, Chenghong and Li, Shuangqi and Ju, Xin and Alhazmi, Mohannad},
  journal={Energy Reports},
  volume={13},
  pages={4647--4660},
  year={2025},
  publisher={Elsevier}
}

@article{zhou5292390scalable,
  title={Scalable Quantum-Assisted Optimization Framework for Three-Stage Resilience Improvement in Distribution Networks},
  author={Zhou, Jingxian and Zhu, Ziqing and Hu, Chenxi and Bu, Siqi},
  journal={Available at SSRN 5292390}
}

@article{xie2025quantum,
  title={Quantum-Assisted Resilience Enhancement for Distribution Systems With Networked Microgrids Considering Full-Potential Failure Risks},
  author={Xie, Haipeng and Fu, Wei},
  journal={IET Generation, Transmission \& Distribution},
  volume={19},
  number={1},
  pages={e70158},
  year={2025},
  publisher={Wiley Online Library}
}

@article{salman2025quircy,
  title={QUIRCY: QUantum Integrated ResilienCY for Power Systems},
  author={Salman, Umar T and Wang, Zongjie and Hansen, Timothy M},
  journal={IEEE Access},
  year={2025},
  publisher={IEEE}
}

@article{lin2025reforming,
  title={Reforming Quantum Microgrid Formation},
  author={Lin, Chaofan and Zhang, Peng and Bragin, Mikhail A and Shamash, Yacov A},
  journal={IEEE Transactions on Power Systems},
  volume={40},
  number={2},
  pages={1977--1980},
  year={2025},
  publisher={IEEE}
}

@inproceedings{jing2024integrating,
  title={Integrating Quantum Computing into Optimal Control for Optimality and Stability in Microgrids},
  author={Jing, Hang and Li, Yan},
  booktitle={2024 IEEE Power \& Energy Society General Meeting (PESGM)},
  pages={1--5},
  year={2024},
  organization={IEEE}
}

@inproceedings{ngo2022evaluate,
  title={Evaluate quantum combinatorial optimization for distribution network reconfiguration},
  author={Ngo, Anh Phuong and Thomas, Christan and Nguyen, Hieu and Eroglu, Abdullah and Oikonomou, Konstantinos},
  booktitle={2022 North American Power Symposium (NAPS)},
  pages={1--6},
  year={2022},
  organization={IEEE}
}

@inproceedings{qu2024quantum,
  title={A quantum-classical hybrid optimization method for restoration of active distribution networks},
  author={Qu, Kaiping and Chen, Yue and Zhao, Changhong and Huang, Shihan and Xu, Yan},
  booktitle={2024 IEEE 8th Conference on Energy Internet and Energy System Integration (EI2)},
  pages={5374--5379},
  year={2024},
  organization={IEEE}
}

@article{yan2022multiagent,
  title={A multiagent quantum deep reinforcement learning method for distributed frequency control of islanded microgrids},
  author={Yan, Rudai and Wang, Yu and Xu, Yan and Dai, Jiahong},
  journal={IEEE Transactions on Control of Network Systems},
  volume={9},
  number={4},
  pages={1622--1632},
  year={2022},
  publisher={IEEE}
}

@inproceedings{babahajiani2022quantum,
  title={Quantum distributed microgrid control},
  author={Babahajiani, Pouya and Zhang, Peng},
  booktitle={2022 IEEE Power \& Energy Society General Meeting (PESGM)},
  pages={1--5},
  year={2022},
  organization={IEEE}
}

@article{babahajiani2022employing,
  title={Employing interacting qubits for distributed microgrid control},
  author={Babahajiani, Pouya and Zhang, Peng and Wei, Tzu-Chieh and Liu, Ji and Lu, Xiaonan},
  journal={IEEE Transactions on Power Systems},
  volume={38},
  number={4},
  pages={3123--3135},
  year={2023},
  publisher={IEEE}
}

@article{deng2023quantum,
  title={Quantum computing for future real-time building HVAC controls},
  author={Deng, Zhipeng and Wang, Xuezheng and Dong, Bing},
  journal={Applied Energy},
  volume={334},
  pages={120621},
  year={2023},
  publisher={Elsevier}
}

@article{gao2024quantum,
  title={Quantum grover search-inspired global maximum power point tracking for photovoltaic systems under partial shading conditions},
  author={Gao, Fang and Hu, Rongzhao and Yin, Linfei and Cao, Huibin and Yu, Jun and Shuang, Feng},
  journal={IEEE Transactions on Sustainable Energy},
  volume={15},
  number={3},
  pages={1601--1613},
  year={2024},
  publisher={IEEE}
}

@article{jing2024hhl,
  title={{HHL} algorithm with mapping function and enhanced sampling for model predictive control in microgrids},
  author={Jing, Hang and Li, Yan and Brandsema, Matthew J and Chen, Yousu and Yue, Meng},
  journal={Applied Energy},
  volume={361},
  pages={122878},
  year={2024},
  publisher={Elsevier}
}

@article{luo2024quantum,
  title={Quantum model prediction for frequency regulation of novel power systems which includes a high proportion of energy storage},
  author={Luo, Wenbo and Xu, Yufan and Du, Wanlin and Wang, Shilong and Fan, Ziwei},
  journal={Frontiers in Energy Research},
  volume={12},
  pages={1354262},
  year={2024},
  publisher={Frontiers Media SA}
}

@article{paterakis2025quantum,
  title={Quantum Computing in the Computational Landscape of Power Electronics: Vision and Reality},
  author={Paterakis, Nikolaos G and Karamanakos, Petros and O'Meara, Corey and Papafotiou, Georgios},
  journal={IEEE Transactions on Power Electronics},
  volume={41},
  number={8},
  pages={12478--12496},
  year={2026},
  publisher={IEEE}
}

@article{jahed2025quantum,
  title={Quantum Machine Learning for Secondary Frequency Control},
  author={Jahed, Younes Ghazagh and Khatiri, Alireza},
  journal={arXiv preprint arXiv:2512.02065},
  year={2025}
}

@inproceedings{liu2025cooperative,
  title={Cooperative Frequency Control of Microgrids Based on Quantum Reinforcement Learning},
  author={Liu, Wei and Zhang, Peng},
  booktitle={2025 4th Conference on Fully Actuated System Theory and Applications (FASTA)},
  pages={2115--2119},
  year={2025},
  organization={IEEE}
}

@inproceedings{sharmiladevi2026quantum,
  title={Quantum Annealing Algorithm Based Power Optimization for Electric Vehicle Charging Using a Buck Converter},
  author={Sharmiladevi, B and Bhuvanasundar, T and Bose, NP and Dharunkumar, J and Gokul, M},
  booktitle={2026 5th International Conference on Communication, Computing and Electronics Systems (ICCCES)},
  pages={758--763},
  year={2026},
  organization={IEEE}
}

@inproceedings{pan2026constrained,
  title={Constrained Quantum Approximate Optimization for Modular Multilevel Converter Cell Selection},
  author={Pan, Yitong and Li, Yan and Du, Liang and Ismail, Muhammad and Saad, Walid},
  booktitle={2026 IEEE Transportation Electrification Conference \& Expo (ITEC) \& Electric Aircraft Technologies Symposium (EATS)(ITEC+ EATS)},
  pages={1--5},
  year={2026},
  organization={IEEE}
}

@article{ren2024esqfl,
  title={{ESQFL}: Digital twin-driven explainable and secured quantum federated learning for voltage stability assessment in smart grids},
  author={Ren, Chao and Dong, Zhao Yang and Yu, Han and Xu, Minrui and Xiong, Zehui and Niyato, Dusit},
  journal={IEEE Journal of Selected Topics in Signal Processing},
  volume={18},
  pages={964--878},
  year={2024},
  publisher={IEEE}
}

@article{li2025study,
  title={A study on the dynamic optimization strategy of energy routers in zero-carbon ports based on digital twin technology},
  author={Li, Shun and Fan, Xingda and Qi, Zhaoyu},
  journal={International Journal of Electrical Power \& Energy Systems},
  volume={170},
  pages={110897},
  year={2025},
  publisher={Elsevier}
}

@inproceedings{saber2025potential,
  title={Potential of Quantum Computing Applications for Smart Grid Digital Twins and Future Directions},
  author={Saber, Ahmad Mohammad and Kundur, Deepa and Skorek, Adam W and others},
  booktitle={2025 IEEE PES Conference on Innovative Smart Grid Technologies-Middle East (ISGT Middle East)},
  pages={1--5},
  year={2025},
  organization={IEEE}
}

@article{hammadia2026optimal,
  title={Optimal Quantum Variational Circuits for Detecting False Data Injection Attacks on Transmission Substations},
  author={Hammadia, Taha and Saber, Ahmad Mohammad and Kassouf, Marthe and Kundur, Deepa and Khalaf, Mohsen},
  journal={IEEE Transactions on Industry Applications},
  year={2026},
  publisher={IEEE}
}

@article{naderi2025securing,
  title={Securing the future: Integrating quantum computing and digital twin technologies into modern power \& transportation systems for resilient smart cities against false data injection cyberattacks},
  author={Naderi, Ehsan},
  journal={International Journal of Critical Infrastructure Protection},
  pages={100807},
  year={2025},
  publisher={Elsevier}
}

@inproceedings{lemo2025towards,
  title={Towards a digital twin of medium-voltage circuit breakers using quantum algorithms and deep learning},
  author={Lemo, A and Thiam, M and Pierre, KI and Cossette, C and Skorek, AW},
  booktitle={2025 IEEE Canadian Conference on Electrical and Computer Engineering (CCECE)},
  pages={1--5},
  year={2025}
}

@article{ma2026quandt,
  title={QuanDT: Quantum Digital Twin with Applications to Smart Grid},
  author={Ma, Wenxuan and Liu, Mengxiang and Yu, Shang and Chen, Kuan-cheng and Zhou, Yitian and Guo, Rong and Liu, Yifan and Deng, Ruilong and Cheng, Peng and Chen, Jiming},
  year={2026}
}

@article{jiang2026digital,
  title={Digital twin system for collaborative optimization of reactive power and voltage of new energy grid-connected distribution based on quantum computing enhancement},
  author={Jiang, Chenda and Li, Jin},
  journal={AIP Advances},
  volume={16},
  number={2},
  year={2026},
  publisher={AIP Publishing}
}

@article{zhou2022noisy,
  title={Noisy-intermediate-scale quantum electromagnetic transients program},
  author={Zhou, Yifan and Zhang, Peng and Feng, Fei},
  journal={IEEE Transactions on Power Systems},
  volume={38},
  number={2},
  pages={1558--1571},
  year={2023},
  publisher={IEEE}
}

@inproceedings{tran2023applying,
  title={Applying Quantum Computing to Simulate Power System Dynamics' Differential-Algebraic Equations},
  author={Tran, Huynh TT and Nguyen, Hieu T and Vu, Long Thanh and Ojetola, Samuel T},
  booktitle={2023 North American Power Symposium (NAPS)},
  pages={1--6},
  year={2023},
  organization={IEEE}
}

@inproceedings{chanda2023architecture,
  title={Architecture for quantum-in-the loop real-time simulations for designing resilient smart grids},
  author={Chanda, Sayonsom and Mohanpurkar, Manish and Hovsapian, Rob},
  booktitle={2023 13th International Symposium on Advanced Topics in Electrical Engineering (ATEE)},
  pages={1--6},
  year={2023},
  organization={IEEE}
}

@inproceedings{vereno2023exploiting,
  title={Exploiting quantum power flow in smart grid co-simulation},
  author={Vereno, Dominik and Khodaei, Amin and Neureiter, Christian and Lehnhoff, Sebastian},
  booktitle={2023 11th Workshop on Modelling and Simulation of Cyber-Physical Energy Systems (MSCPES)},
  pages={1--6},
  year={2023},
  organization={IEEE}
}

@inproceedings{anjimoon2023quantum,
  title={Quantum Computing Approaches to Time-Domain Simulation of Electromagnetic Transients in Interconnected Power Systems},
  author={Anjimoon, Shaik and Basawaraiu, Swathi and Sobti, Rajeev and Kumar, Ashwani and Chauhan, Shilpi and Alkhafaji, Mohammed Ayad},
  booktitle={2023 International Conference on Power Energy, Environment \& Intelligent Control (PEEIC)},
  pages={461--466},
  year={2023},
  organization={IEEE}
}

@article{vereno2023quantum,
  title={Quantum--classical co-simulation for smart grids: A proof-of-concept study on feasibility and obstacles},
  author={Vereno, Dominik and Khodaei, Amin and Neureiter, Christian and Lehnhoff, Sebastian},
  journal={Energy Informatics},
  volume={6},
  number={Suppl 1},
  pages={25},
  year={2023},
  publisher={Springer}
}

@article{tran2024solving,
  title={Solving differential-algebraic equations in power system dynamic analysis with quantum computing},
  author={Tran, Huynh TT and Nguyen, Hieu T and Vu, Long T and Ojetola, Samuel T},
  journal={Energy Conversion and Economics},
  volume={5},
  number={1},
  pages={28--41},
  year={2024},
  publisher={Wiley Online Library}
}

@article{lou2025matrix,
  title={Matrix Low-dimensional Qubit Casting Based Quantum Electromagnetic Transient Network Simulation Program},
  author={Lou, Qi and Xu, Yijun and Gu, Wei},
  journal={arXiv preprint arXiv:2502.11728},
  year={2025}
}

@article{hartmann2025quantum,
  title={Quantum annealing based power grid partitioning for parallel simulation},
  author={Hartmann, Carsten and Zhang, Junjie and Calaza, Carlos D Gonzalez and Pesch, Thiemo and Michielsen, Kristel and Benigni, Andrea},
  journal={IEEE Transactions on Power Systems},
  volume={40},
  number={6},
  pages={4958--4970},
  year={2025},
  publisher={IEEE}
}

@inproceedings{lange2025quantum,
  title={Quantum Computing for Analyzing Microgrid Systems With Uncertainties},
  author={Lange, Joseph Maxwell and Chen, Jianing and Li, Yan and Du, Liang},
  booktitle={2025 IEEE/AIAA Transportation Electrification Conference and Electric Aircraft Technologies Symposium (ITEC+ EATS)},
  pages={1--5},
  year={2025},
  organization={IEEE}
}

@article{kaseb2026quantum,
  title={Quantum hardware-in-the-loop for optimal power flow in renewable-integrated power systems},
  author={Kaseb, Zeynab and Rane, Rahul and Leki{\'c}, Aleksandra and M{\"o}ller, Matthias and Khodaei, Amin and Palensky, Peter and Vergara, Pedro P},
  journal={IEEE Transactions on Power Systems},
  pages={1--1},
  year={2026},
  publisher={IEEE}
}

@article{soltaninia2025quantum,
  title={Quantum neural networks for solving power system transient simulation problem},
  author={Soltaninia, Mohammadreza and Zhan, Junpeng},
  journal={Energies},
  volume={18},
  number={10},
  pages={2525},
  year={2025},
  publisher={MDPI}
}

@article{saber2026quantum,
  title={Quantum-Enhanced Deep Learning for Resilient Cyberattack Detection in Smart Grids},
  author={Saber, Ahmad Mohammad and Jafari, Saeed and Kassouf, Marthe and Kundur, Deepa},
  journal={IEEE Transactions on Industrial Informatics},
  volume={22},
  number={7},
  pages={6140--6151},
  year={2026},
  publisher={IEEE}
}

@article{sakhnenko2022hybrid,
  title={Hybrid classical-quantum autoencoder for anomaly detection},
  author={Sakhnenko, Alona and O’Meara, Corey and Ghosh, Kumar JB and Mendl, Christian B and Cortiana, Giorgio and Bernab{\'e}-Moreno, Juan},
  journal={Quantum Machine Intelligence},
  volume={4},
  number={2},
  pages={27},
  year={2022},
  publisher={Springer}
}

@article{lakshmi2023quantum,
  title={A quantum-based approach for offensive security against cyber attacks in electrical infrastructure},
  author={Lakshmi, D and Nagpal, Neelu and Chandrasekaran, S and others},
  journal={Applied Soft Computing},
  volume={136},
  pages={110071},
  year={2023},
  publisher={Elsevier}
}

@article{said2023quantum,
  title={Quantum computing and machine learning for cybersecurity: Distributed denial of service (DDoS) attack detection on smart micro-grid},
  author={Said, Dhaou},
  journal={Energies},
  volume={16},
  number={8},
  pages={3572},
  year={2023},
  publisher={MDPI}
}

@article{said2024quantum,
  title={Quantum entropy and reinforcement learning for distributed denial of service attack detection in smart grid},
  author={Said, Dhaou and Bagaa, Miloud and Oukaira, Aziz and Lakhssassi, Ahmed},
  journal={IEEE Access},
  volume={12},
  pages={129858--129869},
  year={2024},
  publisher={IEEE}
}

@inproceedings{jafari2024quantum,
  title={Quantum Leaps: Dynamic Event Identification using Phasor Measurement Units in Power Systems},
  author={Jafari, Hossein and Aghababa, Hossein and Barati, Masoud},
  booktitle={2024 International Conference on Smart Grid Synchronized Measurements and Analytics (SGSMA)},
  pages={1--6},
  year={2024},
  organization={IEEE}
}

@inproceedings{hangun2025classical,
  title={A classical-quantum transfer learning model for Disturbance Detection in Power Systems},
  author={Hangun, Batuhan and Akpinar, Emine and Altun, Oguz and Eyecioglu, Onder},
  booktitle={2025 13th International Conference on Smart Grid (icSmartGrid)},
  pages={700--705},
  year={2025},
  organization={IEEE}
}

@article{addo2025federated,
  title={Federated quantum machine learning for distributed cybersecurity in multi-agent energy systems},
  author={Addo, Kwabena and Kabeya, Musasa and Ojo, Evans Eshiemogie},
  journal={Energies},
  volume={18},
  number={20},
  pages={5418},
  year={2025},
  publisher={MDPI}
}

@inproceedings{nguyen2024integrating,
  title={Integrating Quantum LSTM and Quantum Detector for Forecasting and Anomaly Detection for Load Data in Low-Voltage Power Distribution Networks},
  author={Nguyen, Thanh-Hoan and Truong, Viet-Anh and Nguyen, Huu-Vinh and Le, Tien-Thuong and Nguyen, Phuoc-Tin and Nguyen, Thanh-Duy},
  booktitle={International Conference on Intelligent and Fuzzy Systems},
  pages={750--757},
  year={2024},
  organization={Springer}
}

@article{blazakis2025power,
  title={Power theft detection in smart grids using quantum machine learning},
  author={Blazakis, Konstantinos and Schetakis, Nikolaos and Badr, Mahmoud M and Aghamalyan, Davit and Stavrakakis, Konstantinos and Stavrakakis, Georgios},
  journal={IEEE Access},
  volume={13},
  pages={61511--61525},
  year={2025},
  publisher={IEEE}
}

@article{cirillo2025quantum,
  title={Quantum Autoencoder-Based Anomaly Detection for Cyberattacks in Smart Power Systems},
  author={Cirillo, Franco and Esposito, Christian},
  year={2025}
}

@article{ogiesoba2025quantum,
  title={Quantum Machine Learning Approaches for Coordinated Stealth Attack Detection in Distributed Generation Systems},
  author={Ogiesoba-Eguakun, Osasumwen Cedric and Rath, Suman},
  journal={arXiv preprint arXiv:2601.00873},
  year={2025}
}

@inproceedings{hammadia2025quantum,
  title={Quantum variational circuits for detection of false data injection against power transformers},
  author={Hammadia, Taha and Saber, Ahmad Mohammad and Kundur, Deepa},
  booktitle={2025 IEEE Industry Applications Society Annual Meeting (IAS)},
  pages={1--8},
  year={2025},
  organization={IEEE}
}

@inproceedings{phassadawongse2025quantum,
  title={Quantum-Classical Dual Kernel SVMs for Power Quality Classification},
  author={Phassadawongse, Dhana and Turner, Stephen John},
  booktitle={International Conference on Computational Science},
  pages={165--179},
  year={2025},
  organization={Springer}
}

@article{mutua2025quantum,
  title={Quantum-enhanced battery anomaly detection in smart transportation systems},
  author={Mutua, Alexander Mutiso and de Frein, Ruairi},
  journal={Applied Sciences},
  volume={15},
  number={17},
  pages={9452},
  year={2025},
  publisher={MDPI}
}

@inproceedings{sivakumar2025real,
  title={Real-Time Quantum Machine Learning-Based Anomaly Detection for Lithium-Ion Battery Packs and Identification of Defective Cell},
  author={Sivakumar, P and Jagadeesh, V and Sundararajan, G},
  booktitle={2025 7th International Conference on Energy, Power and Environment (ICEPE)},
  pages={1--6},
  year={2025},
  organization={IEEE}
}

@article{wu2025stochastic,
  title={Stochastic Power Flow and Transmission System Vulnerability Analysis Against FDIAs based on Quantum Computing},
  author={Wu, Zhuoheng},
  year={2025}
}

@inproceedings{yao2025synchrophasor,
  title={Synchrophasor Data Anomaly Detection Via Quantum Generative Adversarial Networks},
  author={Yao, Yongbing and Xu, Yijun and Gu, Wei and Mili, Lamine and Lu, Shuai and Zheng, Zongsheng},
  booktitle={2025 IEEE China International Youth Conference on Electrical Engineering (CIYCEE)},
  pages={1--6},
  year={2025},
  organization={IEEE}
}

@inproceedings{mohapatra2025voltage,
  title={Voltage Sag Analysis Using Quantum Computing Based on Gram Matrix Method},
  author={Mohapatra, Bhabasis and Dash, Ritesh and Sahu, Binod Kumar and Sharma, Renu},
  booktitle={2025 International Conference on Innovations in Intelligent Systems: Advancements in Computing, Communication, and Cybersecurity (ISAC3)},
  pages={1--6},
  year={2025},
  organization={IEEE}
}

@article{huang2026aligning,
  title={Aligning quantum kernels for detecting false data injection attacks in power systems},
  author={Huang, Bin and Wang, Jianhui and Huang, Xiaoge},
  journal={Applied Energy},
  volume={407},
  pages={127332},
  year={2026},
  publisher={Elsevier}
}

@article{ngo2026qupid,
  title={QUPID: A Partitioned Quantum Neural Network for Anomaly Detection in Smart Grid},
  author={Ngo, Hoang M and Jeter, Tre'R and Seo, Jung Taek and Thai, My T},
  journal={arXiv preprint arXiv:2601.11500},
  year={2026}
}

@article{li2025quantum,
  title={A quantum neural network-based approach to power quality disturbances detection and recognition},
  author={Li, Guo-Dong and He, Hai-Yan and Li, Yue and Li, Xin-Hao and Liu, Hao and Wang, Qing-Le and Cheng, Long},
  journal={Physica Scripta},
  volume={100},
  number={7},
  pages={075102},
  year={2025},
  publisher={IOP Publishing}
}

@article{li2025hybrid,
  title={Hybrid Quantum-Classical Convolutional Neural Network for Detection and Identification of Power Quality Disturbance},
  author={Li, Yue and Li, Xinhao and Jia, Haopeng and Liu, Anjiang and Wang, Qingle and Hao, Shuqing and Liu, Hao},
  journal={IET Quantum Communication},
  volume={6},
  number={1},
  pages={e70013},
  year={2025},
  publisher={Wiley Online Library}
}

@article{perdomo2015quantum,
  title={A quantum annealing approach for fault detection and diagnosis of graph-based systems},
  author={Perdomo-Ortiz, Alejandro and Fluegemann, Joseph and Narasimhan, Sriram and Biswas, Rupak and Smelyanskiy, Vadim N},
  journal={The European Physical Journal Special Topics},
  volume={224},
  number={1},
  pages={131--148},
  year={2015},
  publisher={Springer}
}

@incollection{ajagekar2021fault,
  title={Fault diagnosis of electrical power systems with hybrid quantum-classical deep learning},
  author={Ajagekar, Akshay and You, Fengqi},
  booktitle={Computer Aided Chemical Engineering},
  volume={50},
  pages={1173--1179},
  year={2021},
  publisher={Elsevier}
}

@inproceedings{uehara2021quantum,
  title={Quantum neural network parameter estimation for photovoltaic fault detection},
  author={Uehara, Glen and Rao, Sunil and Dobson, Mathew and Tepedelenlioglu, Cihan and Spanias, Andreas},
  booktitle={2021 12th International Conference on Information, Intelligence, Systems \& Applications (IISA)},
  pages={1--7},
  year={2021},
  organization={IEEE}
}

@article{xie2024fault,
  title={Fault location in active distribution network under hybrid classical quantum computing architecture},
  author={Xie, Yuzhe and Yang, Xiaoting and Zhu, Qinran and Bi, Zhongqin},
  journal={IEEE Access},
  volume={12},
  pages={163924--163937},
  year={2024},
  publisher={IEEE}
}

@inproceedings{gbashi2024hybrid,
  title={Hybrid quantum convolutional neural network for defect detection in a wind turbine gearbox},
  author={Gbashi, Samuel M and Olatunji, Obafemi O and Adedeji, Paul A and Madushele, Nkosinathi},
  booktitle={2024 IEEE PES/IAS PowerAfrica},
  pages={01--06},
  year={2024},
  organization={IEEE}
}

@article{fei2024power,
  title={Power system fault diagnosis with quantum computing and efficient gate decomposition},
  author={Fei, Xiang and Zhao, Huan and Zhou, Xiyuan and Zhao, Junhua and Shu, Ting and Wen, Fushuan},
  journal={Scientific reports},
  volume={14},
  number={1},
  pages={16991},
  year={2024},
  publisher={Nature Publishing Group UK London}
}

@article{he2025power,
  title={A Power Transformer Fault Diagnosis Method Based on Improved Variational Quantum Shadow Learning},
  author={He, Hongying and Yu, Jiangchun and Lee, Wei-Jen and Luo, Diansheng and Liang, Wenju},
  journal={IEEE Transactions on Power Delivery},
  volume={40},
  number={3},
  pages={1331--1343},
  year={2025},
  publisher={IEEE}
}

@article{lin2025quantumfault,
  title={Quantum parallel Transformer model for detecting multi-class insulator faults},
  author={Lin, Yijin and Yin, Linfei},
  journal={IEEE Transactions on Instrumentation and Measurement},
  pages={1--1},
  year={2025},
  publisher={IEEE}
}

@article{prasad2025weather,
  title={Weather-Aware Fault Prediction and Budgeted Sensing for Power Distribution: A Hybrid {RL} and Quantum Optimization Approach},
  author={Prasad, Yalla Jnan Devi Satya and Mahadev, Tejaswi},
  journal={Authorea Preprints},
  year={2025},
  publisher={Authorea}
}

@inproceedings{zhang2025wind,
  title={Wind Turbine Fault Detection Using Quantum Long-Short Term Memory Network},
  author={Zhang, Zhefeng and Ma, Xiandong},
  booktitle={2025 30th International Conference on Automation and Computing (ICAC)},
  pages={1--6},
  year={2025},
  organization={IEEE}
}

@article{bowden2026machine,
  title={Machine Failure Detection Based on Projected Quantum Models},
  author={Bowden, Larry and Chu, Qi and Cena, Bernard and Ohno, Kentaro and Parney, Bob and Sharma, Deepak and Takeori, Mitsuharu},
  journal={arXiv preprint arXiv:2601.15641},
  year={2026}
}

@article{li2026qftd,
  title={QFTD: An efficient quantum federated learning for transformer fault diagnosis with minimal gated unit in smart grid},
  author={Li, Guodong and Luo, Junjie and Wang, Qingle and Liu, Lin and Wei, Chunyan and Wang, Huawei and Zhang, Zhichao},
  journal={Engineering Applications of Artificial Intelligence},
  volume={163},
  pages={112974},
  year={2026},
  publisher={Elsevier}
}

@article{haris2021lifetime,
  title={Lifetime Prediction of {LiFePO4} Batteries Using Multilayer Classical-Quantum Hybrid Classifier},
  author={Haris, Muhammad and Hasan, Muhammad Noman and Basit, Abdul and Qin, Shiyin},
  journal={Journal of Quantum Computing},
  volume={3},
  number={3},
  pages={89},
  year={2021},
  publisher={Tech Science Press}
}

@inproceedings{ngo2023quantum,
  title={A quantum neural network regression for modeling lithium-ion battery capacity degradation},
  author={Ngo, Anh Phuong and Le, Nhat and Nguyen, Hieu T and Eroglu, Abdullah and Nguyen, Duong T},
  booktitle={2023 IEEE Green Technologies Conference (GreenTech)},
  pages={164--168},
  year={2023},
  organization={IEEE}
}

@inproceedings{akash2023quantum,
  title={Quantum annealing-based machine learning for battery health monitoring robust to adversarial attacks},
  author={Akash, Alve Rahman and Khot, Amey and Kim, Taesic},
  booktitle={2023 IEEE Energy Conversion Congress and Exposition (ECCE)},
  pages={6307--6311},
  year={2023},
  organization={IEEE}
}

@inproceedings{grandhi2024quantum,
  title={A quantum variational classifier for predictive maintenance and monitoring of battery health in electric vehicles},
  author={Grandhi, Sri Harsha and Al-Jawahry, Hassan M and Kumar, Bura Vijay and Padhi, Manoj Kumar and others},
  booktitle={2024 International Conference on Intelligent Algorithms for Computational Intelligence Systems (IACIS)},
  pages={1--4},
  year={2024},
  organization={IEEE}
}

@article{wang2024lithium,
  title={Lithium-ion battery state of health estimation method based on variational quantum algorithm optimized stacking strategy},
  author={Wang, Longze and Jiang, Siyu and Mao, Yuteng and Li, Zhehan and Zhang, Yan and Li, Meicheng},
  journal={Energy Reports},
  volume={11},
  pages={2877--2891},
  year={2024},
  publisher={Elsevier}
}

@article{soon2025hybrid,
  title={A hybrid quantum neural network and classical gated recurrent unit for battery state of health forecasting incorporating SHAP analysis},
  author={Soon, Kian Lun and Soon, Lam Tatt},
  journal={Journal of Energy Storage},
  volume={136},
  pages={118596},
  year={2025},
  publisher={Elsevier}
}

@article{soon2025quantum2,
  title={A quantum-enhanced ensemble model to forecast battery state of health under varying discharging load},
  author={Soon, Kian Lun and Lai, Nai Shyan and Chow, Chee-Onn and Kanesan, Jeevan and Yen, Kin Sam and Soon, Lam Tatt},
  journal={Measurement},
  pages={118318},
  year={2025},
  publisher={Elsevier}
}

@article{sheilla2025framework,
  title={Framework for Early Prediction of Lithium-Ion Battery Lifetime: A Hybrid Quantum-Classical Approach},
  author={Sheilla Rully, Anggita and Muhamad, Akrom},
  journal={Journal of Multiscale Materials Informatics},
  volume={2},
  number={2},
  pages={40--47},
  year={2025}
}

@inproceedings{mutua2025predicting,
  title={Predicting SoH in Lithium-ion Batteries using a Variational Quantum Neural Network},
  author={Mutua, Alexander Mutiso and De Fr{\'e}in, Ruair{\'\i} and Kimeli, Kangogo},
  booktitle={2025 International Conference on Computer, Information and Telecommunication Systems (CITS)},
  pages={1--8},
  year={2025},
  organization={IEEE}
}

@article{mutiso2025quantum,
  title={Quantum Machine Learning for Battery Health and Thermal Risk Prediction},
  author={Mutiso Mutua, Alexander and de Fr{\'e}in, Ruair{\'\i}},
  year={2025},
  publisher={Technological University Dublin}
}

@inproceedings{khot2025quantum,
  title={Quantum Restricted Boltzmann Machines-Based Feature Selection for Electric Vehicle Battery Health Monitoring},
  author={Khot, Ameya and Kim, Taesic and Akash, Alve R},
  booktitle={2025 IEEE/AIAA Transportation Electrification Conference and Electric Aircraft Technologies Symposium (ITEC+ EATS)},
  pages={1--5},
  year={2025},
  organization={IEEE}
}

@inproceedings{tran2025quantum,
  title={Quantum Soft Actor-Critic-Based Energy Management Strategy for Fuel Cell Hybrid Electric Vehicles},
  author={Tran, Van-Linh and Ahn, Kyoung Kwan},
  booktitle={2025 28th International Conference on Mechatronics Technology (ICMT)},
  pages={67--70},
  year={2025},
  organization={IEEE}
}

@article{wang2025quantum,
  title={Quantum State Estimation for Real-Time Battery Health Monitoring in Photovoltaic Storage Systems},
  author={Wang, Dawei and Wang, Liyong and Zhang, Baoqun and Liu, Chang and Zhao, Yongliang and Luo, Shanna and Feng, Jun},
  journal={Energies},
  volume={18},
  number={11},
  pages={2727},
  year={2025},
  publisher={MDPI}
}

@inproceedings{khot2025quantum2,
  title={Quantum-Inspired Machine Learning for Energy Efficient and Secure Battery Health Monitoring},
  author={Khot, Ameya and Kim, Taesic and Akash, Alve R and Kim, Chris H and Moy, William},
  booktitle={2025 IEEE Power \& Energy Society General Meeting (PESGM)},
  pages={1--5},
  year={2025},
  organization={IEEE}
}

@inproceedings{wang2025state,
  title={State of Health Prediction for Lithium-Ion Battery Pack Using a Shared Embedding Layer-Based Quantum Long Short-Term Memory with Transfer Learning},
  author={Wang, Fu-Kwun and Kebede, Getnet Awoke and Woldegiorgis, Bereket Haile},
  booktitle={Journal of Physics: Conference Series},
  volume={2968},
  number={1},
  pages={012010},
  year={2025},
  organization={IOP Publishing}
}

@article{liang2025stochastic,
  title={Stochastic state of health estimation for lithium-ion batteries with automated feature fusion using quantum convolutional neural network},
  author={Liang, Chen and Tao, Shengyu and Huang, Xinghao and Wang, Yezhen and Xia, Bizhong and Zhang, Xuan},
  journal={Journal of Energy Chemistry},
  volume={106},
  pages={205--219},
  year={2025},
  publisher={Elsevier}
}

@article{komarsofla2026quantum,
  title={Quantum machine learning approaches to state-of-health prediction and optimization in energy storage devices},
  author={Komarsofla, Mojtaba Khakpour and Kiani, Amirkianoosh},
  journal={Journal of Energy Storage},
  volume={153},
  pages={120939},
  year={2026},
  publisher={Elsevier}
}

@article{sushmit2023forecasting,
  title={Forecasting solar irradiance with hybrid classical--quantum models: A comprehensive evaluation of deep learning and quantum-enhanced techniques},
  author={Sushmit, Mushrafi Munim and Mahbubul, Islam Mohammed},
  journal={Energy Conversion and Management},
  volume={294},
  pages={117555},
  year={2023},
  publisher={Elsevier}
}

@article{yu2023prediction,
  title={Prediction of solar irradiance one hour ahead based on quantum long short-term memory network},
  author={Yu, Yunjun and Hu, Guoping and Liu, Caicheng and Xiong, Junjie and Wu, Ziyang},
  journal={IEEE Transactions on Quantum Engineering},
  volume={4},
  pages={1--15},
  year={2023},
  publisher={IEEE}
}

@article{jeong2024short,
  title={Short-term photovoltaic power forecasting based on hybrid quantum gated recurrent unit},
  author={Jeong, Seon-Geun and Do, Quang Vinh and Hwang, Won-Joo},
  journal={Ict Express},
  volume={10},
  number={3},
  pages={608--613},
  year={2024},
  publisher={Elsevier}
}

@article{oliveira2024application,
  title={Application of quantum neural network for solar irradiance forecasting: A case study using the Folsom Dataset, California},
  author={Oliveira Santos, Victor and Marinho, Felipe Pinto and Costa Rocha, Paulo Alexandre and Th{\'e}, Jesse Van Griensven and Gharabaghi, Bahram},
  journal={Energies},
  volume={17},
  number={14},
  pages={3580},
  year={2024},
  publisher={MDPI}
}

@article{khan2024quantum,
  title={Quantum long short-term memory (QLSTM) vs. classical LSTM in time series forecasting: a comparative study in solar power forecasting},
  author={Khan, Saad Zafar and Muzammil, Nazeefa and Ghafoor, Salman and Khan, Haibat and Zaidi, Syed Mohammad Hasan and Aljohani, Abdulah Jeza and Aziz, Imran},
  journal={Frontiers in Physics},
  volume={12},
  pages={1439180},
  year={2024},
  publisher={Frontiers Media SA}
}

@inproceedings{phan2025hybrid,
  title={Hybrid Long-Short Term Memory with Variational Quantum Eigensolver for Photovoltaic Power Forecasting: A Novel Approach},
  author={Phan, Ha-Vu and Phan, Quoc-Thang and Wu, Yuan-Kang and Phan, Quoc Dung},
  booktitle={2025 IEEE Industry Applications Society Annual Meeting (IAS)},
  pages={1--6},
  year={2025},
  organization={IEEE}
}

@inproceedings{phan2025modified,
  title={Modified Quantum Long-Short Term Memory with Variational Quantum Circuits for PV Power Forecasting},
  author={Phan, Ha-Vu and Pham, Tan-Hung and Tran, Khang B and Phan, Quoc-Thang and Phan, Quoc Dung and Wu, Yuan-Kang},
  booktitle={2025 IEEE Industry Applications Society Annual Meeting (IAS)},
  pages={1--7},
  year={2025},
  organization={IEEE}
}

@article{sagingalieva2025photovoltaic,
  title={Photovoltaic power forecasting using quantum machine learning},
  author={Sagingalieva, Asel and Komornyik, Stefan and Senokosov, Arsenii and Joshi, Ayush and Mansell, Christopher and Tsurkan, Olga and Pinto, Karan and Pflitsch, Markus and Melnikov, Alexey},
  journal={Solar Energy},
  volume={302},
  pages={114016},
  year={2025},
  publisher={Elsevier}
}

@article{mechiche2025quantum,
  title={Quantum Fourier Transform Based Kernel for Solar Irrandiance Forecasting},
  author={Mechiche-Alami, Nawfel and Rodriguez, Eduardo and Cardemil, Jose M and Droguett, Enrique Lopez},
  journal={arXiv preprint arXiv:2511.17698},
  year={2025}
}

@article{wang5336951quantum,
  title={Quantum-Enhanced Forecasting and Optimal Scheduling for Energy Management in Photovoltaic Greenhouses},
  author={Weng, Zhenhao and Xu, Zhanpeng and Li, Ruichong and Wang, Longze and Li, Zhehan and Zhang, Yan and Li, Meicheng},
  journal={Available at SSRN: https://ssrn.com/abstract=5382333}
}

@article{hong2023robust,
  title={A robust hybrid classical and quantum model for short-term wind speed forecasting},
  author={Hong, Ying-Yi and Arce, Christine Joy E and Huang, Tsung-Wei},
  journal={IEEE Access},
  volume={11},
  pages={90811--90824},
  year={2023},
  publisher={IEEE}
}

@inproceedings{hangun2024hybrid,
  title={A hybrid quantum-classical machine learning approach to offshore wind farm power forecasting},
  author={Hangun, Batuhan and Akpinar, Emine and Oduncuoglu, Murat and Altun, Oguz and Eyecioglu, Onder},
  booktitle={2024 13th international Conference on renewable energy research and applications (ICRERA)},
  pages={1105--1110},
  year={2024},
  organization={IEEE}
}

@article{pires2025quantum,
  title={A quantum neural network model for short term wind speed forecasting using weather data},
  author={Pires, Otto Menegasso and Nascimento, Erick Giovani Sperandio and Moret, Marcelo A},
  journal={Energy and AI},
  pages={100588},
  year={2025},
  publisher={Elsevier}
}

@inproceedings{hangun2025comparative,
  title={Comparative Analysis of QNN Architectures for Wind Power Prediction: Feature Maps and Ansatz Configurations},
  author={Hangun, Batuhan and Akpinar, Emine and Altun, Oguz and Eyecioglu, Onder},
  booktitle={2025 IEEE Computer Society Annual Symposium on VLSI (ISVLSI)},
  volume={1},
  pages={1--6},
  year={2025},
  organization={IEEE}
}

@article{hong2025implementing,
  title={Implementing a Hybrid Quantum Neural Network for Wind Speed Forecasting: Insights from Quantum Simulator Experiences},
  author={Hong, Ying-Yi and Santos, Jay Bhie D},
  journal={Energies},
  volume={18},
  number={7},
  pages={1771},
  year={2025},
  publisher={MDPI}
}

@article{hangun2025quantum,
  title={Quantum Neural Networks for Wind Energy Forecasting: A Comparative Study of Performance and Scalability with Classical Models},
  author={Hangun, Batuhan and Altun, Oguz and Eyecioglu, Onder},
  journal={arXiv preprint arXiv:2506.22845},
  year={2025}
}

@article{kumar2026wind,
  title={Wind speed forecasting using hybrid classical-quantum-infused model},
  author={Kumar, Ajay and Singh, AJ and Kumar, Sanjay},
  journal={Electrical Engineering},
  volume={108},
  number={2},
  pages={96},
  year={2026},
  publisher={Springer}
}

@article{kumar2023quantum,
  title={A quantum controlled-not neural network-based load forecast and management model for smart grid},
  author={Kumar, Jatinder and Saxena, Deepika and Singh, Ashutosh Kumar and Vasilakos, Athanasios V},
  journal={IEEE Systems Journal},
  volume={17},
  number={4},
  pages={5714--5725},
  year={2023},
  publisher={IEEE}
}

@inproceedings{arvanitidis2023quantum,
  title={A quantum machine learning methodology for precise appliance identification in smart grids},
  author={Arvanitidis, Athanasios Ioannis and Valdez, Luis Arturo and Alamaniotis, Miltiadis},
  booktitle={2023 14th International Conference on Information, Intelligence, Systems \& Applications (IISA)},
  pages={1--6},
  year={2023},
  organization={IEEE}
}

@article{lalindeautomatic,
  title={Automatic Electrical Meter Forecasting: a Benchmarking Between Quantum Machine Learning and Classical Machine learning},
  author = {Montes C., Jonathan J. and Sierra-So, Danie and Lalinde-Pulido, Juan G},
  year={2024},
  pages={1--46},
  organization={{EAFIT} University}
}

@article{nutakki2024quantum,
  title={Quantum support vector machine for forecasting house energy consumption: a comparative study with deep learning models},
  author={Nutakki, Mounica and Koduru, Suprabhath and Mandava, Srihari and others},
  journal={Journal of Cloud Computing},
  volume={13},
  number={1},
  pages={1--12},
  year={2024},
  publisher={Springer}
}

@article{habibi2025electrical,
  title={Electrical load forecasting in power systems based on quantum computing using time series-based quantum artificial intelligence},
  author={Habibi, Mohammad Reza and Golestan, Saeed and Wu, Yanpeng and Guerrero, Josep M and Vasquez, Juan C},
  journal={Scientific Reports},
  volume={15},
  number={1},
  pages={7429},
  year={2025},
  publisher={Nature Publishing Group UK London}
}

@article{wang5114742power,
  title={Power Load Time Series Forecasting Based on Quantum Annealing Optimized Quantum Neural Network},
  author={Wang, Baonan and Fan, Youdian and Li, Fengyong and Zhang, Weina and Zhang, Dan},
  journal={Available at SSRN 5114742}
}

@article{zhang2025privacy,
  title={Privacy-Shielded Federated Quantum Learning Approach for Charging Load Forecasting},
  author={Zhang, Yanyu and Zhao, Ke and Jiao, Feixiang and Li, Hengji and Ning, Nianwen and Zhang, Xibeng and Zhou, Yi},
  journal={IEEE Transactions on Industry Applications},
  volume={62},
  number={2},
  pages={7429},
  year={2026},
  publisher={IEEE}
}

@article{dash2025dynamic,
  title={Dynamic modeling and Quantum-Enhanced forecasting of Multi-Seasonal energy prices in simulated microgrid environments},
  author={Dash, Ritesh and Sinha, Anupa and Reddy, K Jyotheeswara and Dhanamjayulu, C and Kamwa, Innocent},
  journal={IEEE access},
  year={2025},
  publisher={IEEE}
}

@article{feng2021quantum,
  title={Quantum power flow},
  author={Feng, Fei and Zhou, Yifan and Zhang, Peng},
  journal={IEEE Transactions on Power Systems},
  volume={36},
  number={4},
  pages={3810--3812},
  year={2021},
  publisher={IEEE}
}

@article{feng2023noise,
  title={Noise-resilient quantum power flow},
  author={Feng, Fei and Zhou, Yi-Fan and Zhang, Peng},
  journal={iEnergy},
  volume={2},
  number={1},
  pages={63--70},
  year={2023},
  publisher={TUP}
}

@article{gao2023solving,
  title={Solving DC power flow problems using quantum and hybrid algorithms},
  author={Gao, Fang and Wu, Guojian and Guo, Suhang and Dai, Wei and Shuang, Feng},
  journal={Applied Soft Computing},
  volume={137},
  pages={110147},
  year={2023},
  publisher={Elsevier}
}

@inproceedings{kaseb2024adiabatic,
  title={Adiabatic Computing for Power Flow Analysis},
  author={Kaseb, Zeynab and M{\"o}ller, Matthias and Kirsch, Markus and Palensky, Peter and Vergara, Pedro P},
  booktitle={2024 IEEE International Conference on Quantum Computing and Engineering (QCE)},
  volume={2},
  pages={543--544},
  year={2024},
  organization={IEEE}
}

@inproceedings{zheng2024early,
  title={Early exploration of a flexible framework for efficient quantum linear solvers in power systems},
  author={Zheng, Muqing and Chen, Yousu and Yang, Xiu and Li, Ang},
  booktitle={2024 IEEE Power \& Energy Society General Meeting (PESGM)},
  pages={1--5},
  year={2024},
  organization={IEEE}
}

@inproceedings{kaseb2024hybrid,
  title={Hybrid Quantum Physics-Informed Neural Networks for Power Flow Analysis},
  author={Kaseb, Zeynab and Palensky, Peter and Vergara, Pedro P},
  booktitle={2024 IEEE PES Innovative Smart Grid Technologies Europe (ISGT EUROPE)},
  pages={1--5},
  year={2024},
  organization={IEEE}
}

@inproceedings{el2024newton,
  title={Newton-Raphson Method Using {HHL} Algorithm for Power Flow Quantum Computing},
  author={El-Khatib, Ziad and Moussa, Sherif},
  booktitle={2024 Third International Conference on Sustainable Mobility Applications, Renewables and Technology (SMART)},
  pages={1--4},
  year={2024},
  organization={IEEE}
}

@article{kaseb2024power,
  title={Power flow analysis using quantum and digital annealers: a discrete combinatorial optimization approach},
  author={Kaseb, Zeynab and M{\"o}ller, Matthias and Vergara, Pedro P and Palensky, Peter},
  journal={Scientific reports},
  volume={14},
  number={1},
  pages={23216},
  year={2024},
  publisher={Nature Publishing Group UK London}
}

@article{kaseb2024quantum,
  title={Quantum neural networks for power flow analysis},
  author={Kaseb, Zeynab and M{\"o}ller, Matthias and Balducci, Giorgio Tosti and Palensky, Peter and Vergara, Pedro P},
  journal={Electric Power Systems Research},
  volume={235},
  pages={110677},
  year={2024},
  publisher={Elsevier}
}

@article{liu2024quantum,
  title={Quantum power flows: From theory to practice},
  author={Liu, Junyu and Zheng, Han and Hanada, Masanori and Setia, Kanav and Wu, Dan},
  journal={Quantum Machine Intelligence},
  volume={6},
  number={2},
  pages={55},
  year={2024},
  publisher={Springer}
}

@article{zheng2025early,
  title={An early investigation of the {HHL} quantum linear solver for scientific applications},
  author={Zheng, Muqing and Liu, Chenxu and Stein, Samuel and Li, Xiangyu and M{\"u}lmenst{\"a}dt, Johannes and Chen, Yousu and Li, Ang},
  journal={Algorithms},
  volume={18},
  number={8},
  pages={491},
  year={2025},
  publisher={MDPI}
}

@article{zhu2025bayesian,
  title={Bayesian Quantum Neural Network for Renewable-Rich Power Flow with Training Efficiency and Generalization Capability Improvements},
  author={Zhu, Ziqing and Zhu, Shuyang and Bu, Siqi},
  journal={IEEE Transactions on Power Systems},
  year={2025},
  publisher={IEEE}
}

@inproceedings{kaseb2025combinatorial,
  title={Combinatorial Power Flow Analysis using Adiabatic Quantum Algorithms},
  author={Kaseb, Zeynab and M{\"o}ller, Matthias and Kirsch, Markus and Palensky, Peter and Vergara, Pedro P},
  booktitle={2025 IEEE Kiel PowerTech},
  pages={1--6},
  year={2025},
  organization={IEEE}
}

@inproceedings{le2025learning,
  title={Learning AC Power Flow Solutions using a Data-Dependent Variational Quantum Circuit},
  author={Le, Thinh Viet and Rahman, Md Obaidur and Kekatos, Vassilis},
  booktitle={2025 IEEE International Conference on Communications, Control, and Computing Technologies for Smart Grids (SmartGridComm)},
  pages={1--7},
  year={2025},
  organization={IEEE}
}

@article{pareek2025limitations,
  title={Limitations of Fault-Tolerant Quantum Linear System Solvers for Quantum Power Flow},
  author={Pareek, Parikshit and Jayakumar, Abhijith and Coffrin, Carleton and Misra, Sidhant},
  journal={IEEE Transactions on Power Systems},
  volume={41},
  number={2},
  pages={811--820},
  year={2025},
  publisher={IEEE}
}

@article{kaseb2025quantumRL,
  title={Quantum-Enhanced Reinforcement Learning for Accelerating Newton-Raphson Convergence with Ising Machines: A Case Study for Power Flow Analysis},
  author={Kaseb, Zeynab and Moller, Matthias and Spoor, Lindsay and Guo, Jerry J and Xiang, Yu and Palensky, Peter and Vergara, Pedro P},
  journal={arXiv preprint arXiv:2511.20237},
  year={2025}
}

@article{fan2026quantum,
  title={Quantum Newtonian Power Flow},
  author={Fan, Ruoyan and Lin, Chaofan and Zhang, Peng},
  journal={IEEE Transactions on Power Systems},
  volume={41},
  number={1},
  pages={777--780},
  year={2026},
  publisher={IEEE}
}

@article{morstyn2022annealing,
  title={Annealing-based quantum computing for combinatorial optimal power flow},
  author={Morstyn, Thomas},
  journal={IEEE Transactions on Smart Grid},
  volume={14},
  number={2},
  pages={1093--1102},
  year={2022},
  publisher={IEEE}
}

@inproceedings{amani2023quantum,
  title={Quantum-enhanced dc optimal power flow},
  author={Amani, Farshad and Mahroo, Reza and Kargarian, Amin},
  booktitle={2023 IEEE Texas Power and Energy Conference (TPEC)},
  pages={1--6},
  year={2023},
  organization={IEEE}
}

@inproceedings{cao2024differentially,
  title={A differentially private quantum neural network for probabilistic optimal power flow},
  author={Cao, Yuji and Chen, Yue and Xu, Yan},
  booktitle={4th Energy Conversion and Economics Annual Forum (ECE Forum 2024)},
  volume={2024},
  pages={601--608},
  year={2024},
  organization={IET}
}

@inproceedings{amani2024quantum,
  title={Quantum-inspired optimal power flow},
  author={Amani, Farshad and Kargarian, Amin},
  booktitle={2024 IEEE Texas Power and Energy Conference (TPEC)},
  pages={1--6},
  year={2024},
  organization={IEEE}
}

@article{hu2024advancing,
  title={Advancing hybrid quantum neural network for alternative current optimal power flow},
  author={Hu, Ze and Zhu, Ziqing and Zhu, Linghua and Wei, Xiang and Bu, Siqi and Chan, Ka Wing},
  journal={arXiv preprint arXiv:2410.20275},
  year={2024}
}

@article{prasad2025constraint,
  title={Constraint-Preserving QAOA for Real-Time Optimal Power Flow in Renewable-Rich Distribution Networks},
  author={Prasad, Yalla Jnan Devi Satya and Chatrati, Sahil Ram and Masthan, Shaik Fathima},
  journal={Authorea Preprints},
  year={2025},
  publisher={Authorea}
}

@article{amani2025optimal,
  title={Optimal power flow solution via noise-resilient quantum interior-point methods},
  author={Amani, Farshad and Kargarian, Amin},
  journal={Electric Power Systems Research},
  volume={240},
  pages={111216},
  year={2025},
  publisher={Elsevier}
}

@inproceedings{hafshejani2025quantum,
  title={Quantum algorithms for optimal power flow},
  author={Hafshejani, S Fathi and Uddin, Md Mohsin and Neufeld, David and Benkoczi, Robert and Gaur, Daya},
  booktitle={2025 15th International Conference on Power, Energy, and Electrical Engineering (CPEEE)},
  pages={295--300},
  year={2025},
  organization={IEEE}
}

@inproceedings{magar2025quantum,
  title={Quantum Inspired DC Optimal Power Flow Under Frequency Constraint},
  author={Magar, Madan Rana and Nguyen, Nga},
  booktitle={2025 IEEE Texas Power and Energy Conference (TPEC)},
  pages={1--6},
  year={2025},
  organization={IEEE}
}

@inproceedings{carrillo2025quantum,
  title={Quantum optimization approach based on {Qibo} framework for DC-OPF},
  author={Carrillo-Mu{\~n}oz, Marc and Martinez-Hermida, Sergio and Barja-Martinez, Sara and Salda{\~n}a-Gonz{\'a}lez, Antonio E and Pe{\~n}alba, M{\'o}nica Arag{\"u}{\'e}s},
  booktitle={2025 IEEE PES Innovative Smart Grid Technologies Conference Europe (ISGT Europe)},
  pages={1--5},
  year={2025},
  organization={IEEE}
}

@article{amani2025quantum,
  title={Quantum Optimization for Optimal Power Flow: CVQLS-Augmented Interior Point Method},
  author={Amani, Farshad and Kargarian, Amin},
  journal={IEEE Transactions on Smart Grid},
  volume={16},
  number={6},
  pages={5040--5052},
  year={2025},
  publisher={IEEE}
}

@article{le2026solving,
  title={Solving Conic Programs over Sparse Graphs using a Variational Quantum Approach: The Case of the Optimal Power Flow},
  author={Le, Thinh Viet and Wilde, Mark M and Kekatos, Vassilis},
  journal={arXiv preprint arXiv:2509.00341},
  year={2026}
}

@inproceedings{koretsky2021adapting,
  title={Adapting quantum approximation optimization algorithm (QAOA) for unit commitment},
  author={Koretsky, Samantha and Gokhale, Pranav and Baker, Jonathan M and Viszlai, Joshua and Zheng, Honghao and Gurung, Niroj and Burg, Ryan and Paaso, Esa Aleksi and Khodaei, Amin and Eskandarpour, Rozhin and others},
  booktitle={2021 IEEE International Conference on Quantum Computing and Engineering (QCE)},
  pages={181--187},
  year={2021},
  organization={IEEE}
}

@article{nikmehr2022quantumUC,
  title={Quantum distributed unit commitment: An application in microgrids},
  author={Nikmehr, Nima and Zhang, Peng and Bragin, Mikhail A},
  journal={IEEE transactions on power systems},
  volume={37},
  number={5},
  pages={3592--3603},
  year={2022},
  publisher={IEEE}
}

@article{paterakis2023hybrid,
  title={Hybrid quantum-classical multi-cut benders approach with a power system application},
  author={Paterakis, Nikolaos G},
  journal={Computers \& Chemical Engineering},
  volume={172},
  pages={108161},
  year={2023},
  publisher={Elsevier}
}

@article{mahroo2023learning,
  title={Learning infused quantum-classical distributed optimization technique for power generation scheduling},
  author={Mahroo, Reza and Kargarian, Amin},
  journal={IEEE Transactions on Quantum Engineering},
  volume={4},
  pages={1--14},
  year={2023},
  publisher={IEEE}
}

@article{feng2022novel,
  title={Novel resolution of unit commitment problems through quantum surrogate Lagrangian relaxation},
  author={Feng, Fei and Zhang, Peng and Bragin, Mikhail A and Zhou, Yifan},
  journal={IEEE Transactions on Power Systems},
  volume={38},
  number={3},
  pages={2460--2471},
  year={2023},
  publisher={IEEE}
}

@article{zheng2024fast,
  title={A fast quantum algorithm for searching the quasi-optimal solutions of unit commitment},
  author={Zheng, Xiaodong and Wang, Jianhui and Yue, Meng},
  journal={IEEE Transactions on Power Systems},
  volume={39},
  number={2},
  pages={4755--4758},
  year={2024},
  publisher={IEEE}
}

@article{salgado2024hybrid,
  title={A hybrid classical-quantum approach to highly constrained unit commitment problems},
  author={Salgado, Bruna and Sequeira, Andr{\'e} and Santos, Luis Paulo},
  journal={arXiv preprint arXiv:2412.11312},
  year={2024}
}

@inproceedings{magar2024dc,
  title={{DC} optimal power flow in unit commitment using quantum computing: An {ADMM} approach},
  author={Magar, Madan Rana and Pandit, Dilip and Nguyen, Duong and Nguyen, Nga},
  booktitle={2024 56th North American Power Symposium (NAPS)},
  pages={1--6},
  year={2024},
  organization={IEEE}
}

@inproceedings{christeson2024quantum,
  title={Quantum-compatible unit commitment modeling through logarithmic discretization},
  author={Christeson, Tyler and Khodaei, Amin and Eskandarpour, Rozhin},
  booktitle={2024 IEEE PES Innovative Smart Grid Technologies Europe (ISGT EUROPE)},
  pages={1--5},
  year={2024},
  organization={IEEE}
}

%\vspace{-30pt}
%\begin{IEEEbiography}[{\includegraphics[width=1.0in,height=1.25in,clip,keepaspectratio]{Figures/ullah-mdhabib.jpg}}]
%{Md Habib Ullah} (Member, IEEE) is currently an Assistant Professor of Electrical Engineering at Penn State Harrisburg, Middletown, PA, USA. He received the B.S. degree from the Ahsanullah University of Science and Technology, Dhaka, Bangladesh, in 2012, the M.S. degree from South Dakota State University, Brookings, SD, USA, in 2016, and the Ph.D. degree in electrical engineering from the University of Colorado Denver, Denver, CO, USA, in 2021. He was a Postdoctoral Research Associate in the Department of Electrical and Computer Engineering at the University of Denver, Denver, CO, USA, and an Affiliate Faculty member at Metropolitan State University of Denver, Denver, CO, USA. He currently serves as an Associate Editor for IEEE Transactions on Industrial Informatics. 

%His research interests include artificial intelligence, quantum computing, optimization and control, transactive energy management, and energy markets. 
%\end{IEEEbiography}

\end{document}